\documentclass[12pt, preprint]{aastex}
\usepackage{graphicx, subfigure, lscape, float}
\usepackage{natbib}
\shorttitle{The Solar Neighborhood {\it XXXII}}
\shortauthors{Dieterich et al.}
\begin{document}

\title{The Solar Neighborhood {\it XXXII}: The Hydrogen Burning Limit\footnote{Based in part on observations
 obtained at the Southern Astrophysical Research (SOAR) telescope, which is a
 joint project of the Minist\'{e}rio da Ci\^{e}ncia, Tecnologia, e Inova\c{c}\~{a}o (MCTI)
 da Rep\'{u}blica Federativa do Brasil, the U.S. National Optical Astronomy Observatory (NOAO),
the University of North Carolina at Chapel Hill (UNC), and Michigan State University (MSU).} \footnote{
Based in part on observations obtained via the Cerro Tololo Inter-American Observatory Parallax
Investigation ({\it CTIOPI}), at the Cerro Tololo 0.9m telescope. {\it CTIOPI} began under the
auspices of the NOAO Surveys Program in 1999, and continues via the SMARTS Consortium.}}

\author{Sergio B. Dieterich, Todd J. Henry, Wei-Chun Jao, Jennifer G. Winters,
Altonio D. Hosey}
\affil{Georgia State University, Atlanta, GA 30302-4106;
 Visiting astronomer, Cerro Tololo Inter-American Observatory}
\email{dieterich@chara.gsu.edu}

\author{Adric R. Riedel}
\affil{American Museum of Natural History, New York, NY 10024; 
Visiting astronomer, Cerro Tololo Inter-American Observatory}

\author{John P. Subasavage}
\affil{United States Naval Observatory, Flagstaff, AZ 86001;
 Visiting astronomer, Cerro Tololo Inter-American Observatory}

\begin{abstract}
We construct a Hertzsprung-Russell diagram for the stellar/substellar
boundary based on a sample of 63 objects ranging in spectral type from
M6V to L4. We report newly observed {\it VRI} photometry for all 63
objects and new trigonometric parallaxes for 37 objects. The remaining
26 objects have trigonometric parallaxes from the literature.  We
combine our optical photometry and trigonometric parallaxes with {\it
2MASS} and {\it WISE} photometry and employ a novel {\it SED} fitting
algorithm to determine effective temperatures, bolometric
luminosities, and radii.  Our uncertainties range from $\sim$20K to
$\sim$150K in temperature, $\sim$0.01 to $\sim$0.06 in
$log(L/L_{\odot}$) and $\sim$3\% to $\sim$10\% in radius. We check
our methodology by comparing our calculated radii to radii directly
measured via long baseline optical interferometry.  We find evidence
 for the local minimum in the radius$-$temperature and
radius$-$luminosity trends that signals the end of the stellar main
sequence and the start of the brown dwarf sequence at
$T_{eff}\sim2075K$, $log(L/L_{\sun})\sim-3.9$, and
$(R/R_{\sun})\sim0.086$. The existence of this local minimum is
predicted by evolutionary models, but at temperatures $\sim$400K
cooler. The minimum radius happens near the locus of 2MASS J0523-1403,
an L2.5 dwarf with $V-K=9.42$. We make qualitative arguments as to
why the effects of
the recent revision in solar abundances accounts for the discrepancy
between our findings and the evolutionary models. We also report new
color-absolute magnitude relations for optical and infrared colors
useful for estimating photometric distances. We study the optical
variability of all 63 targets and find an overall variability fraction
of 36$^{+9}_{-7}$\% at a threshold of 15 milli-magnitudes in the {\it I} band,
 in agreement with previous studies.
\end{abstract}


\section{Introduction}
\label{sec:intro}

The first comprehensive stellar structure and evolution models for the
low mass end of the main sequence were published in the late 20$^{th}$
century \citep[e.g.,][]{Burrowsetal1993,Baraffeetal1995}. While the
predictions of these models are widely accepted today, they remain
largely unconstrained by observations. The problem is particularly
noteworthy when it comes to the issue of distinguishing stellar
objects from the substellar brown dwarfs.  While the internal physics
of stars and brown dwarfs is  different, their atmospheric
properties overlap in the late M and early L spectral types, thus
making them difficult to distinguish based on photometric and
spectroscopic features alone. One test used to identify substellar
objects $-$ the lithium test \citep{Reboloetal1992}$-$ relies on the
fact that lithium undergoes nuclear burning at temperatures slightly
lower than hydrogen, and therefore should be totally consumed in fully
convective hydrogen burning objects at time scales $\ll$ than their
evolutionary time scales. Detection of the Li$\lambda$6708 line would
therefore signal the substellar nature of an object. This is a
powerful test, but it fails us when we most need it.  While
evolutionary models predict the minimal stellar mass to be anywhere
from 0.070$M_{\sun}$ to 0.077$M_{\sun}$ (see $\S$\ref{subsec:models}),
the lithium test only works for masses $\lesssim$0.060$M_{\sun}$ due
to the lower mass at which core temperatures are sufficient to
fuse lithium.

The models for low mass stars and brown dwarfs in current usage
\citep{Burrowsetal1993,Burrowsetal1997,Baraffeetal1998,
Chabrieretal2000,Baraffeetal2003} predict the end of the stellar main
sequence at temperatures ranging from 1550$-$1750K, corresponding
roughly to spectral type L4. These models have achieved varying
degrees of success, but as we discuss in $\S$\ref{subsec:models}, are
mutually inconsistent when it comes to determining the properties of
the smallest possible star. The inconsistency is not surprising given
that none of these decade-old evolutionary models incorporates the
state-of-the-art in  atmospheric models, nor do they account for the
recent 22\% downward revision in solar abundances
\citep{Caffauetal2011}, which are in agreement with the results
of solar astero-seismology\footnote{A review of the history of revisions
to solar abundances, including issues related to solar astero-seismology, is
given in \citet{Allardetal2013}.}.

 Over the last ten years few changes were made to
models for {\it VLM} stars and brown
dwarfs in large part because the  models provide predictions that are not directly
observable. Whereas an atmospheric model can be fully tested against
an observed spectrum, testing an evolutionary model requires accurate
knowledge of mass, age, and metallicity as well as an accurate
atmospheric model that serves as a boundary condition.

The problem of understanding the stellar/substellar boundary can
essentially be formulated by posing two questions. The first one is:
{\it ``What do objects close to either side of the stellar/substellar
boundary look like to an observer?''} The second question is: {\it
``What are the masses and other structural parameters  of objects on either side of the
stellar/substellar boundary?''} While it is the second question that
usually gets the most attention, we note that any attempt to determine
masses at the stellar/substellar boundary assumes an inherently model
dependent (and therefore possibly flawed) answer to the first
question. What is needed is an observational test that relies as
little on evolutionary models as possible.  Because brown dwarfs are
supported by electron degeneracy pressure, their internal physics is
fundamentally different from that of stars. One manifestation of this
difference is the reversal of the mass$-$radius relation at the
hydrogen burning limit
\citep{Chabrieretal2009,Burrowsetal2011}. Whereas for stars radius
increases as a function of mass, the opposite is true for brown
dwarfs. The result is a pronounced minimum in radius at the hydrogen
burning limit (see $\S$\ref{sec:discussion}).

To reveal the end of the stellar main sequence by identifying the minimum in the radius trend,
we construct a {\it bona fide} {\it HR} diagram for the
stellar/substellar boundary based on wide photometric coverage from
$\sim0.4$ $\micron$ to $\sim$ 17$\micron$, trigonometric parallaxes,
and the new {\it BT-Settl} model atmospheres
\citep{Allardetal2012,Allardetal2013}, which have been shown to be
highly accurate for M and L dwarfs (see $\S$\ref{sec:tefflum}).  We
also employ a new custom-made iterative Spectral Energy Distribution
({\it SED}) fitting code that interpolates between model grid points to
determine effective temperatures and performs small adjustments to the
{\it SED} templates based on observed photometry to better determine
luminosities.  The results of our calculations are effective temperatures
($T_{eff}$), luminosities ($log(L/L_{\sun})$), and radii
($R/R_{\sun}$), which we then use to construct the {\it HR} diagram as
well as temperature$-$radius and luminosity$-$radius diagrams. The
latter two diagrams provide the same essential information as the {\it
HR} diagram, but facilitate the inspection of radius trends.

The paper is organized as follows. We include all of our observed and
derived quantities in Table 1. The data presented in this table form the
basis for subsequent discussions in the paper\footnote{An expanded version of 
Table 1 that also includes {\it 2MASS} and
{\it WISE} photometry is available in machine-readable format as an online supplement.}.
We discuss our observed sample in $\S$\ref{sec:sample} and
discuss the methodology of our photometric and astrometric
observations in $\S$\ref{sec:photometricobs}
 and $\S$\ref{sec:astrometricobs} respectively. We discuss our {\it SED} fitting algorithms
and check their results against radii measured with long baseline
optical interferometry in $\S$\ref{sec:tefflum}. We discuss our new optical photometric results,
trigonometric parallaxes, effective temperatures, color-magnitude
relations, and optical variability in $\S\S$6.1$-$6.5.  We discuss the
newly discovered astrometric binary DENIS J1454-6604AB in
$\S$\ref{subsec:den1454}.  In $\S$\ref{sec:discussion} we discuss the
end of the stellar main sequence based on radius trends. We discuss
individual objects in $\S$\ref{sec:individual} and make concluding
remarks and discuss future work in $\S$\ref{sec:conclusion}.

\clearpage
\addtolength{\voffset}{1in}
\addtolength{\hoffset}{-0.5in}
\textheight=8.0in
\textwidth=7.6in
\begin{deluxetable}{ccclccccrclrlllcccccl}
\tabletypesize{\scriptsize}
\rotate
\tablecaption{Observed and Derived Properties}
\setlength{\tabcolsep}{0.02in}
\label{tab:master}
\tablewidth{0pt}
\tablehead{
	   \colhead{ID }                          &
	   \colhead{R. A.}               	  &
	   \colhead{Dec.}               	  &
	   \colhead{Name}                   	  &
	   \colhead{Spct.}               	  &
	   \colhead{ref.\tablenotemark{a,b}}  	  &
	   \colhead{$\mu$}                  	  &
	   \colhead{P.A.}                   	  &
	   \colhead{Parallax}               	  &
	   \colhead{ref.\tablenotemark{b}}  	  &
	   \colhead{Distance}               	  &
	   \colhead{ V$_{tan}$}              	  &
	   \colhead{{\it V}}            	  &
	   \colhead{{\it R}}            	  &
	   \colhead{{\it I}}            	  &
	   \colhead{Tel.\tablenotemark{c}}	  &
           \colhead{{\it VRI}}              	  &
	   \colhead{$T_{eff}$  }              	  &
	   \colhead{Luminosity}                   &
	   \colhead{Radius}                       &
           \colhead{Notes}                        \\
	   \colhead{   }                          &
	   \colhead{2000}                	  &
	   \colhead{2000}                    	  &
	   \colhead{   }                    	  &
	   \colhead{Type}                    	  &
	   \colhead{   }                 	  &
	   \colhead{$\arcsec$/yr}             	  &
	   \colhead{deg }               	  &
	   \colhead{mas}                	  &
	   \colhead{   }                 	  &
	   \colhead{pc}                 	  &
	   \colhead{$Km/s$}              	  &
	   \colhead{   }                	  &
	   \colhead{   }                    	  &
	   \colhead{   }                    	  &
	   \colhead{   }                       	  &
           \colhead{Epochs}                 	  &
	   \colhead{K  }                 	  &
	   \colhead{$Log(L/L_{\odot})$}           &
	   \colhead{$R_{\odot}$}                  &
           \colhead{   }                          \\
	   \colhead{(1)}               &
	   \colhead{(2)}	              &
	   \colhead{(3)}	              &
	   \colhead{(4)}	              &
	   \colhead{(5)}	              &
	   \colhead{(6)}	              &
	   \colhead{(7)}	              &
	   \colhead{(8)}	              &
	   \colhead{(9)}	              &
	   \colhead{(10)}              &
	   \colhead{(11)}              &
	   \colhead{(12)}              &
	   \colhead{(13)}              &
	   \colhead{(14)}              &
	   \colhead{(15)}              &
	   \colhead{(16)}              &
           \colhead{(17)}              &
	   \colhead{(18)}              &
	   \colhead{(19)}              &
	   \colhead{(20)}              &
           \colhead{(21)}              }
\startdata
1  &  00:04:34.9 &   $-$40:44:06 &  GJ 1001BC          &   L4.5  &   14     &     1.643 &  156.0  &   77.02$\pm$2.07 &     1      &   13.15$^{+0.36}_{-0.34}$ & 102.4 &  22.77$\pm$.025 &  19.24$\pm$.045 &  16.76$\pm$.012 &  S &  2  &    1725$\pm$21  &  -4.049$\pm$.048 &   0.105$\pm$.005   & d,e,f    \\        
2  &  00:21:05.8 &   $-$42:44:49 &  LEHPM1-0494B       &   M9.5  &    9     &     0.253 &  089.1  &   39.77$\pm$2.10 &     1      &   25.14$^{+1.40}_{-1.26}$ &  30.1 &  21.61$\pm$.085 &  19.08$\pm$.010 &  16.69$\pm$.001 &  S &  2  &    2305$\pm$57	 &  -3.506$\pm$.046 &   0.110$\pm$.008   & d,e    \\     
3  &  00:21:10.7 &   $-$42:45:40 &  LEHPM1-0494A       &   M6.0  &   24     &     0.253 &  086.5  &   37.20$\pm$1.99 &     1      &   26.88$^{+1.51}_{-1.36}$ &  32.2 &  17.28$\pm$.045 &  15.67$\pm$.005 &  13.84$\pm$.005 &  S &  2  &    2918$\pm$21	 &  -2.818$\pm$.046 &   0.152$\pm$.008   & d,e    \\       
4  &  00:24:24.6 &   $-$01:58:20 &  BRI B0021-0214     &   M9.5  &   20     &     0.155 &  328.8  &   86.60$\pm$4.00 &     3      &   11.54$^{+0.55}_{-0.50}$ &   8.4 &  20.01$\pm$.051 &  17.45$\pm$.016 &  15.00$\pm$.025 &  S &  1  &    2315$\pm$54	 &  -3.505$\pm$.042 &   0.109$\pm$.007   & \nodata    \\     
5  &  00:36:16.0 &   $+$18:21:10 &  2MASS J0036+1821   &   L3.5  &   10     &     0.907 &  082.4  &  114.20$\pm$0.80 &    11      &    8.75$^{+0.06}_{-0.06}$ &  37.6 &  21.43$\pm$.024 &  18.32$\pm$.016 &  15.92$\pm$.022 &  S &  1  &    1796$\pm$33	 &  -3.950$\pm$.011 &   0.109$\pm$.004   & \nodata    \\     
6  &  00:52:54.7 &   $-$27:06:00 &  RG 0050-2722       &   M9.0  &   18     &     0.098 &  026.0  &   41.00$\pm$4.00 &     4      &   24.39$^{+2.63}_{-2.16}$ &  11.3 &  21.54$\pm$.051 &  19.14$\pm$.017 &  16.57$\pm$.026 &  S &  1  &    2402$\pm$34	 &  -3.599$\pm$.078 &   0.091$\pm$.008   & \nodata    \\     
7  &  01:02:51.2 &   $-$37:37:44 &  LHS  132           &   M8.0  &   22     &     1.479 &  079.8  &   81.95$\pm$2.73 &    17      &   12.20$^{+0.42}_{-0.39}$ &  85.5 &  18.53$\pm$.021 &  16.30$\pm$.008 &  13.88$\pm$.012 &  C &  3  &    2513$\pm$29	 &  -3.194$\pm$.030 &   0.133$\pm$.005   & \nodata    \\     
8  &  02:48:41.0 &   $-$16:51:22 &  LP 771-021         &   M8.0  &   15     &     0.274 &  175.7  &   61.60$\pm$5.40 &     5      &   16.23$^{+1.55}_{-1.30}$ &  21.0 &  19.97$\pm$.050 &  17.70$\pm$.015 &  15.27$\pm$.015 &  S &  2  &    2512$\pm$19	 &  -3.507$\pm$.076 &   0.093$\pm$.008   & \nodata    \\     
9  &  02:53:00.5 &   $+$16:52:58 &  SO0253+1652        &   M6.5  &   21     &     5.050 &  137.9  &  259.41$\pm$0.89 &    21,30   &    3.85$^{+0.01}_{-0.01}$ &  92.2 &  15.14$\pm$.006 &  13.03$\pm$.004 &  10.65$\pm$.003 &  C &  3  &    2656$\pm$37	 &  -3.137$\pm$.013 &   0.127$\pm$.004   & \nodata    \\     
10 &  03:06:11.5 &   $-$36:47:53 &  DENIS J0306-3647   &   M8.5  &   18     &     0.690 &  196.0  &   76.46$\pm$1.42 &     1      &   13.07$^{+0.24}_{-0.23}$ &  42.7 &  19.38$\pm$.002 &  16.98$\pm$.023 &  14.49$\pm$.009 &  C &  2  &    2502$\pm$40	 &  -3.366$\pm$.017 &   0.110$\pm$.004   & \nodata    \\     
11 &  03:39:35.2 &   $-$35:25:44 &  LP 944-020         &   M9.0  &    7     &     0.408 &  048.5  &  155.89$\pm$1.03 &     1      &    6.41$^{+0.04}_{-0.04}$ &  12.4 &  18.70$\pm$.026 &  16.39$\pm$.007 &  14.01$\pm$.011 &  C &  3  &    2312$\pm$71	 &  -3.579$\pm$.010 &   0.101$\pm$.006   & \nodata    \\     
12 &  03:51:00.0 &   $-$00:52:45 &  LHS 1604           &   M7.5  &   15     &     0.526 &  176.0  &   68.10$\pm$1.80 &     4      &   14.68$^{+0.39}_{-0.37}$ &  36.6 &  18.11$\pm$.053 &  16.08$\pm$.020 &  13.80$\pm$.017 &  C &  2  &    \nodata        &  \nodata         &   \nodata          & d     \\
13 &  04:28:50.9 &   $-$22:53:22 &  2MASS J0428-2253   &   L0.5  &   16     &     0.189 &  038.1  &   38.48$\pm$1.85 &     1      &   25.98$^{+1.31}_{-1.19}$ &  23.2 &  21.68$\pm$.050 &  19.18$\pm$.025 &  16.79$\pm$.017 &  S &  2  &    2212$\pm$57	 &  -3.441$\pm$.042 &   0.129$\pm$.009   & \nodata    \\     
14 &  04:35:16.1 &   $-$16:06:57 &  LP 775-031         &   M7.0  &   25     &     0.356 &  028.0  &   95.35$\pm$1.06 &     1      &   10.48$^{+0.11}_{-0.11}$ &  17.6 &  17.67$\pm$.005 &  15.49$\pm$.029 &  13.08$\pm$.030 &  C &  2  &    2532$\pm$25	 &  -3.033$\pm$.012 &   0.157$\pm$.003   & \nodata    \\     
15 &  04:51:00.9 &   $-$34:02:15 &  2MASS J0451-3402   &   L0.5  &   15     &     0.158 &  036.5  &   47.46$\pm$1.51 &     1      &   21.07$^{+0.69}_{-0.64}$ &  15.7 &  22.11$\pm$.052 &  19.38$\pm$.015 &  16.84$\pm$.024 &  S &  1  &    2146$\pm$41	 &  -3.676$\pm$.029 &   0.104$\pm$.005   & d    \\     
16 &  05:00:21.0 &   $+$03:30:50 &  2MASS J0500+0330   &   L4.0  &   29     &     0.350 &  177.9  &   73.85$\pm$1.98 &     1      &   13.54$^{+0.37}_{-0.35}$ &  22.4 &  23.01$\pm$.037 &  19.77$\pm$.026 &  17.32$\pm$.032 &  S &  1  &    1783$\pm$19	 &  -4.010$\pm$.024 &   0.103$\pm$.003   & \nodata    \\     
17 &  05:23:38.2 &   $-$14:03:02 &  2MASS J0523-1403   &   L2.5  &   15     &     0.195 &  032.5  &   80.95$\pm$1.76 &     1      &   12.35$^{+0.27}_{-0.26}$ &  11.4 &  21.05$\pm$.112 &  18.71$\pm$.021 &  16.52$\pm$.012 &  C &  2  &    2074$\pm$27	 &  -3.898$\pm$.021 &   0.086$\pm$.003   & d    \\     
18 &  06:52:19.7 &   $-$25:34:50 &  DENIS J0652-2534   &   L0.0  &   28     &     0.250 &  289.3  &   63.76$\pm$0.94 &     1      &   15.68$^{+0.23}_{-0.22}$ &  18.5 &  20.77$\pm$.050 &  18.38$\pm$.005 &  15.85$\pm$.016 &  S &  2  &    2313$\pm$56	 &  -3.600$\pm$.015 &   0.098$\pm$.005   & \nodata    \\     
19 &  07:07:53.3 &   $-$49:00:50 &  ESO 207-61         &   M8.0  &   31     &     0.405 &  005.0  &   60.93$\pm$3.02 &     4      &   16.41$^{+0.85}_{-0.77}$ &  31.5 &  21.09$\pm$.035 &  18.74$\pm$.023 &  16.19$\pm$.032 &  S &  1  &    2403$\pm$31	 &  -3.625$\pm$.039 &   0.088$\pm$.004   & \nodata    \\     
20 &  07:46:42.5 &   $+$20:00:32 &  2MASS J0746+2000AB &   L0.0J &   33     &     0.377 &  261.9  &   81.84$\pm$0.30 &   11,34,35 &   12.21$^{+0.04}_{-0.04}$ &  21.8 &  20.05$\pm$.038 &  17.42$\pm$.037 &  14.90$\pm$.038 &  S &  1  &    2310$\pm$51	 &  -3.413$\pm$.009 &   0.122$\pm$.005   & f    \\     
21 &  07:51:16.4 &   $-$25:30:43 &  DENIS J0751-2530   &   L2.5  &   28     &     0.889 &  279.2  &   59.15$\pm$0.84 &     1      &   16.90$^{+0.24}_{-0.23}$ &  71.2 &  21.66$\pm$.045 &  18.86$\pm$.020 &  16.39$\pm$.005 &  S &  2  &    2186$\pm$32	 &  -3.732$\pm$.013 &   0.094$\pm$.003   & \nodata    \\     
22 &  08:12:31.7 &   $-$24:44:42 &  DENIS J0812-2444   &   L1.5  &   28     &     0.196 &  135.5  &   45.47$\pm$0.96 &     1      &   21.99$^{+0.47}_{-0.45}$ &  20.4 &  21.89$\pm$.053 &  19.45$\pm$.016 &  17.05$\pm$.025 &  S &  1  &    2295$\pm$47	 &  -3.696$\pm$.021 &   0.089$\pm$.004   & \nodata    \\     
23 &  08:28:34.1 &   $-$13:09:19 &  SSSPM J0829-1309   &   L1.0  &   28     &     0.578 &  273.0  &   87.96$\pm$0.78 &     1      &   11.36$^{+0.10}_{-0.09}$ &  31.1 &  21.19$\pm$.023 &  18.41$\pm$.025 &  16.01$\pm$.026 &  S &  1  &    2117$\pm$37	 &  -3.845$\pm$.011 &   0.088$\pm$.003   & d    \\     
24 &  08:29:49.3 &   $+$26:46:33 &  GJ 1111            &   M6.5  &   31     &     1.290 &  242.2  &  275.80$\pm$3.00 &     4      &    3.62$^{+0.03}_{-0.03}$ &  22.1 &  14.94$\pm$.033 &  12.88$\pm$.005 &  10.58$\pm$.018 &  C &  2  &    2690$\pm$27	 &  -3.107$\pm$.022 &   0.128$\pm$.004   & \nodata    \\     
25 &  08:40:29.7 &   $+$18:24:09 &  GJ 316.1           &   M6.0  &   32     &     0.908 &  240.0  &   71.10$\pm$1.00 &     4      &   14.06$^{+0.20}_{-0.19}$ &  60.5 &  17.67$\pm$.042 &  15.72$\pm$.017 &  13.44$\pm$.006 &  C &  2  &    2683$\pm$30	 &  -3.039$\pm$.013 &   0.139$\pm$.003   & \nodata    \\     
26 &  08:47:28.7 &   $-$15:32:37 &  2MASS J0847-1532   &   L2.0  &   15     &     0.240 &  146.1  &   58.96$\pm$0.99 &     1      &   16.96$^{+0.28}_{-0.28}$ &  19.2 &  21.93$\pm$.068 &  19.16$\pm$.028 &  16.86$\pm$.024 &  C &  2  &    1922$\pm$66	 &  -3.798$\pm$.017 &   0.113$\pm$.008   & \nodata    \\     
27 &  08:53:36.0 &   $-$03:29:28 &  LHS 2065           &   M9.0  &   31     &     0.550 &  249.4  &  117.98$\pm$0.76 &     1      &    8.47$^{+0.05}_{-0.05}$ &  22.0 &  18.94$\pm$.032 &  16.74$\pm$.015 &  14.44$\pm$.029 &  C &  3  &    2324$\pm$27	 &  -3.516$\pm$.010 &   0.107$\pm$.002   & \nodata    \\     
28 &  09:00:23.6 &   $+$21:50:04 &  LHS 2090           &   M6.0  &   21     &     0.774 &  221.2  &  156.87$\pm$2.67 &    21      &    6.37$^{+0.11}_{-0.10}$ &  23.3 &  16.11$\pm$.032 &  14.12$\pm$.020 &  11.84$\pm$.010 &  C &  3  &    2680$\pm$24	 &  -3.084$\pm$.016 &   0.132$\pm$.003   & \nodata    \\     
29 &  09:49:22.2 &   $+$08:06:45 &  LHS 2195           &   M8.0  &    6     &     0.887 &  177.4  &   60.32$\pm$1.67 &     1      &   16.57$^{+0.47}_{-0.44}$ &  69.7 &  19.76$\pm$.152 &  17.66$\pm$.035 &  15.20$\pm$.036 &  C &  1  &    2481$\pm$36	 &  -3.399$\pm$.025 &   0.107$\pm$.004   & \nodata    \\     
30 &  10:48:12.8 &   $-$11:20:11 &  LHS 292            &   M6.0  &   31     &     1.645 &  158.0  &  220.30$\pm$3.60 &     4      &    4.53$^{+0.07}_{-0.07}$ &  35.3 &  15.78$\pm$.057 &  13.63$\pm$.002 &  11.25$\pm$.025 &  C &  3  &    2588$\pm$32	 &  -3.166$\pm$.016 &   0.129$\pm$.004   & \nodata    \\     
31 &  10:49:03.4 &   $+$05:02:23 &  LHS 2314           &   M6.0  &    2     &     0.624 &  217.0  &   41.10$\pm$2.30 &     4      &   24.33$^{+1.44}_{-1.28}$ &  71.9 &  19.14$\pm$.033 &  17.13$\pm$.033 &  14.91$\pm$.025 &  C &  2  &    2691$\pm$13	 &  -3.169$\pm$.049 &   0.119$\pm$.006   & \nodata    \\     
32 &  10:56:29.2 &   $+$07:00:53 &  GJ 406             &   M6.0  &   31     &     4.696 &  235.0  &  419.10$\pm$2.10 &     4      &    2.38$^{+0.01}_{-0.01}$ &  53.1 &  13.58$\pm$.008 &  11.64$\pm$.028 &   9.44$\pm$.014 &  C &  2  &    2700$\pm$56	 &  -3.036$\pm$.044 &   0.138$\pm$.009   & \nodata    \\     
33 &  10:58:47.9 &   $-$15:48:17 &  DENIS J1058-1548   &   L3.0  &   12     &     0.290 &  288.1  &   57.70$\pm$1.00 &   11,34,35 &   17.33$^{+0.30}_{-0.29}$ &  23.8 &  23.01$\pm$.005 &  20.01$\pm$.045 &  17.66$\pm$.027 &  S &  2  &    1804$\pm$13	 &  -3.997$\pm$.019 &   0.102$\pm$.002   & \nodata    \\     
34 &  11:06:18.9 &   $+$04:28:32 &  LHS 2351           &   M7.0  &  \nodata &     0.460 &  129.1  &   48.1$\pm$3.1   &     5      &   20.79$^{+1.43}_{-1.25}$ &  45.3 &  19.49$\pm$.049 &  17.27$\pm$.017 &  14.87$\pm$.017 &  C &  2  &    2619$\pm$27	 &  -3.218$\pm$.056 &   0.119$\pm$.008   & \nodata    \\     
35 &  11:21:49.0 &   $-$13:13:08 &  LHS 2397aAB        &   M8.5J &    8     &     0.507 &  264.7  &   65.83$\pm$2.02 &     1      &   15.19$^{+0.48}_{-0.45}$ &  36.5 &  19.43$\pm$.036 &  17.33$\pm$.048 &  14.84$\pm$.040 &  S &  1  &    2376$\pm$25	 &  -3.291$\pm$.028 &   0.133$\pm$.005   & d,f  \\     
36 &  11:26:39.9 &   $-$50:03:55 &  2MASS J1126-5003   &   L4.5  &   27     &     1.646 &  286.2  &   59.38$\pm$1.64 &     1      &   16.84$^{+0.47}_{-0.45}$ & 131.3 &  23.75$\pm$.010 &  20.11$\pm$.020 &  17.51$\pm$.005 &  S &  2  &    1797$\pm$49	 &  -4.035$\pm$.025 &   0.098$\pm$.006   & \nodata    \\     
37 &  11:53:52.7 &   $+$06:59:56 &  LHS 2471           &   M6.5  &  \nodata &     0.955 &  160.0  &   70.30$\pm$2.60 &     4      &   14.22$^{+0.54}_{-0.50}$ &  64.3 &  18.10$\pm$.009 &  16.02$\pm$.035 &  13.77$\pm$.005 &  C &  2  &    2611$\pm$22	 &  -3.113$\pm$.032 &   0.135$\pm$.005   & \nodata    \\     
38 &  11:55:42.9 &   $-$22:24:58 &  LP 851-346         &   M7.5  &   18     &     0.409 &  244.0  &   89.54$\pm$1.77 &     1      &   11.16$^{+0.22}_{-0.21}$ &  21.6 &  18.18$\pm$.027 &  15.97$\pm$.030 &  13.50$\pm$.031 &  C &  2  &    2595$\pm$28	 &  -3.194$\pm$.018 &   0.125$\pm$.003   & \nodata    \\     
39 &  12:24:52.2 &   $-$12:38:36 &  BRI B1222-1222     &   M9.0  &   31     &     0.322 &  234.4  &   58.60$\pm$3.80 &     5      &   17.06$^{+1.18}_{-1.03}$ &  26.0 &  20.41$\pm$.039 &  17.99$\pm$.036 &  15.54$\pm$.038 &  S &  1  &    2398$\pm$38	 &  -3.454$\pm$.057 &   0.108$\pm$.007   & \nodata    \\     
40 &  12:50:52.2 &   $-$21:21:09 &  LEHPM2-0174        &   M6.5  &  \nodata &     0.566 &  125.8  &   57.77$\pm$1.72 &     1      &   17.31$^{+0.53}_{-0.50}$ &  46.4 &  18.36$\pm$.063 &  16.15$\pm$.005 &  13.78$\pm$.027 &  C &  2  &    2598$\pm$25	 &  -2.909$\pm$.026 &   0.173$\pm$.006   & \nodata    \\     
41 &  13:05:40.2 &   $-$25:41:06 &  Kelu-1AB           &   L2.0J &   19     &     0.285 &  272.2  &   52.00$\pm$1.54 &    11      &   19.23$^{+0.58}_{-0.55}$ &  25.9 &  22.03$\pm$.060 &  19.14$\pm$.050 &  16.80$\pm$.001 &  S &  2  &    2026$\pm$45	 &  -3.616$\pm$.033 &   0.126$\pm$.007   & d,f  \\     
42 &  13:09:21.9 &   $-$23:30:33 &  CE 303             &   M7.0  &   13     &     0.381 &  176.0  &   69.33$\pm$1.33 &     1      &   14.42$^{+0.28}_{-0.27}$ &  26.0 &  19.37$\pm$.026 &  17.00$\pm$.013 &  14.58$\pm$.008 &  C &  2  &    2508$\pm$35	 &  -3.309$\pm$.018 &   0.117$\pm$.004   & \nodata    \\     
43 &  14:25:27.9 &   $-$36:50:22 &  DENIS J1425-3650   &   L3.0  &   29     &     0.544 &  211.6  &   86.45$\pm$0.83 &     1      &   11.56$^{+0.11}_{-0.11}$ &  29.8 &  22.81$\pm$.060 &  19.67$\pm$.041 &  17.35$\pm$.034 &  S &  1  &    1752$\pm$69	 &  -4.029$\pm$.009 &   0.104$\pm$.008   & \nodata    \\     
44 &  14:39:28.4 &   $+$19:29:15 &  2MASS J1439+1929   &   L1.0  &   11     &     1.295 &  288.3  &   69.60$\pm$0.50 &    11      &   14.36$^{+0.10}_{-0.10}$ &  88.1 &\nodata          &  18.45$\pm$.056 &  15.97$\pm$.052 &  S &  1  &    2186$\pm$100 &  -3.703$\pm$.010 &   0.098$\pm$.009   & g  \\
45 &  14:40:22.9 &   $+$13:39:23 &  2MASS J1440+1339   &   M8.0  &   25     &     0.331 &  204.7  &   45.00$\pm$1.11 &     1      &   22.22$^{+0.56}_{-0.53}$ &  34.8 &  18.95$\pm$.026 &  17.04$\pm$.080 &  14.81$\pm$.010 &  C &  2  &    2624$\pm$22	 &  -3.163$\pm$.022 &   0.126$\pm$.003   & \nodata    \\     
46 &  14:54:07.9 &   $-$66:04:47 &  DENIS J1454-6604   &   L3.5  &   28     &     0.565 &  125.1  &   84.88$\pm$1.71 &     1      &   11.78$^{+0.24}_{-0.23}$ &  31.5 &\nodata          &  19.22$\pm$.034 &  16.89$\pm$.022 &  C &  1  &    1788$\pm$100 &  -3.931$\pm$.019 &   0.112$\pm$.012   & g  \\     
47 &  14:56:38.5 &   $-$28:09:51 &  LHS 3003           &   M7.0  &   17     &     0.965 &  210.0  &  152.49$\pm$2.02 &     4      &    6.55$^{+0.08}_{-0.08}$ &  29.9 &  16.95$\pm$.014 &  14.90$\pm$.006 &  12.53$\pm$.008 &  C &  2  &    2581$\pm$17	 &  -3.266$\pm$.013 &   0.116$\pm$.002   & \nodata    \\     
48 &  15:01:07.9 &   $+$22:50:02 &  2MASS J1501+2250   &   M9.0  &   31     &     0.074 &  211.7  &   94.40$\pm$0.60 &    11      &   10.59$^{+0.06}_{-0.06}$ &   3.7 &  19.63$\pm$.021 &  17.39$\pm$.006 &  15.02$\pm$.018 &  C &  2  &    2398$\pm$36	 &  -3.602$\pm$.009 &   0.091$\pm$.002   & \nodata    \\     
49 &  15:39:41.9 &   $-$05:20:43 &  DENIS J1539-0520   &   L3.5  &   25     &     0.602 &  079.9  &   61.25$\pm$1.26 &     1      &   16.32$^{+0.34}_{-0.32}$ &  46.5 &\nodata          &  19.69$\pm$.035 &  17.56$\pm$.046 &  C &  1  &    1835$\pm$100 &  -4.006$\pm$.019 &   0.098$\pm$.010   & g  \\      
50 &  15:52:44.4 &   $-$26:23:07 &  LHS 5303           &   M6.0  &   18     &     0.495 &  155.1  &   94.63$\pm$0.70 &     1      &   10.56$^{+0.07}_{-0.07}$ &  24.7 &  16.53$\pm$.039 &  14.66$\pm$.021 &  12.49$\pm$.007 &  C &  2  &    2718$\pm$12	 &  -2.972$\pm$.008 &   0.147$\pm$.001   & \nodata    \\     
51 &  15:55:15.7 &   $-$09:56:05 &  2MASS J1555-0956   &   L1.0  &   13     &     1.217 &  129.9  &   74.53$\pm$1.21 &     1      &   13.41$^{+0.22}_{-0.21}$ &  77.4 &  21.04$\pm$.150 &  18.28$\pm$.019 &  15.82$\pm$.019 &  C &  1  &    2194$\pm$27	 &  -3.712$\pm$.015 &   0.096$\pm$.002   & \nodata    \\     
52 &  16:07:31.3 &   $-$04:42:06 &  SIPS J1607-0442    &   M8.0  &   13     &     0.415 &  180.2  &   63.90$\pm$1.47 &     1      &   15.64$^{+0.36}_{-0.35}$ &  30.7 &  19.49$\pm$.024 &  17.19$\pm$.014 &  14.78$\pm$.014 &  C &  1  &    2466$\pm$30	 &  -3.271$\pm$.021 &   0.126$\pm$.004   & \nodata    \\     
53 &  16:32:58.8 &   $-$06:31:45 &  SIPS J1632-0631    &   M7.0  &   13     &     0.342 &  176.3  &   53.31$\pm$1.48 &     1      &   18.75$^{+0.53}_{-0.50}$ &  30.4 &  20.23$\pm$.042 &  18.01$\pm$.014 &  15.58$\pm$.005 &  C &  2  &    2485$\pm$26	 &  -3.459$\pm$.025 &   0.100$\pm$.003   & \nodata    \\     
54 &  16:45:22.1 &   $-$13:19:51 &  2MASS J1645-1319   &   L1.5  &   13     &     0.874 &  203.8  &   90.12$\pm$0.82 &     1      &   11.09$^{+0.10}_{-0.10}$ &  45.9 &  20.96$\pm$.045 &  17.99$\pm$.001 &  15.65$\pm$.014 &  S &  2  &    1925$\pm$66	 &  -3.793$\pm$.011 &   0.113$\pm$.008   & \nodata    \\     
55 &  16:55:35.3 &   $-$08:23:40 &  GJ 644C            &   M7.0  &   17     &     1.202 &  223.4  &  154.96$\pm$0.52 &    26      &    6.45$^{+0.02}_{-0.02}$ &  36.7 &  16.85$\pm$.059 &  14.64$\pm$.015 &  12.25$\pm$.015 &  C &  3  &    2611$\pm$43	 &  -3.214$\pm$.007 &   0.120$\pm$.004   & e    \\     
56 &  17:05:48.3 &   $-$05:16:46 &  2MASS J1705-0516AB &   L0.5  &   23     &     0.165 &  132.5  &   55.07$\pm$1.76 &     1      &   18.15$^{+0.59}_{-0.56}$ &  14.2 &  21.67$\pm$.032 &  19.04$\pm$.009 &  16.67$\pm$.006 &  C &  1  &    2207$\pm$62	 &  -3.695$\pm$.029 &   0.097$\pm$.006   & d,f  \\     
57 &  19:16:57.6 &   $+$05:09:02 &  GJ 752B            &   M8.0  &   31     &     1.434 &  203.8  &  171.20$\pm$0.50 &    26      &    5.84$^{+0.01}_{-0.01}$ &  39.7 &  17.68$\pm$.029 &  15.21$\pm$.032 &  12.76$\pm$.026 &  S &  1  &    2478$\pm$29	 &  -3.340$\pm$.009 &   0.115$\pm$.003   & e     \\     
58 &  20:45:02.3 &   $-$63:32:05 &  SIPS J2045-6332    &   M9.0  &   25     &     0.218 &  158.0  &   41.72$\pm$1.50 &     1      &   23.96$^{+0.89}_{-0.83}$ &  24.7 &  21.14$\pm$.155 &  18.49$\pm$.036 &  16.04$\pm$.008 &  C &  2  &    2179$\pm$111 &  -3.129$\pm$.032 &   0.190$\pm$.020   & d     \\     
59 &  21:04:14.9 &   $-$10:37:37 &  2MASS J2104-1037   &   L3.0  &   15     &     0.662 &  116.0  &   53.00$\pm$1.71 &     1      &   18.86$^{+0.62}_{-0.58}$ &  59.2 &  22.37$\pm$.023 &  19.46$\pm$.023 &  17.18$\pm$.021 &  S &  1  &    1851$\pm$53	 &  -3.812$\pm$.030 &   0.120$\pm$.008   & \nodata    \\     
60 &  22:24:43.8 &   $-$01:58:52 &  2MASS J2224-0158   &   L4.5  &   11     &     0.984 &  152.3  &   86.70$\pm$0.69 &    11      &   11.53$^{+0.09}_{-0.09}$ &  53.7 &  23.82$\pm$.039 &  20.26$\pm$.028 &  17.77$\pm$.022 &  S &  1  &    1567$\pm$88	 &  -4.185$\pm$.013 &   0.109$\pm$.012   & \nodata    \\     
61 &  23:06:58.7 &   $-$50:08:58 &  SSSPM J2307-5009   &   M9.0  &   20     &     0.458 &  082.7  &   46.59$\pm$1.57 &     1      &   21.46$^{+0.74}_{-0.69}$ &  46.5 &  21.36$\pm$.050 &  18.90$\pm$.005 &  16.46$\pm$.019 &  S &  2  &    2347$\pm$48	 &  -3.593$\pm$.030 &   0.096$\pm$.005   & \nodata    \\     
62 &  23:54:09.3 &   $-$33:16:25 &  LHS 4039C          &   M9.0  &   29     &     0.505 &  218.3  &   44.38$\pm$2.09 &     1      &   22.53$^{+1.11}_{-1.01}$ &  53.9 &  20.96$\pm$.015 &  18.45$\pm$.001 &  15.98$\pm$.001 &  S &  2  &    2412$\pm$40	 &  -3.423$\pm$.041 &   0.111$\pm$.006   & d,e  \\     
63 &  23:56:10.8 &   $-$34:26:04 &  SSSPM J2356-3426   &   M9.0  &   20     &     0.312 &  167.1  &   52.37$\pm$1.71 &     1      &   19.09$^{+0.64}_{-0.60}$ &  28.2 &  20.81$\pm$.055 &  18.34$\pm$.001 &  15.89$\pm$.001 &  S &  2  &    2438$\pm$42	 &  -3.542$\pm$.029 &   0.094$\pm$.004   & \nodata    \\        
\enddata
\tablenotetext{a}{Unfortunately many papers do not cite references for spectral types. We have made an effort to track down primary sources.
                 The references listed here are either primary sources or, if a primary source could not be found, secondary sources 
                 that discuss spectral typing. In a few cases several papers list the same spectral type with no reference and do not
                 discuss the spectral type. In these cases this column was left blank.}
\tablenotetext{b}{References: (1) This work; (2) \citet{Reidetal1995}; (3) \citet{Tinneyetal1995}; (4) \citet{vanAltenaetal1995}; (5) \citet{Tinneyetal1996};
                 (6) \citet{GizisandReid1997}; (7) \citet{Kirkpatricketal1997}; (8) \citet{Martinetal1999}; (9) \citet{Basrietal2000};
                 (10) \citet{Gizisetal2000}; (11) \citet{Dahnetal2002}; (12) \citet{Geballeetal2002}; (13) \citet{Gizisetal2002}; (14) \citet{Leggetetal2002};
                 (15) \citet{Cruzetal2003}; (16) \citet{Kendalletal2003}; (17) \citet{Costaetal2005}; (18) \citet{Crifoetal2005}; (19) \citet{LiuandLeggett2005};
                 (20) \citet{Lodieuetal2005}; (21) \citet{Henryetal2006}; (22) \citet{Reyleetal2006}; (23) \citet{Reidetal2006}; (24) \citet{Caballero2007};
                 (25) \citet{Schmidtetal2007}; (26) \citet{vanLeeuwen2007}; (27) \citet{Looperetal2008}; (28) \citet{Phan-Baoetal2008};
                 (29) \citet{Reidetal2008}; (30) \citet{GatewoodandCoban2009}; (31) \citet{Jenkinsetal2009}; (32) \citet{Shkolniketal2009};
                 (33) \citet{Konopackyetal2010}; (34) \citet{DupuyandLiu2012}; (35) \citet{Fahertyetal2012} }
\tablenotetext{c}{S - {\it SOAR}; C - {\it CTIO} 0.9m}
\tablenotetext{d}{See notes in $\S$\ref{sec:individual}.}
\tablenotetext{e}{Member of resolved multiple system. Parallaxes for 1, 55, and 57 are for brighter components.}
\tablenotetext{f}{Unresolved multiple}
\tablenotetext{g}{No $V$ photometry is available. {\it SED} fit and $T_{eff}$ excludes $V$.}

\end{deluxetable}

\clearpage
\addtolength{\voffset}{-1in}
\addtolength{\hoffset}{0.5in}


\section{The Observed Sample}
\label{sec:sample}
Table 1 lists our observed sample.  The goal of our target selection was
to obtain an observing list that samples the color continuum between
spectral types M6V to L4, corresponding to $V-K$ ranging from 6.2 to 11.8,
for the nearby Galactic disk
population. Targets with known spectral types were selected from the
literature, with at least eight targets in each spectral subclass, for
a total of 82 targets.  Because the differences between stellar and
substellar objects become more pronounced at ages $>$ 1 Gyr, we
avoided objects with known youth signatures. All targets have original
distance estimates within 25 pc, and are located south of declination
+30$\degr$. This declination requirement makes all targets observable
from {\it CTIO}. Of these 82 targets, 26 have previously established
trigonometric parallaxes. The remaining 56 were placed on our parallax
observing list. In this paper we report new trigonometric parallaxes
for 37 targets and new {\it VRI} photometry for all 63 targets that
either have trigonometric parallaxes from the literature or have new
trigonometric parallaxes reported here.  Parallax observations for 19
targets are still ongoing and will be described in a future
publication. Figure 1 is a histogram showing the spectral type
distribution of the observed sample for this paper. There are more M
dwarfs than L dwarfs in Figure 1 because more M dwarfs had
trigonometric parallaxes from the literature.  Once parallax
observations for the 19 ongoing targets are finished the spectral type
distribution will become nearly even. 


\begin{figure}[ht]
 \begin{center}
  \includegraphics[scale=0.6, angle=0]{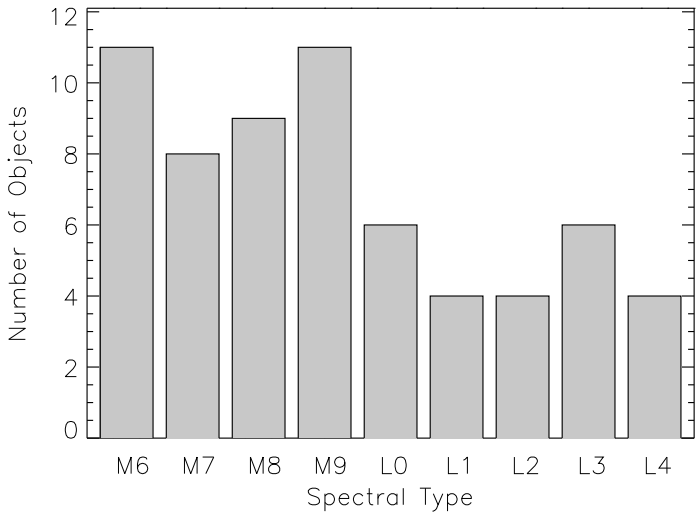}
 \figcaption[Figure 1]{\scriptsize Spectral type distribution for the observed sample.
M dwarfs are more heavily sampled because most M dwarfs already had
trigonometric parallaxes at the beginning of the study. 
Several L dwarf parallaxes are still in progress.}
\end{center}
\end{figure}

\clearpage


\section{Photometric Observations}
\label{sec:photometricobs}

{\it VLM} stars and brown dwarfs have traditionally been studied in
the near infrared where they emit most of their flux. However, as
discussed in detail in $\S$\ref{sec:tefflum}, optical photometry is
essential for determining the effective temperatures and the bolometric
fluxes of these very red objects.  We obtained {\it VRI} photometry for
all targets in our sample using the {\it CTIO} 0.9m telescope for the
brighter targets and the {\it SOAR Optical Imager}
camera on the {\it SOAR} 4.1m telescope for fainter targets. {\it
SOAR} observations were conducted between September 2009 and December
2010 during six observing runs comprising {\it NOAO} programs
2009B-0425, 2010A-0185, and 2010B-0176. A total of
17 nights on {\it SOAR} were used for optical photometry. Column 16 of
Table 1 indicates which telescope was used for each target. The
division between the 0.9m telescope and {\it SOAR} fell roughly along
the M/L divide. To ensure consistency, 28 targets were observed on
both telescopes.

Essentially the same observing procedure was used for both photometry programs.
After determining that a night was likely to be entirely cloudless in
the late afternoon, three or four photometric standard fields were
chosen and an observing schedule was constructed so that each field was
observed at three different airmasses, typically around 2.0, 1.5, and
the lowest possible airmass given the standard field's declination. We used the
photometric standards compiled by Arlo Landolt
\citep{Landolt1992,Landolt2007,Landolt2009} as well as standards from
\citet{Bessel1990} and \citet{Graham1982}. In each night, at least two
standards were red standards with $V-I>3.0$.
Details of the transformation equations used to derive the nightly
photometric solution from the observation of photometric standards are
given in \citet{Jaoetal2005}.

After flat fielding, bias subtraction, and mosaic integration in the
case of {\it SOAR/SOI} images, we performed aperture photometry using
the {\it IRAF apphot} package. Landolt standards are reduced using an
aperture 7\arcsec\~ in radius. Ideally, we would perform aperture
photometry on our targets using the same size aperture (7\arcsec) as Landolt used
to compile the standards we are using, but the faintness of our
targets required us to use a smaller aperture for two reasons. First,
the depth of our exposures (as faint as $V\sim24$ at {\it SOAR} 
and $V\sim21$ at the {\it CTIO} 0.9m, see $\S$\ref{subsec:photresults})
means that the science target is often not
more than 7\arcsec\~ apart from another resolved source.  Second, the
signal-to-noise error associated with a photometric observation is a
combination of the Poisson error and the sky subtraction error. The
latter's contribution is proportional to the area of the photometric
aperture and is particularly problematic in deep exposures where the
sky annulus may contain diffuse background sources.  It therefore
makes sense to use a smaller aperture and apply an aperture correction
based on the curve of growth of bright stars in the same exposure. We
used a 3\arcsec\~ aperture with an aperture correction to 7\arcsec. The
uncertainty associated with this aperture correction depends strongly
on the seeing, but is typically on the order of 1\% to 3\%. The final
photometric error is the sum in quadrature of the signal-to-noise
error, the error due to the aperture correction, and the error from
the nightly photometric solution, which is typically on the order of
1\% to 2\%. Each photometric night had at least two targets in overlap
with another night in order to check the validity of the the night's
photometric solution. We discuss optical variability in
$\S$\ref{subsec:variability}, where we show that the variability is
usually less than the formal uncertainty in the photometry, thus
justifying the use of only one epoch of photometry in cases where we
were unable to obtain a second epoch due to time constraints on {\it
SOAR}.

Several different {\it UBVRI} photometric systems are in current
usage.  While the photometry taken on the {\it CTIO} 0.9m telescope
used filters in the Johnson-Kron-Cousins system, data taken on {\it SOAR} used
Bessell filters. Descriptions of both systems, as well as conversion
relations, are given in \citet{Bessell1995}. The {\it V} filter is
photometrically identical between both systems. The {\it R} and {\it
I} filters have color dependent differences that reach a few percent
in the color regime explored by \citet{Bessell1995}, which considered
stars as red as $(V - R) = 1.8$ and $(V - I) = 4.0$. The targets in
this study are significantly redder, with $(V - I)$ as red as
5.7. In $\S$\ref{subsec:photresults} we derive new relations relevant to
the very red regime considered in this study. The values listed on Table 1
are on the system used on each telescope (see $\S$6.1).

\section{Astrometric Observations}
\label{sec:astrometricobs}
 The Cerro Tololo Inter-American Observatory Parallax Investigation
({\it CTIOPI}), \citep{Jaoetal2005,Henryetal2006} is a large and
versatile astrometric monitoring program targeting diverse types of
stellar and substellar objects in the solar neighborhood.
Observations are taken using the {\it CTIO} 0.9m telescope and its
sole instrument, a 2048$\times$2048 Tektronix imaging CCD detector with
a plate scale of 0\farcs401 pixel$^{-1}$. We use the central quarter
of the CCD chip, yielding a 6\farcm8$\times$6\farcm8  field of view. Details of
the observing procedures and data reduction pipeline are given in
\citet{Jaoetal2005}. A brief description of the aspects most relevant
for the observation of very red and faint targets is given here.

Each target was typically observed for five ``evening'' epochs
(i.e. before the midpoint of a given night) and five ``morning''
epochs over the course of at least two years. Observations were typically taken
in sets of three consecutive 600 s exposures always within $\pm$60 minutes of
target transiting, and in most cases within $\pm$30 minutes of transiting.
This restriction in hour angle means that the target is always
observed very close to its lowest possible airmass, which minimizes
the effects of differential atmospheric refraction. All but one target were
observed in the {\it I} band, where their optical spectrum is the
brightest and also where atmospheric refraction is minimized. The sole exception
is GJ 1001 A-BC, for which the parallax of the A component was measured in the 
{\it R} band to avoid saturation. The long
exposures caused the fields to be rich with background stars, which
greatly facilitated the selection of parallax reference stars. In most
cases we were able to setup the parallax field with the ideal
configuration of $\sim$10 reference stars symmetrically distributed
around the science target.  Care was taken to position the reference
fields using the same pixel coordinates for all epochs. Our experience
shows that this consistency of positioning the reference fields helps
reduce the final parallax error faster, but is not absolutely
required. There have been instances when a missaligned epoch was added
to the parallax reduction, and having an additional epoch, although not
perfectly positioned, still reduced the parallax error. Such
instances were considered on an individual basis.

{\it VRI} photometry (see $\S$\ref{sec:photometricobs}) of the reference field
was used to transform the relative
parallaxes into absolute parallaxes using photometric distance
relations. This transformation accounts for the fact that the parallax
reference stars are not located at infinite distances and therefore have a
finite, albeit much smaller, parallax.  Any original reference star
later found to be closer than 100 pc was discarded.  The {\it VRI}
photometry of the reference field and the science star was also used
to correct for small shifts in the apparent positions of the stars due
to atmospheric differential color refraction.

\section{Methodology for Calculating Effective Temperatures and Luminosities}
\label{sec:tefflum}
Determining the effective temperatures ({\it T$_{eff}$})\footnote{The
{\it effective temperature (T$_{eff}$)} of a surface is {\it defined} as the
temperature at which a perfect blackbody would emit the same flux (energy per time per area) as
the surface in question according to the Stephan-Boltzmann law: $F =
\sigma_{SB} T^4$. This quantity often
differs from the stellar atmosphere's actual temperature, 
which is a function of optical depth as well as other factors.}
of M and L dwarfs has
traditionally been difficult due to the complex nature of
radiative transfer in cool stellar atmospheres.  The task is
particularly challenging in the L dwarf regime, where inter-phase
chemistry between solid grains and the same substances in the gas
phase becomes relevant. Significant progress has occurred recently
with the publication of the {\it BT-Settl} family of model atmospheres
\citep{Allardetal2012, Allardetal2013}. The {\it BT-Settl} models are
the first to include a comprehensive cloud model based on
non-equilibrium chemistry between grains and the gas phase and the
rate of gravitational settling of solid grains. They have also been
computed using the latest revised solar metallicities
\citep{Caffauetal2011}. The authors \citep[e.g.,][]{Allardetal2012}
have demonstrated unprecedented agreement between observed M and L spectra
and the {\it BT-Settl} model atmospheres.

We determined {\it T$_{eff}$} for each object in our sample by
comparing observed photometric colors to synthetic colors derived from
the {\it BT-Settl} model grid using custom made {\it IDL} procedures.
Our procedure exploits the fact that synthetic colors can be computed 
from synthetic spectra and those colors can then be directly compared 
to observed colors. How well the synthetic colors match the observed colors
is then a measure of how well the input properties of a given synthetic spectrum
($T_{eff}$, {\it log g}, and $[M/H]$) match the real properties of the
object in question. The best matching $T_{eff}$ can then be found by interpolating
$T_{eff}$ as a function of the residuals of the color comparison
(observed color $-$ synthetic color) to the point of zero residual. The technique can
be applied independently to each available photometric color, and
the standard deviation of the resulting ensemble of $T_{eff}$ values
is the measure of the uncertainty in $T_{eff}$.

In our implementation of this technique, we first combined our {\it VRI} photometry (Bessel system)
 with {\it 2MASS JHK$_s$}
\citep{Skrutskieetal2006} and {\it WISE W1}, {\it W2}, and {\it W3}
photometry \citep{Wrightetal2010} to derive a total of 36 different colors for each object
covering the spectral range from $\sim0.4\micron$  to $\sim16.7\micron$\footnote{
We did not use the {\it WISE W4} band centered at $\sim22\micron$ because it produces mostly null detections
and upper limits for late M and L dwarfs.}.
We then calculated the same 36 colors for each spectrum in the {\it BT-Settl}
model grid using the photometric properties for each band
listed in Table 2\footnote{A thorough review of photometric
quantities, terminology, and procedures for deriving synthetic colors
is given in the appendix of \citet{BessellandMurphy2012}.}.

\begin{deluxetable}{lccccl}
\tabletypesize{\scriptsize}
\tablecaption{Photometric Properties of Individual Bands}
\setlength{\tabcolsep}{0.09in}
\label{tab:photproperties}
\tablewidth{0pt}
\tablehead{
	   \colhead{Band}                            &
	   \colhead{Blue Limit \tablenotemark{a}}    &
           \colhead{Red  Limit \tablenotemark{a}}    &
 	   \colhead{Effective Isophotal $\lambda$}     & 
	   \colhead{Mag. Zero Point}                 &
           \colhead{Reference}                       \\
           \colhead{    }                            &
           \colhead{$\micron$}                         &
           \colhead{$\micron$}                         &
	   \colhead{$\micron$}                         &
	   \colhead{{\it photon s$^{-1}$ cm$^{-2}$}} &
	   \colhead{     }                              }
\startdata
{\it V    } & 0.485 & 0.635  & 0.545 & 1.0146$\times$10$^{11}$ & \citet{BessellandMurphy2012}   \\ 
{\it R    } & 0.554 & 0.806  & 0.643 & 7.1558$\times$10$^{10}$ & \citet{BessellandMurphy2012}   \\
{\it I    } & 0.710 & 0.898  & 0.794 & 4.7172$\times$10$^{10}$ & \citet{BessellandMurphy2012}   \\
{\it J    } & 1.102 & 1.352  & 1.235 & 1.9548$\times$10$^{10}$ & \citet{Cohenetal2003}   \\
{\it H    } & 1.494 & 1.804  & 1.662 & 9.4186$\times$10$^{9}$  & \citet{Cohenetal2003}   \\  
{\it K$_s$} & 1.977 & 2.327  & 2.159 & 4.6692$\times$10$^{9}$  & \citet{Cohenetal2003}   \\
{\it W1   } & 2.792 & 3.823  & 3.353 & 1.4000$\times$10$^{9}$  & \citet{Jarettetal2011}   \\
{\it W2   } & 4.037 & 5.270  & 4.603 & 5.6557$\times$10$^{8}$  & \citet{Jarettetal2011}   \\  
{\it W3   } & 7.540 & 16.749 &11.560 & 3.8273$\times$10$^{7}$  & \citet{Jarettetal2011}   \\
\enddata
\tablenotetext{a}{10\% transmission normalized to band's peak transmission}
\end{deluxetable}
\clearpage

For each color, we then tabulated the residuals of (observed color
$-$ synthetic color) as a function of the synthetic spectrum's temperature.
The residuals are negative if the synthetic spectrum's temperature is too cold, 
approach zero for spectra with the right temperature, and are positive
for models hotter than the science object. 
For each color, we then interpolated the residuals as a function of temperature
to the point of zero residual. The temperature value of this point was taken
as the object's effective temperature  according to the color in question.
We then repeated the procedure for all 36 color combinations, thus providing 36 independent 
determinations of  $T_{eff}$. The adopted $T_{eff}$ for each object is 
the mean of the $T_{eff}$ values from each color.
The uncertainty in $T_{eff}$ is the standard deviation of the values used to compute the mean.
After performing this procedure we noted that the majority of colors produced 
$T_{eff}$ results that converged in a Gaussian fashion about a central value,
while other colors produced outliers that were a few hundred Kelvin
away from the Gaussian peak. Further inspection showed that colors 
for which the bluest band was an optical band ({\it VRI}) were producing 
the convergent results while colors in which both bands were infrared bands
tended to produce erratic values with no apparent systemic trend.
 We therefore performed the calculations a second
time using only the colors involving the {\it VRI} bands and excluding $I-J$,
which also did not converge well, for a total of twenty colors.
Occasionally, a color combination still produced an outlier at $T_{eff} >> 2\sigma$
from the adopted value. These outliers were excluded as well; however, the majority
of objects had their effective temperatures computed using all twenty colors.
The fact that none of the colors composed of infrared bands alone had
good convergence emphasizes the need to include optical photometry when studying
{\it VLM} stars and brown dwarfs.

 The model grid we used was a 3-dimensional grid with a
{\it T$_{eff}$} range from 1300K to 4500K in steps of 100K, {\it log g}
range from 3.0 to 5.5 in steps of 0.5 dex, and  metallicity, $[M/H]$,
range of $-$2.0 to 0.5 in steps of 0.5 dex.  The procedure was repeated for
each different combination of {\it log g} and $[M/H]$.  The final adopted
{\it T$_{eff}$} was the one from the combination of gravity and
metallicity that yielded the lowest {\it T$_{eff}$} dispersion amongst
the colors. As expected for {\it VLM} stars and brown dwarfs in
the Solar Neighborhood, the vast majority of objects had their best
fit effective temperatures at $log \: g = 5.0$ and $[M/H] = 0.0$. The
color$-${\it T$_{eff}$} interpolations often did not converge for grid
points where {\it log g} or $[M/H]$ was more than 1.0 dex away from the
final adopted value.  We reserve a comprehensive discussion of
metallicity and gravity issues in our observed sample for a future
publication reporting our spectroscopic observations.

The {\it IDL} procedure for determining effective temperatures also
indicates which model spectrum in the {\it BT-Settl} grid provides the
overall best fit to the observed photometry. We used the indicated
best fit spectrum as a template for an object's Spectral Energy
Distribution ({\it SED}) in order to calculate an object's
luminosity. 
Because the model spectra are spaced in a discrete grid,
and because no model spectrum can be expected to provide a perfect
match to observations, significant differences may still remain between the 
best fit synthetic spectrum and the real {\it SED}.
We devised an iterative procedure that applies
small modifications to the chosen {\it SED} template in order to
provide a better match to the photometry. We first calculated
synthetic photometry from the {\it SED} template for all nine bands listed in
Table 2 using a procedure identical to the one used
for calculating synthetic colors for the purpose of {\it T$_{eff}$}
determination. We then did a band by band comparison of the synthetic
photometry to the observed photometry and computed a corrective flux
factor by dividing the flux corresponding to the observed photometry
by the synthetic flux.  Next, we paired the corrective flux factors to
the corresponding isophotal effective wavelength for each band and fit
those values to a 9$^{th}$ order polynomial using the {\it IDL}
function {\it poly\_fit}, thus creating a continuous corrective
function with the same wavelength coverage as the {\it SED}
template. While it may seem unusual to fit nine bands of photometry
to a 9$^{th}$ order polynomial, we note that the purpose of the fit is
not to follow the general trend in the data, but rather to provide corrections
to each individual band while still preserving the continuity
of the {\it SED}. It therefore makes sense to use a function
with the same order as the number of data points.
To facilitate computations, {\it poly\_fit} was run on a
logarithmic wavelength scale that was then transformed back to a
linear scale.  The original {\it SED} template was then multiplied by
the corrective function and the process was iterated until residuals
for all bands fell below 2\%. Because the {\it W3} band is much broader than the other bands,
two additional points were used to compute the corrective function at the blue
and red ends of the band as well as at the isophotal wavelength. 
Figure 2 describes the process graphically. 
 The first iteration typically produced mean color shifts
of 0.1 to 0.25 magnitudes, depending on how well the real {\it SED} of
a given object matched the closest point in the spectral template
grid. 

The {\it BT-Settl} models are published with flux units as they appear
at the stellar surface.  These are very high fluxes when compared to
observed fluxes on Earth. To facilitate computations the model spectra
were first normalized to a value that is comparable in magnitude to
the observed photometric fluxes that are used to calibrate the
spectrum. Given the range of magnitudes of our objects, we found that
normalizing the model spectra so that their bolometric flux is
$10^{-10}\;erg\;s^{-1}\;cm^{-2}$ works well.  The first iteration
corrected for the bulk of the flux mismatch between the real target
and this arbitrary normalization, thus causing a much larger
correction than the subsequent iterations. The number of iterations
necessary for conversion varied greatly, ranging anywhere from three
to twenty or more. Table 3 shows the overall corrective factor for
each band for the three examples shown in Figure 2, as well as the
number of iterations that were necessary. The flux factors in Table 3
were normalized to 1.000 at the {\it H} band for ease of comparison.
To check that our corrective polynomial approach was producing 
consistent results, we computed the luminosities for the objects
listed in Table 3 using 9$^{th}$ order polynomials as well as 8$^{th}$
order polynomials. The results of dividing the luminosity obtained
using the 8$^{th}$ order polynomials by that obtained using 9$^{th}$
order polynomials were 1.00052, 1.00077, and 0.99451
respectively for LHS 3003, 2MASS J1501+2250, and 2MASS J2104-1037.
The uncertainties associated with the adopted 9$^{th}$ order solution are
3.08\%, 1.91\%, and 6.97\% respectively for LHS 3003, 2MASS J1501+2250, and 2MASS J2104-1037.
This test shows that so long as the polynomials used are
of high enough order, varying the order of the corrective polynomial causes
changes to the resulting luminosities that are well within the formal uncertainties.

\clearpage
\begin{figure}[ht]
 \begin{center}
   \subfigure[]
      {\includegraphics[scale=0.4]{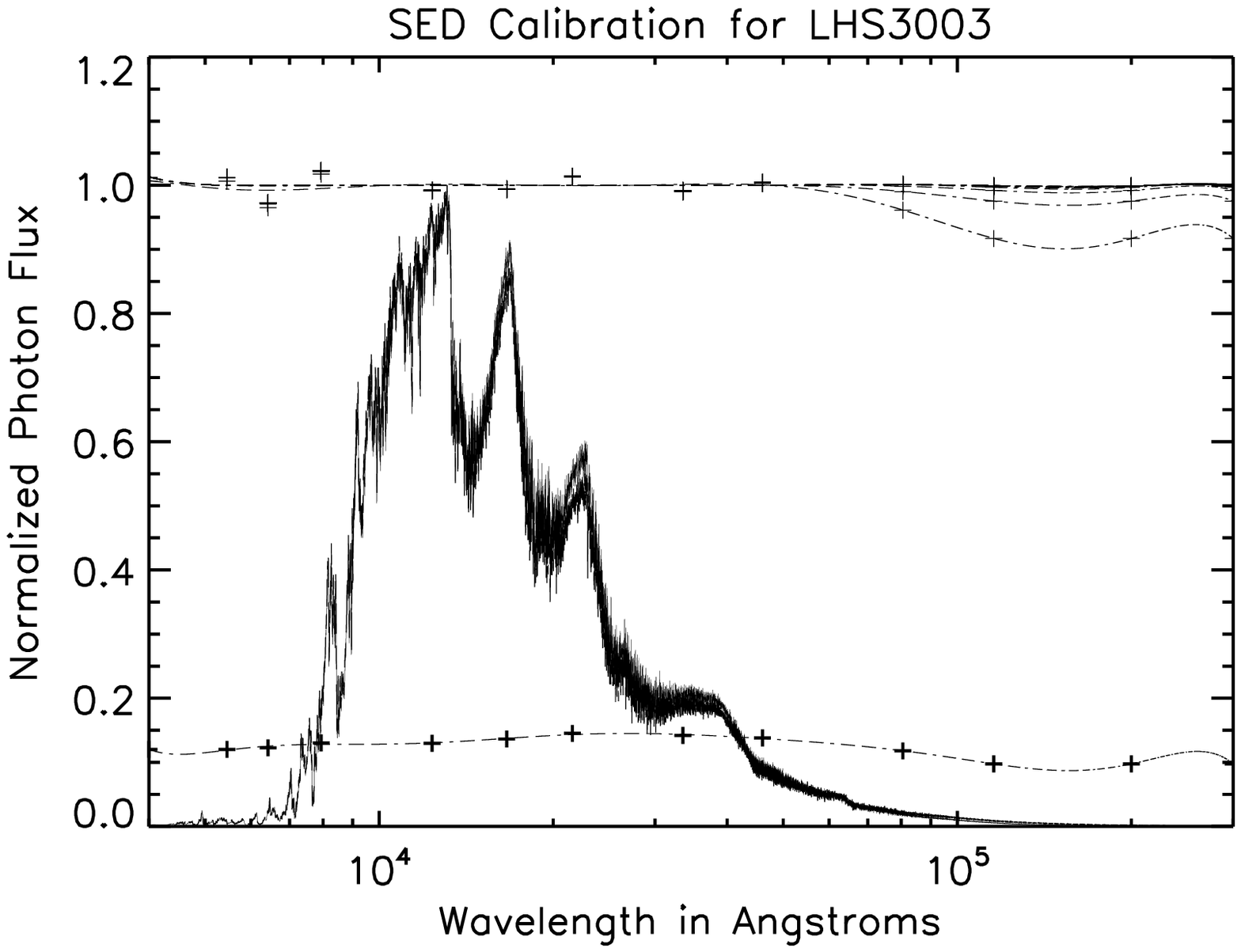}}
   \subfigure[]
      {\includegraphics[scale=0.4]{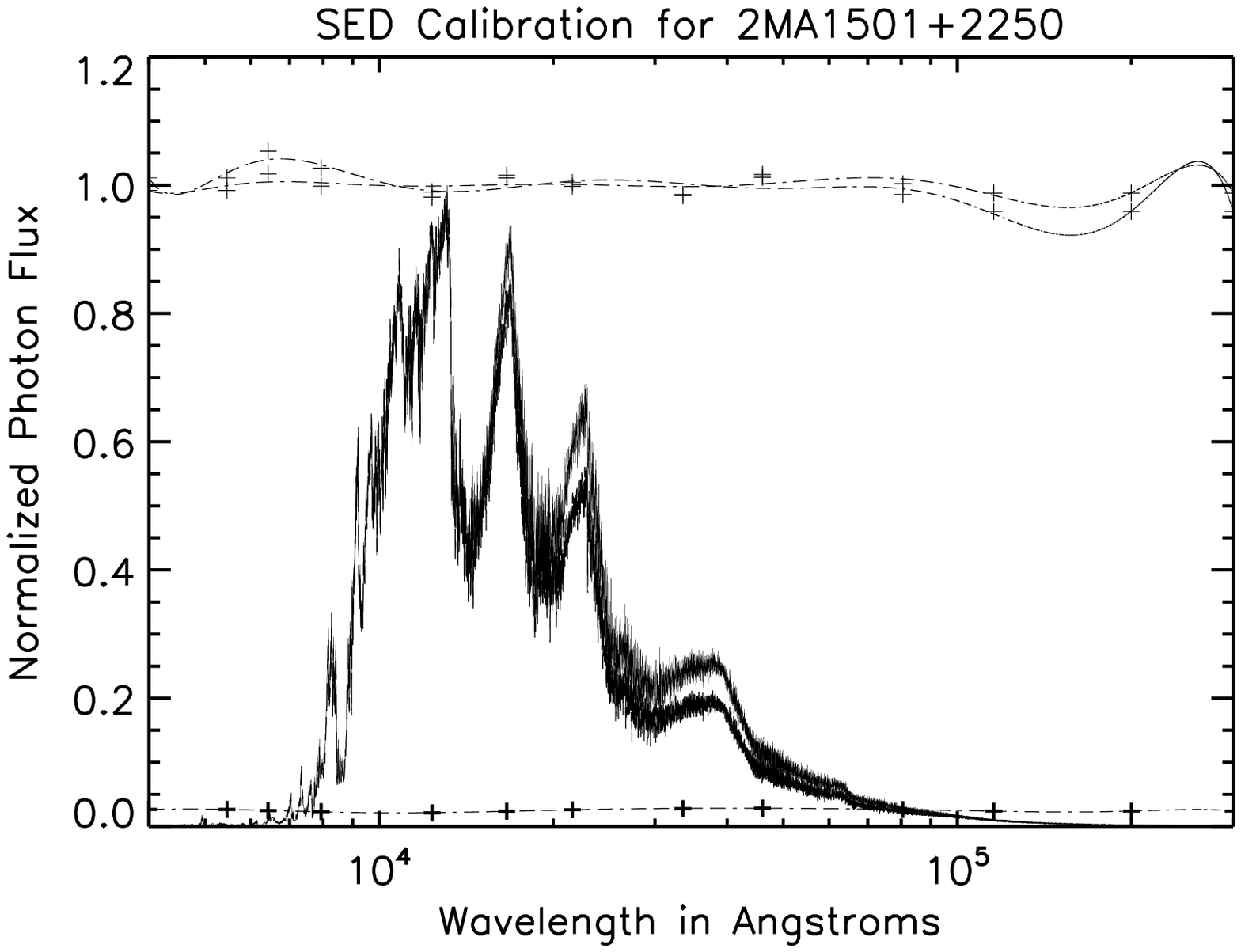}}
   \subfigure[]
      {\includegraphics[scale=0.4]{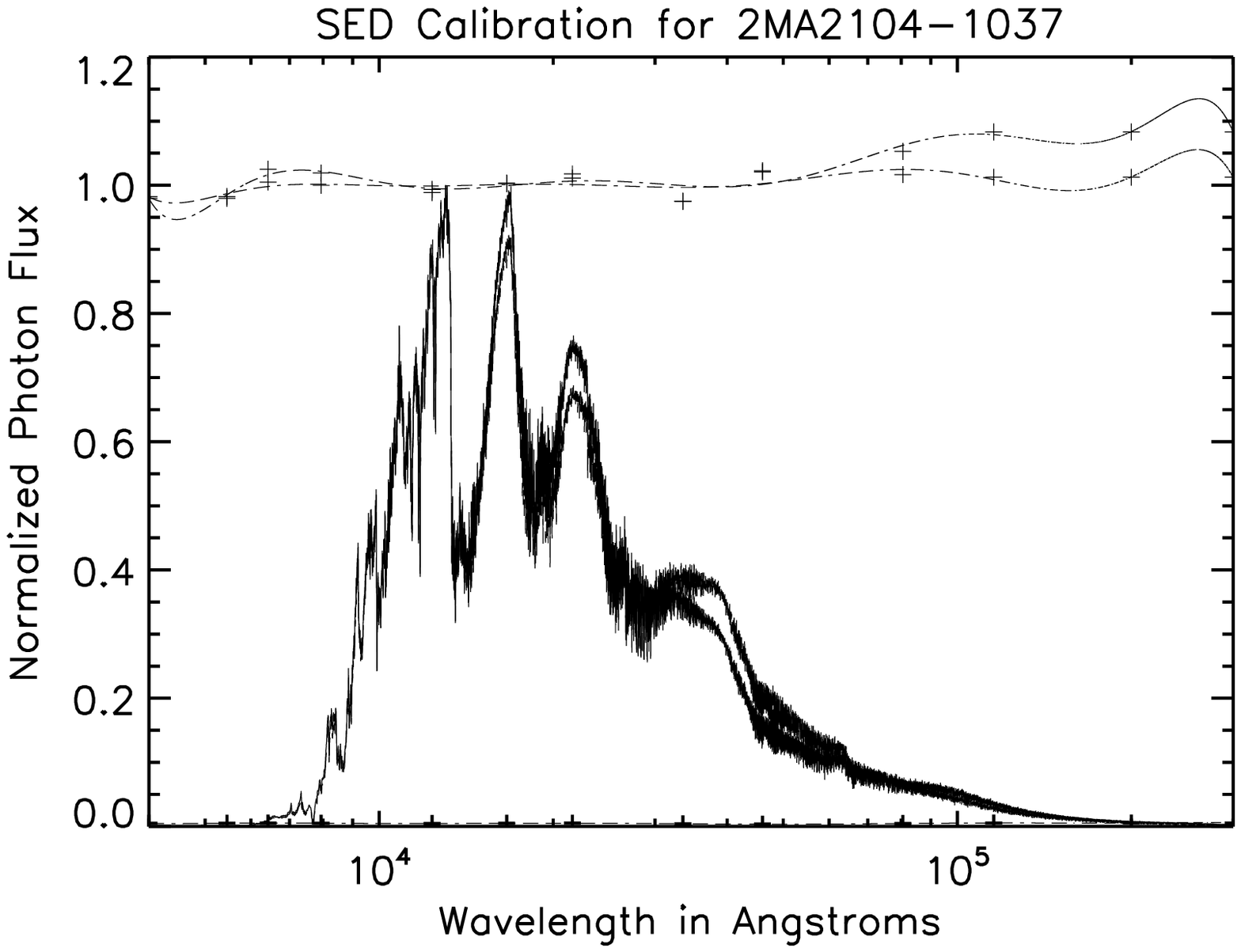}} 

\figcaption[figure2]{\scriptsize{\it SED} calibrations for (a) LHS 3003 (M7V), (b) 2MASS J1501+2250 (M9V)
                  and (c) 2MASS J2104-1037 (L3). The corrective polynomial functions
                  are shown by dashed lines and are fits to the corrective factors
		  shown by plus signs. In the first two cases the polynomial generated in the
                  first iteration stands out at the bottom of the graph due to the flux
		  mismatch caused by the distance modulus between the object's real distance
		  and the distance at which the {\it SED} template was calculated.
		  The first iteration is too close to the wavelength axis to be noticeable in (c).
                  The following iterations
                  then produce corrective functions that differ only slightly from
                  a flat 1.0 function and perform a ``fine tuning'' of the modifications
		  caused by the first iteration. Both the original {\it SED} template
                  and the final fit are plotted normalized to 1 at their maximum values.
		  In the cases of 2MASS J1501+2250 and 2MASS J2104-1037 the end result is an {\it SED} slightly redder
		  than the template. The template used for LHS 3003 was a very good
		  fit and the resulting {\it SED} almost entirely overlaps the initial
		  template. Table 3 lists the cumulative correction factors applied to each band
                  for the three objects in this figure. }
\end{center}
\end{figure}

\clearpage


\begin{deluxetable}{lcccccccccccc}
\tabletypesize{\scriptsize}
\tablecaption{Corrective Factors for {\it SEDs} Shown in Figure 2\tablenotemark{a}}
\setlength{\tabcolsep}{0.09in}
\label{tab:photproperties}
\tablewidth{0pt}
\tablehead{
           \colhead{Object}         &
	   \colhead{Iterations}     &
           \colhead{{\it V}}        &
           \colhead{{\it R}}        &
           \colhead{{\it I}}        &
           \colhead{{\it J}}        &
           \colhead{{\it H}}        &
           \colhead{{\it K$_s$}}    &
           \colhead{{\it W1}}       &
           \colhead{{\it W2}}       &
           \colhead{{\it W3}}       }

\startdata
LHS 3003          & 22 & 0.865 & 0.917 & 0.927 & 0.952 & 1.000 & 1.044 & 1.042 & 1.002 & 0.711 \\
2MASS J1501+2250  &  3 & 1.144 & 1.037 & 0.986 & 0.911 & 1.000 & 1.113 & 1.203 & 1.201 & 1.032 \\
2MASS J2104-1037  &  3 & 1.031 & 1.066 & 1.028 & 0.931 & 1.000 & 1.033 & 0.842 & 0.724 & 1.075 \\
\enddata
\tablenotetext{a}{All values are normalized to 1.000 in the {\it H} band.}
\end{deluxetable}  
\clearpage

The uncertainty
in the final flux under the {\it SED} was calculated by propagating
the uncertainty in the observed photometry and the residuals of the
final {\it SED} fit for each band and summing the results in
quadrature. Finally, the total flux was divided by the fraction of a
blackbody's total flux covered by the {\it SED} template given the
effective temperature of the object in question. This correction
accounted for the finite wavelength range of the {\it SED} and was
typically on the order of 1.5\%.

Once the effective temperatures and the observed bolometric fluxes were
determined by the procedures described above, determining the radii
of stars or brown dwarfs with a known trigonometric parallax followed
easily from the Stephan-Boltzmann law:
$$L = 4 \pi R^2 \sigma_{SB} T_{eff}^4$$ where $L$ is the object's
luminosity, $R$ is its radius, $ \sigma_{SB} = 5.6704
\times 10^{-5} erg \; cm^{-2} s^{-1} K^{-4}$ is the Stephan-Boltzmann
constant, and $T_{eff}$ is the effective temperature.

In order to check the accuracy of our procedures for determining effective
temperatures and luminosities, we applied our methodology to seven M
dwarfs that have direct model-independent radius measurements obtained
using Georgia State University's {\it CHARA} Array Long Baseline
Optical Interferometer \citep{Boyajianetal2012}. Figure 3 shows the
comparison.  The mean absolute residual is 3.4\%.  While it is
currently difficult to directly measure the angular diameters of
late M and L dwarfs using interferometry, the good agreement we obtain
when comparing the results of our {\it SED} fitting procedure to
direct radius measurements for hotter M dwarfs serves as a check on our
technique. We also note that while direct radius measurements exist
for several eclipsing binaries, the individual components of these systems lack
the photometric coverage needed for applying our method and therefore
cannot be used as checks.

\begin{figure}[ht]
 \begin{center}
  \includegraphics[scale=0.6, angle=0]{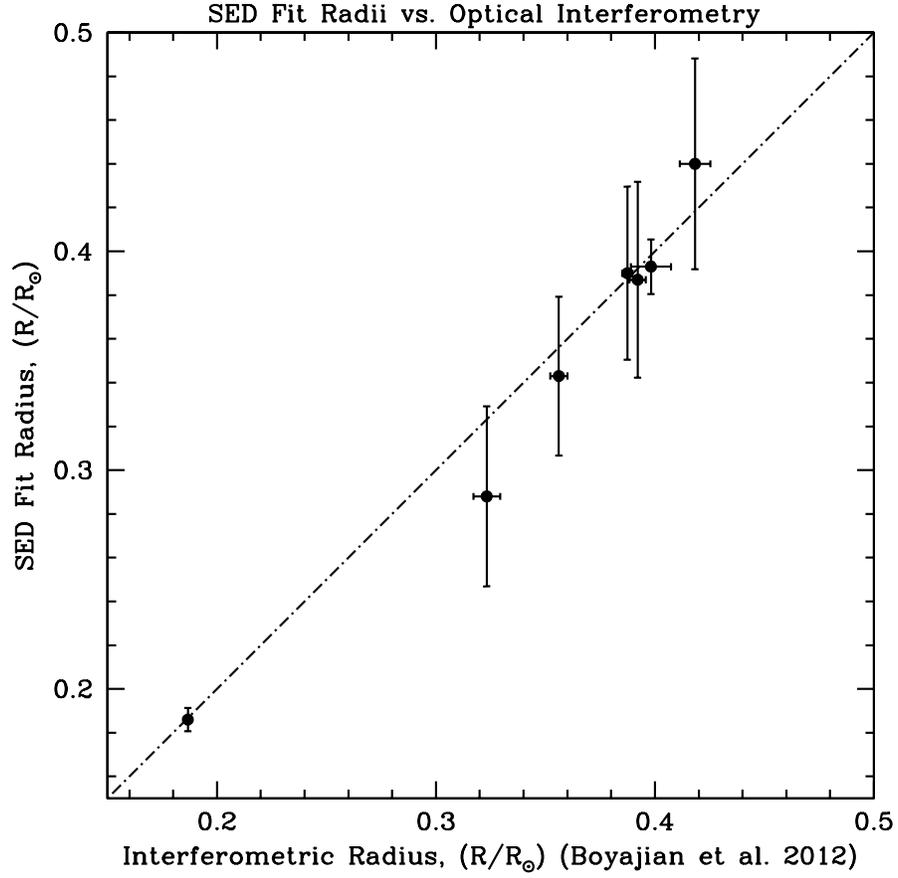}
 \figcaption[Figure 3]{\scriptsize Comparison of M dwarf radii obtained via our {\it SED} fitting
 technique to values based on direct angular diameter measurements
 obtained with Georgia State University's {\it CHARA} Array Optical
 Interferometer \citep{Boyajianetal2012}. From smallest to largest,
 the points correspond to: Barnard's Star (M4.0V), GJ 725B (M3.5V),
 GJ 725A (M3.0V), GJ 15A (M1.5V), GJ 411 (M2.0V), GJ 412A (M1.0V), and
 GJ 678 (M3.0V). The percent residuals in the sense ({\it SED} fit $-$ CHARA)
 are: -0.3\%, -10.9\%, -3.6\%, 0.8\%, -1.3\%, -1.3\%, and 5.3\%,
 respectively. The mean absolute residual is 3.4\%.}
\end{center}
\end{figure}

\clearpage

\setlength{\voffset}{1.5in}
\setlength{\hoffset}{-1.0in}
\textheight=8.0in
\textwidth=8.2in
\begin{deluxetable}{ccccccccccccccccc}
\tabletypesize{\scriptsize}
\rotate
\tablecaption{New Trigonometric Parallaxes, Proper Motions, and Optical Variability}
\setlength{\tabcolsep}{0.06in}
\label{tab:parallaxes}
\tablewidth{0pt}
\tablehead{
           \colhead{ID }         &
           \colhead{Name}        &
           \colhead{R.A.}        &
           \colhead{Decl.}       &
           \colhead{Filt.}       &
           \colhead{$N_{sea}$\tablenotemark{a}}   &
           \colhead{$N_{frm}$\tablenotemark{b}}   &
           \colhead{Coverage}    &
           \colhead{Years}       &
           \colhead{$N_{ref}$\tablenotemark{c}}   &
           \colhead{$\pi$(rel)}  &
           \colhead{$\pi$(corr)} &
           \colhead{$\pi$(abs)}  &
           \colhead{$\mu$}       &
           \colhead{P.A.}        &
           \colhead{$V_{tan}$}   &
           \colhead{Var.\tablenotemark{d}}        \\
           \colhead{ }           &
           \colhead{ }           &
           \colhead{2000.0 }     &
           \colhead{2000.0 }     &
           \colhead{ }           &
           \colhead{ }           &
           \colhead{ }           &
           \colhead{ }           &
           \colhead{ }           &
           \colhead{ }           &
           \colhead{mas}         &
           \colhead{mas}         &
           \colhead{mas}         &
           \colhead{mas yr$^{-1}$}   &
           \colhead{deg E. of N.}    &
           \colhead{Km s$^{-1}$} &
           \colhead{milli-mag}   }
 
 \startdata
1  &  GJ 1001BC          & 00:04:34.9 &   -40:44:06 & {\it R} & 10s & 112  &  1999.64-2011.74 & 12.10 &  6 &  75.99$\pm$2.06 & 1.03$\pm$0.16 &   77.02$\pm$2.07  & 1643.6$\pm$0.6    & 156.0$\pm$.04     &  102.4 & \nodata \\
2  &  LEHPM1-0494B       & 00:21:05.8 &   -42:44:49 & {\it I} &  5s &  24  &  2008.86-2012.83 &  3.97 &  9 &  39.22$\pm$2.10 & 0.55$\pm$0.08 &   39.77$\pm$2.10  &  252.9$\pm$1.6    & 089.1$\pm$.53     &   30.1 &  15.4  \\   
3  &  LEHPM1-0494A       & 00:21:10.7 &   -42:45:40 & {\it I} &  5s &  24  &  2008.86-2012.83 &  3.97 &  9 &  36.65$\pm$1.99 & 0.55$\pm$0.08 &   37.20$\pm$1.99  &  252.8$\pm$1.5    & 086.5$\pm$.53     &   32.2 &  6.5  \\  
10 &  DENIS J0306-3647   & 03:06:11.5 &   -36:47:53 & {\it I} &  4s &  39  &  2009.75-2012.94 &  3.19 &  8 &  75.79$\pm$1.42 & 0.67$\pm$0.08 &   76.46$\pm$1.42  &  690.0$\pm$1.1    & 196.0$\pm$.16     &   42.7 &  8.4  \\
11 &  LP 944-020         & 03:39:35.2 &   -35:25:44 & {\it I} &  8s &  59  &  2003.95-2012.94 &  8.99 & 10 & 154.53$\pm$1.03 & 1.36$\pm$0.10 &  155.89$\pm$1.03  &  408.3$\pm$0.3    & 048.5$\pm$.07     &   12.4 &  8.8  \\
13 &  2MASS J0428-2253   & 04:28:50.9 &   -22:53:22 & {\it I} &  4s &  22  &  2010.01-2012.94 &  2.92 &  9 &  38.04$\pm$1.85 & 0.44$\pm$0.04 &   38.48$\pm$1.85  &  189.3$\pm$1.9    & 038.1$\pm$1.11    &   23.2 &  18.0  \\
14 &  LP 775-031         & 04:35:16.1 &   -16:06:57 & {\it I} &  7c &  74  &  2003.95-2012.88 &  8.94 &  8 &  94.53$\pm$1.05 & 0.82$\pm$0.13 &   95.35$\pm$1.06  &  356.0$\pm$0.4    & 028.0$\pm$.11     &   17.6 &  8.1  \\
15 &  2MASS J0451-3402   & 04:51:00.9 &   -34:02:15 & {\it I} &  5s &  22  &  2008.86-2013.12 &  4.26 &  8 &  46.43$\pm$1.43 & 1.03$\pm$0.47 &   47.46$\pm$1.51  &  157.6$\pm$1.0    & 036.5$\pm$.75     &   15.7 &  50.6  \\
16 &  2MASS J0500+0330   & 05:00:21.0 &   +03:30:50 & {\it I} &  4c &  23  &  2009.75-2012.89 &  3.15 & 13 &  73.38$\pm$1.98 & 0.47$\pm$0.12 &   73.85$\pm$1.98  &  350.2$\pm$1.7    & 177.9$\pm$.41     &   22.4 &  14.8  \\
17 &  2MASS J0523-1403   & 05:23:38.2 &   -14:03:02 & {\it I} &  3c &  24  &  2010.98-2013.12 &  2.14 &  9 &  80.35$\pm$1.76 & 0.60$\pm$0.10 &   80.95$\pm$1.76  &  194.5$\pm$1.6    & 032.5$\pm$.94     &   11.4 &  11.7  \\
18 &  DENIS J0652-2534   & 06:52:19.7 &   -25:34:50 & {\it I} &  4c &  36  &  2010.02-2013.12 &  3.10 & 12 &  63.24$\pm$0.94 & 0.52$\pm$0.04 &   63.76$\pm$0.94  &  249.6$\pm$0.7    & 289.3$\pm$.31     &   18.5 &  10.5  \\
21 &  DENIS J0751-2530   & 07:51:16.4 &   -25:30:43 & {\it I} &  4c &  35  &  2010.15-2013.12 &  2.97 & 10 &  58.65$\pm$0.84 & 0.50$\pm$0.02 &   59.15$\pm$0.84  &  889.1$\pm$0.8    & 279.2$\pm$.09     &   71.2 &  15.3  \\
22 &  DENIS J0812-2444   & 08:12:31.7 &   -24:44:42 & {\it I} &  4s &  28  &  2010.02-2013.25 &  3.24 & 14 &  44.79$\pm$0.96 & 0.68$\pm$0.03 &   45.47$\pm$0.96  &  196.4$\pm$0.7    & 135.5$\pm$.41     &   20.4 &  19.9  \\
23 &  SSSPM J0829-1309   & 08:28:34.1 &   -13:09:19 & {\it I} &  4c &  24  &  2009.94-2013.26 &  3.32 & 10 &  87.24$\pm$0.76 & 0.72$\pm$0.16 &   87.96$\pm$0.78  &  577.3$\pm$0.7    & 273.0$\pm$.10     &   31.1 &  9.2  \\
26 &  2MASS J0847-1532   & 08:47:28.7 &   -15:32:37 & {\it I} &  5s &  35  &  2009.32-2013.27 &  3.95 & 12 &  58.34$\pm$0.99 & 0.62$\pm$0.09 &   58.96$\pm$0.99  &  239.5$\pm$0.7    & 146.1$\pm$.34     &   19.2 &  9.9  \\
27 &  LHS 2065           & 08:53:36.0 &   -03:29:28 & {\it I} & 10s & 101  &  2003.95-2013.26 &  9.31 &  6 & 117.19$\pm$0.76 & 0.79$\pm$0.03 &  117.98$\pm$0.76  &  550.3$\pm$0.2    & 249.4$\pm$.04     &   22.0 &  12.1  \\
29 &  LHS 2195           & 09:49:22.2 &   +08:06:45 & {\it I} &  4s &  36  &  2010.01-2013.10 &  3.09 &  9 &  59.55$\pm$1.66 & 0.77$\pm$0.15 &   60.32$\pm$1.67  &  886.7$\pm$1.2    & 177.4$\pm$.12     &   69.7 &  11.1  \\
35 &  LHS 2397aAB        & 11:21:49.0 &   -13:13:08 & {\it I} &  7c &  68  &  2005.09-2013.26 &  8.16 &  9 &  65.28$\pm$2.02 & 0.55$\pm$0.07 &   65.83$\pm$2.02  &  506.9$\pm$0.6    & 264.7$\pm$.11     &   36.5 &  22.1  \\
36 &  2MASS J1126-5003   & 11:26:39.9 &   -50:03:55 & {\it I} &  5s &  20  &  2009.19-2013.25 &  4.07 & 13 &  58.82$\pm$1.64 & 0.56$\pm$0.12 &   59.38$\pm$1.64  & 1645.7$\pm$1.0    & 286.2$\pm$.06     &  131.3 &  25.3  \\
38 &  LP 851-346         & 11:55:42.9 &   -22:24:58 & {\it I} &  7s &  56  &  2007.18-2013.28 &  6.10 &  9 &  88.92$\pm$1.77 & 0.62$\pm$0.06 &   89.54$\pm$1.77  &  408.6$\pm$0.9    & 244.0$\pm$.23     &   21.6 &  10.4  \\
40 &  LEHPM2-0174        & 12:50:52.2 &   -21:21:09 & {\it I} &  8s &  45  &  2005.14-2013.38 &  8.25 &  9 &  57.33$\pm$1.72 & 0.44$\pm$0.03 &   57.77$\pm$1.72  &  565.7$\pm$0.6    & 125.8$\pm$.15     &   46.4 &  7.8  \\
42 &  CE 303             & 13:09:21.9 &   -23:30:33 & {\it I} &  4s &  47  &  2010.16-2013.27 &  3.11 & 11 &  68.41$\pm$1.32 & 0.92$\pm$0.14 &   69.33$\pm$1.33  &  380.5$\pm$1.1    & 176.0$\pm$.26     &   26.0 &  10.2  \\
43 &  DENIS J1425-3650   & 14:25:27.9 &   -36:50:22 & {\it I} &  5s &  33  &  2009.31-2013.28 &  3.96 & 13 &  85.80$\pm$0.79 & 0.65$\pm$0.24 &   86.45$\pm$0.83  &  543.7$\pm$0.8    & 211.6$\pm$.17     &   29.8 &  15.1  \\
45 &  2MASS J1440+1339   & 14:40:22.9 &   +13:39:23 & {\it I} &  5s &  34  &  2009.25-2013.26 &  4.01 &  8 &  44.13$\pm$1.11 & 0.87$\pm$0.07 &   45.00$\pm$1.11  &  331.1$\pm$0.9    & 204.7$\pm$.28     &   34.8 &  6.9  \\
46 &  DENIS J1454-6604\tablenotemark{e}   & 14:54:07.9 &   -66:04:47 & {\it I} &  5s &  22  &  2009.32-2013.26 &  3.94 & 11 &  84.21$\pm$1.70 & 0.67$\pm$0.17 &   84.88$\pm$1.71  &  564.8$\pm$1.3    & 125.1$\pm$.25     &   31.5 &  19.9  \\
49 &  DENIS J1539-0520\tablenotemark{e}   & 15:39:41.9 &   -05:20:43 & {\it I} &  5c &  29  &  2009.25-2013.25 &  4.00 & 11 &  60.51$\pm$1.26 & 0.74$\pm$0.08 &   61.25$\pm$1.26  &  602.3$\pm$1.1    &  79.9$\pm$.17     &   46.5 &  17.0  \\
50 &  LHS 5303           & 15:52:44.4 &   -26:23:07 & {\it I} &  9s &  85  &  2004.57-2012.59 &  8.02 & 10 &  94.10$\pm$0.70 & 0.53$\pm$0.07 &   94.63$\pm$0.70  &  495.4$\pm$0.2    & 155.1$\pm$.05     &   24.7 &  10.7  \\
51 &  2MASS J1555-0956   & 15:55:15.7 &   -09:56:05 & {\it I} &  4c &  25  &  2010.19-2013.28 &  3.08 & 10 &  73.94$\pm$1.21 & 0.59$\pm$0.05 &   74.53$\pm$1.21  & 1217.0$\pm$1.3    & 129.9$\pm$.12     &   77.4 &  9.9  \\
52 &  SIPS J1607-0442    & 16:07:31.3 &   -04:42:06 & {\it I} &  4c &  32  &  2010.39-2013.26 &  2.87 &  8 &  62.79$\pm$1.47 & 1.11$\pm$0.06 &   63.90$\pm$1.47  &  414.6$\pm$1.2    & 180.2$\pm$.26     &   30.7 &  12.2  \\
53 &  SIPS J1632-0631    & 16:32:58.8 &   -06:31:45 & {\it I} &  3c &  40  &  2010.19-2012.58 &  2.39 & 11 &  52.31$\pm$1.47 & 1.00$\pm$0.14 &   53.31$\pm$1.48  &  342.2$\pm$1.9    & 176.3$\pm$.45     &   30.4 &  17.9  \\
54 &  2MASS J1645-1319   & 16:45:22.1 &   -13:19:51 & {\it I} &  5c &  48  &  2009.32-2013.27 &  3.95 & 15 &  89.19$\pm$0.81 & 0.93$\pm$0.10 &   90.12$\pm$0.82  &  873.8$\pm$0.6    & 203.8$\pm$.08     &   45.9 &  11.6  \\
56 &  2MASS J1705-0516AB & 17:05:48.3 &   -05:16:46 & {\it I} &  5s &  18  &  2009.32-2013.25 &  3.93 & 10 &  53.34$\pm$1.74 & 1.73$\pm$0.26 &   55.07$\pm$1.76  &  164.7$\pm$1.1    & 132.5$\pm$.79     &   14.2 &  40.9  \\
58 &  SIPS J2045-6332    & 20:45:02.3 &   -63:32:05 & {\it I} &  4c &  45  &  2010.59-2013.54 &  2.95 & 11 &  40.65$\pm$1.50 & 1.07$\pm$0.07 &   41.72$\pm$1.50  &  220.4$\pm$1.2    & 158.0$\pm$.88     &   24.7 &  38.9  \\
59 &  2MASS J2104-1037   & 21:04:14.9 &   -10:37:37 & {\it I} &  4c &  22  &  2009.56-2012.58 &  3.02 & 12 &  52.23$\pm$1.70 & 0.77$\pm$0.15 &   53.00$\pm$1.71  &  661.9$\pm$1.3    & 116.0$\pm$.22     &   59.2 &  12.5  \\
61 &  SSSPM J2307-5009   & 23:06:58.7 &   -50:08:58 & {\it I} &  4c &  41  &  2009.55-2012.81 &  3.26 &  9 &  46.21$\pm$1.57 & 0.38$\pm$0.06 &   46.59$\pm$1.57  &  457.8$\pm$1.6    &  82.7$\pm$.32     &   46.5 &  11.5  \\
62 &  LHS 4039C          & 23:54:09.3 &   -33:16:25 & {\it I} &  4c &  58  &  2003.51-2007.74 &  4.23 &  5 &  41.91$\pm$2.08 & 2.47$\pm$0.15 &   44.38$\pm$2.09  &  505.5$\pm$1.8    & 218.3$\pm$.40     &   53.9 &  20.0  \\
63 &  SSSPM J2356-3426   & 23:56:10.8 &   -34:26:04 & {\it I} &  3c &  28  &  2009.56-2011.77 &  2.21 &  9 &  51.80$\pm$1.71 & 0.57$\pm$0.07 &   52.37$\pm$1.71  &  312.5$\pm$2.1    & 167.1$\pm$.67     &   28.2 &  10.2  \\
\enddata
\tablenotetext{a}{Number of seasons observed, where 2$-$3 months of observations count as one season, for seasons having more than three images taken.
The letter ``c'' indicates a continuous set of observations where multiple nights of data were taken in each season, whereas an ``s'' indicates scattered
observations when one or more seasons have only a single night of observations. Generally ``c'' observations are better.}
\tablenotetext{b}{Total number of images used in reduction. Images are typically taken in sets of three consecutive observations.}
\tablenotetext{c}{Number of reference stars used to reduce the parallax.}
\tablenotetext{d}{Photometric variability of the science target.}
\tablenotetext{e}{No $V$ photometry. Correction for differential color refraction based on estimated $V$ from color-magnitude relations.}
\voffset=0in
\hoffset=0in
\end{deluxetable}
\clearpage
\voffset=0in
\hoffset=0in

\addtolength{\hoffset}{-1.1in}
\addtolength{\voffset}{0.7in}
\begin{landscape}
\begin{figure}[ht]
 \begin{center}
 \includegraphics[scale=0.9, angle=-90]{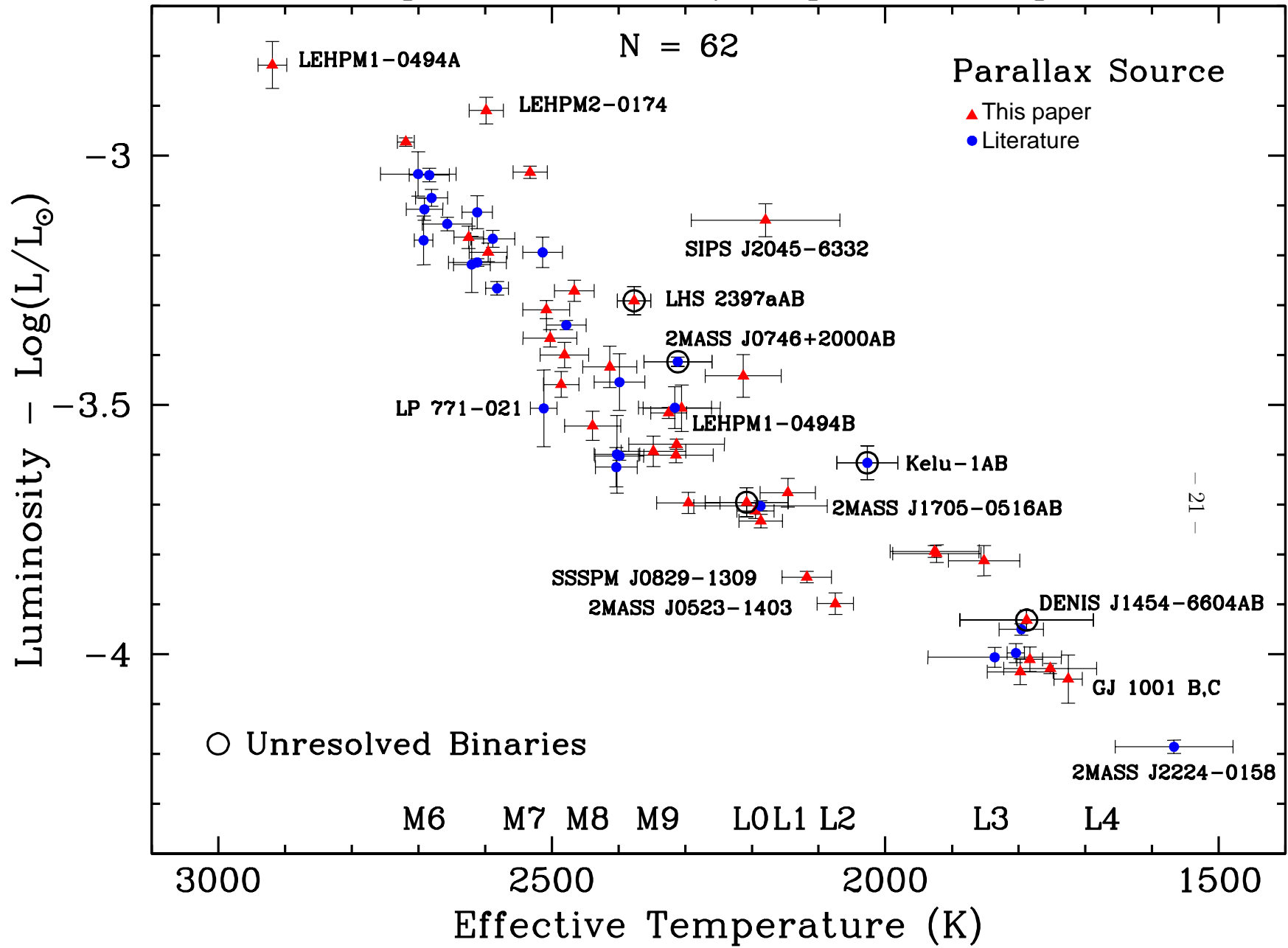}
 \figcaption[Figure 4]{\scriptsize {\it HR} Diagram for objects with spectral types ranging from
             M6V to L4.5. Several representative objects are named. Known binaries
             with joint photometry are enclosed in open circles. A few
             known binaries are clearly over-luminous, denoting their low luminosity
             ratios. The L4.5 binary GJ 1001BC was deconvolved based on the nearly
             identical luminosity of both components \citep{Golimowskietal2004b}.
             As we discuss in $\S$\ref{sec:discussion}, the L2.5 dwarf
             2MASS J0523-1403 lies at a pronounced minimum in the radius-luminosity relation
             and its location likely constitutes the end of the stellar main sequence.
             Versions of this diagram that use the ID labels in Table 1
             and spectral type labels for plotting symbols are available as supplemental online material. }

 \end{center}
\end{figure}
\end{landscape}

\addtolength{\hoffset}{1.1in}
\addtolength{\voffset}{-0.7in}
\clearpage

\section{Results}
\label{sec:results}

Table 1 includes astrometric results (our new values as well as values from the literature),
 our {\it VRI} photometry, and the derived effective
temperatures, luminosities, and radii for all objects. Table 4 reports
detailed astrometric results for the 37 objects for which we report
new trigonometric parallaxes.  All resulting quantities are synthesized
and summarized graphically in Figure 4, a {\it bona fide
Hertzsprung-Russell} diagram for the end of the stellar main
sequence. We discuss several auxiliary results separately here and
save a thorough discussion of the structure of the
stellar/substellar boundary for $\S$\ref{sec:discussion}.

\subsection{Photometric Results}
\label{subsec:photresults}

Columns 13$-$17 of Table 1 list our {\it VRI} photometry, the
telescope in which the photometry was taken, and the number of epochs
for which each target was observed.  For the 28 targets observed on
both telescopes, Table 1 lists the set of observations with the
smallest error or the most epochs, with the number of epochs taking
priority in selecting which data set to adopt. The electronic version
of Table 1 lists both sets of photometry for these objects, along with
{\it 2MASS JHK$_s$} and {\it WISE W1W2W3} photometry for all
objects. We achieved sensitivities of $V=23.75\pm0.01$ on {\it SOAR}
with 90 minute exposures under dark skies and good seeing. The time
demands of the {\it CTIOPI} program at the 0.9m telescope forced us to
limit exposures to 20 minutes for the majority of targets. Under dark
skies and good seeing (i.e. $\lesssim$1\farcs0) 20 minute integrations
yielded results as faint as $V=19.50\pm0.05$. In exceptional cases
when we took longer integrations we were able to achieve
$V=21.93\pm0.07$ in 90 minutes under extraordinary conditions.  The
majority of the measurements had errors $<$ 0.05 magnitudes
(i.e. 5\%); however, for the fainter 0.9m observations the errors are
as large as 0.15 magnitudes.  It was our original intention to
observe all targets for at least two epochs, but this was not possible
for some targets due to time constraints on {\it SOAR}. As discussed
in $\S$\ref{subsec:variability}, the optical variability for the
sample is comparable to the photometric error, meaning that single
epoch photometry should be generally consistent with the values we
would obtain by averaging more observations.

Table 1 shows the photometry in the photometric system used by the
telescope in which the measurements were taken $-$
Johnson-Kron-Cousins for the {\it CTIO} 0.9m telescope and Bessell for
{\it SOAR}.  We have also converted the {\it CTIO} 0.9m values to the
Bessell system, and present these data in the electronic version of
Table 1.  Rather than extrapolating the relations of
\citet{Bessell1995}, we used the 28 objects observed on both
telescopes to derive new relations between the colors $(V - R_B)$ and
$(V - R_C)$ as well as $(V - I_B)$ and $(V - I_C)$ and show the
results in Figure 5.  Given the photometric uncertainties of our {\it
V} and {\it R} observations (typically $\lesssim$5\%, Table 1), we
find no systematic deviation between the two $(V - R)$ colors.  We
therefore adopt $R_B = R_C$ for the purpose of this study. We do
detect a trend in the $(V - I)$ colors, as shown in Figure 5b. Based
on the data shown in Figure 5b, we derive the transformation
$$(V - I_B) = -0.0364(V - I_C)^2 +1.4722(V - I_C) - 1.3563 .$$
We
emphasize that the relations we derive here are based on a small
sample and serve the purposes of our study only.  They should not be
used as general relations analogous to those of \citet{Bessell1995}.
In particular, the difference in the {\it I} band is likely dominated
by the different detector efficiencies between the {\it CTIO} 0.9m and
the {\it SOAR/SOI} CCDs in the far red.
The {\it I} photometry listed in Table 1 is in
the photometric system of the telescope that took the adopted
observations.

We note that the procedure for determining effective temperatures and
luminosities described in $\S$\ref{sec:tefflum} uses photometry in the
Bessell system because the transmission curves for Bessell filters are
well-characterized \citep{BessellandMurphy2012}.

\begin{figure}[ht]
 \begin{center}
    \subfigure[]
	{\includegraphics[scale=0.6,angle=0]{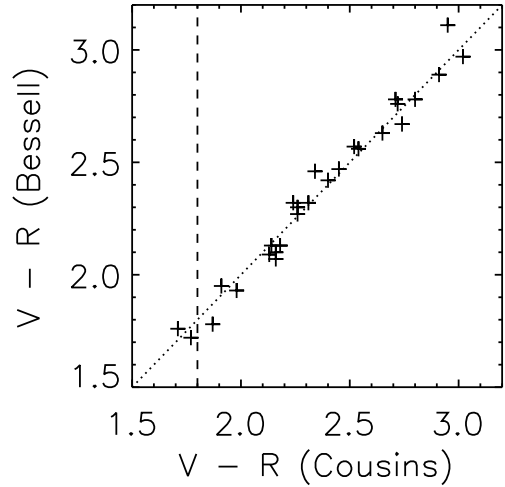}}
    \subfigure[]
	{\includegraphics[scale=0.6,angle=0]{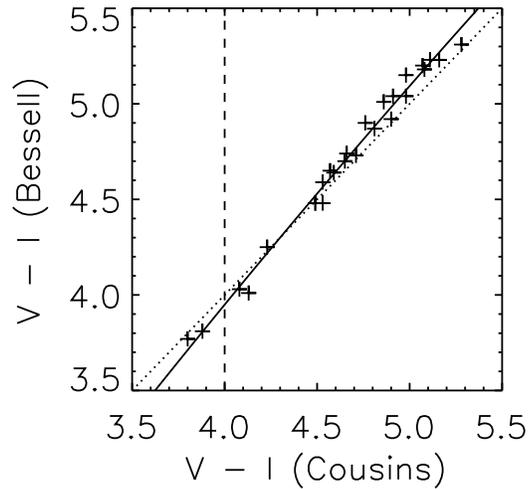}}
 \figcaption[figure5] {\scriptsize  Comparison of photometry for 28 objects observed on both
the {\it CTIO} 0.9m telescope (Kron-Cousins filters) and the {\it SOI} instrument on 
the {\it SOAR} 4.1m telescope (Bessell filters). The dotted line indicates
a 1 to 1 relation. The {\it V} band is photometrically identical on both
systems. Panel ``a'' shows that there is no systematic difference between $R_C$ and
$R_B$. Panel ``b'' shows the trend for the {\it I} band. The solid line
represents the polynomial fit $(V - I_B) = -0.0364(V - I_C)^2 +1.4722(V - I_C) - 1.3563$.
Most of the difference in the {\it I} band is likely due to different sensitivities
between the two detectors in the far red. The dashed vertical lines indicate the red
limit of the \citet{Bessell1995} color relations for the two filter systems.}
 \end{center}
\end{figure}

\clearpage


\subsection{New Trigonometric Parallaxes}
\label{subsec:newpi}
As reported in Table 4, our trigonometric parallax measurements have a
mean uncertainty of 1.43 mas, corresponding to a distance error of
$\sim$1\% at 10 pc and $\sim$3.5\% at our original distance horizon of
25 pc. When comparing our results to other samples observed by {\it
CTIOPI} we found that nearby late M and early L targets tend to be
ideal targets for optical parallax investigations on one meter class
telescopes.  Although the intrinsic faintness of the targets made them
a challenge in nights with poor seeing or a bright moon, the parallax
solution converged with fewer epochs and had smaller errors than what
we experience for brighter samples. We suspect that several factors
contribute to this good outcome. First, the long exposures average out
short atmospheric anomalies that may cause asymmetric Point Spread
Functions ({\it PSFs}). The resulting symmetric {\it PSF} profiles
facilitate centroiding. Second, the long exposures generate images 
rich in background stars that are likely more distant than reference
stars available in shorter exposures.  Because exposure times for
brighter targets are often limited by the time it takes for the
science target to saturate the detector, these faint and distant
reference stars are not available for brighter parallax
targets. Third, as already mentioned, the use of the {\it I}  band
minimizes atmospheric refraction when compared to other optical bands.

From a mathematical point of view, solving a trigonometric parallax
consists of fitting the measured apparent displacements of the science
target to an ellipse whose eccentricity and orientation is
pre-determined by the target's position in the celestial sphere.
At the same time, we deconvolve the constant linear component of motion due to the
object's proper motion. The size of the ellipse's major and minor axes provide a
measure of the object's distance.  Figure 6 shows the parallax
ellipses for our observations. In these plots a parallax factor of 1
or -1 indicates the target's maximum apparent displacement from its
mean position in the right ascension axis. Because we restricted the
hour angle of our observations to $\pm 30$ minutes
($\S$\ref{sec:astrometricobs}), high parallax factor observations
occurred during evening and morning twilight.  As is clear from Figure
6, these twilight observations are essential for determining the
parallax ellipse's major axis. The extent to which observations with 
lower parallax factors constrained
the final parallax solution depended greatly in the parallax ellipse's
eccentricity.  An object with coordinates close to the ecliptic pole
produces a parallax ellipse that is nearly circular, and in that case
low parallax factor observations can still provide significant
constraints to the parallax solution (e.g., object \# 1).  The opposite occurs with the high
eccentricity parallax ellipses for objects lying close to the ecliptic
plane, where low parallax factor observations contribute little
towards the final solution (e.g., object \# 29).

\clearpage
\begin{figure}[ht]
 \includegraphics[scale=0.8, angle=0]{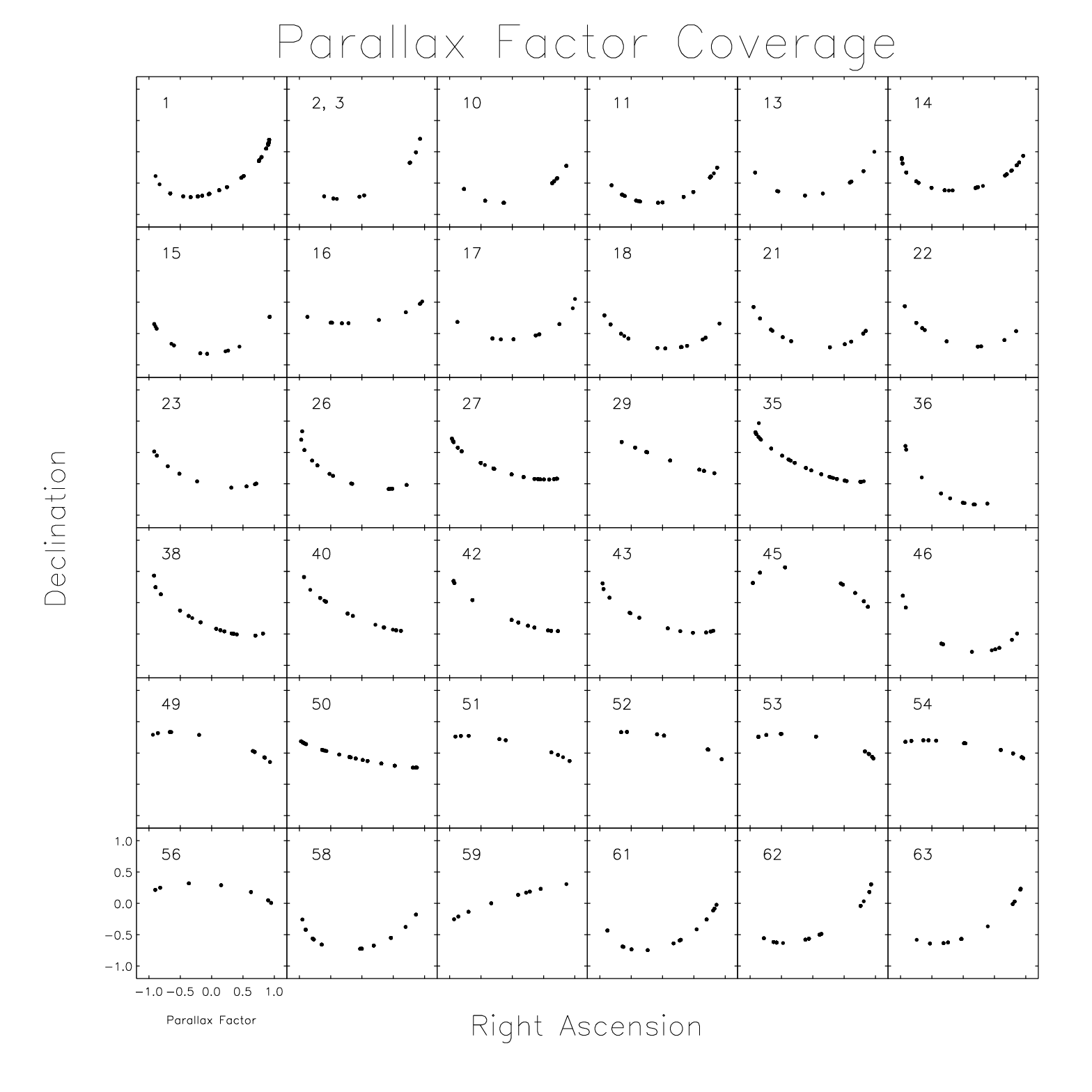}
 \figcaption[Figure 6]{\scriptsize Parallax ellipses for the 37 parallaxes
     reported in this paper (objects 2 and 3 comprise a wide binary and parallaxes are derived
for both components using the same images). The black dots sample the ellipse that each
object appears to trace on the sky as a result of Earth's annual motion. The 
eccentricity of the ellipse is a function of the target's location in the celestial
sphere, with objects close to the ecliptic plane producing the most eccentric ellipses.
Low parallax factor observations provide significant constraints when the ellipse
is not markedly eccentric.}

\end{figure}

\clearpage


Regardless of the target's position on the celestial sphere, we
found out that attempting to fit a parallax ellipse to more than
$\sim$4 epochs but fewer than $\sim$7 epochs will often produce an
erroneous answer whose formal uncertainty is also unrealistically
small. True convergence of a parallax result was best determined by
assuring that the following conditions were met: (1) adding new epochs
caused changes that were small compared to the formal uncertainty; (2)
high parallax factors observations were taken during both evening
twilight and morning twilight; and (3) the parallax ellipses shown in
Figure 6 appeared to be sufficiently sampled so that the points trace out 
a unique ellipse.

Nine of the 37 targets listed in Table 4 have previously published
trigonometric parallaxes. These targets are listed in Table 5 
with our new trigonometric parallax and the previous value. In five
cases trigonometric parallaxes were not yet published at the beginning
of this study in 2009
\citep{Andreietal2011,DupuyandLiu2012,Fahertyetal2012}. 
LHS 4039C \citep{Subasavageetal2009} is a member of a resolved triple
system; we re-reduced our data set with LHS 4039C as the
science target (see $\S$\ref{sec:individual}).  Finally, we
note that LP944-020 \citep{Tinneyetal1996} is no longer a member of
the 5 pc sample and 2MASS J1645-1319 is no longer a member of the
10 pc sample \citep{Henryetal2006}.


\begin{deluxetable}{llccccl}
\tablecolumns{7}
\tabletypesize{\scriptsize}
\tablecaption{Targets with Previously Published Parallaxes}
\label{tab:previousparallaxes}
\tablehead{
           \colhead{ID}           &
           \colhead{Name}         &
           \colhead{New}          &
           \colhead{New}          &
           \colhead{Previous}     &
           \colhead{Previous}     &
           \colhead{Reference}    \\
	   \colhead{ }                   &
           \colhead{ }                   &
           \colhead{$\pi_{abs}$ (mas)}   &
           \colhead{distance (pc)}       &
           \colhead{$\pi_{abs}$ (mas)}   &
           \colhead{distance (pc)}       &
 	   \colhead{   }          }
\startdata
1  & GJ 1001 BC         &   77.02$\pm$2.07 &  12.98$^{+0.36}_{-0.34}$ &  76.86$\pm$3.97 & 13.01$^{+0.71}_{-0.63}$ & \citet{Henryetal2006}   \\
11 & LP 944-020         &  155.89$\pm$1.03 &   6.41$^{+0.04}_{-0.04}$ & 201.40$\pm$4.20 &  4.96$^{+0.11}_{-0.10}$ & \citet{Tinneyetal1996}  \\
26 & 2MASS J0847-1532   &   58.96$\pm$0.99 &  16.96$^{+0.28}_{-0.28}$ &  76.5$\pm$3.5   & 13.07$^{+0.63}_{-0.57}$ & \citet{Fahertyetal2012} \\
27 & LHS 2065           &  117.98$\pm$0.76 &   8.47$^{+0.05}_{-0.05}$ & 117.30$\pm$1.50 &  8.52$^{+0.11}_{-0.11}$ & \citet{vanAltenaetal1995} \\
35 & LHS2397Aab         &   65.83$\pm$2.02 &  15.19$^{+0.48}_{-0.45}$ &  73.0$\pm$2.1   & 13.78$^{+0.41}_{-0.38}$ & \citet{DupuyandLiu2012} \\
49 & DENIS J1539-0520   &   61.25$\pm$1.26 &  16.32$^{+0.34}_{-0.32}$ &  64.5$\pm$3.4   & 15.50$^{+0.86}_{-0.78}$ & \citet{Andreietal2011} \\
54 & 2MASS J1645-1319   &   90.12$\pm$0.82 &  11.09$^{+0.10}_{-0.10}$ & 109.9$\pm$6.1   &  9.01$^{+0.53}_{-0.48}$ & \citet{Fahertyetal2012} \\
56 & 2MASS J1705-0516AB &   55.07$\pm$1.76 &  18.15$^{+0.59}_{-0.56}$ &  45.0$\pm$12.0  & 22.22$^{+8.08}_{-4.68}$ & \citet{Andreietal2011}  \\
62 & LHS 4039C          &   44.38$\pm$2.09 &  22.53$^{+1.11}_{-1.01}$ &  44.24$\pm$1.78 & 22.60$^{+0.95}_{-0.87}$ & \citet{Subasavageetal2009} \\
\enddata

\end{deluxetable}

\clearpage


Table 9 of \citet{DupuyandLiu2012} lists all known ultra-cool dwarfs
with trigonometric parallaxes at the time of that publication. In that
list, 156 objects have spectral types matching the spectral type range
of our study, M6V to L4.  In addition, out of the seventy
trigonometric parallaxes reported by \citet{Fahertyetal2012}, 24 are
first parallaxes for objects in the M6V to L4 spectral type range. The
28 objects for which we publish first parallaxes in this paper
therefore represent a 15.5\% increase in the number of objects with
trigonometric parallaxes in the M6V to L4 spectral type range, for a
total of 208 objects.

\subsection{Effective Temperatures}
\label{subsec:Teff}

While the agreement with interferometric measurements shown in Figure
3 makes us confident that our overall methodology is right, the effective temperatures we
derived based on nine bands of photometry are still essentially model-dependent.
 The uncertainties in temperatures listed in Table 1
and shown by the error bars in Figure 4 can therefore be interpreted
as measures of how accurate the model atmospheres are in a given
temperature range.  Inspection of Figure 4 shows that the models work
very well for temperatures above 2600 K, with uncertainties generally
smaller than 30 K. The uncertainties then progressively increase as
the temperature lowers and can be greater than 100 K for objects
cooler than 2000 K.  The turning point at 2600 K has been explained by
the model authors \citep{Allardetal2012} as a consequence of solid
grain formation starting at that temperature, thus making the
atmosphere significantly more complex.

The year 2012 brought about crucial advances in our ability to
determine effective temperatures for cool stellar (and substellar)
atmospheres.  First, the publication of the {\it WISE} All Sky
Catalog\footnote{http://wise2.ipac.caltech.edu/docs/release/allsky/}
provided uniform photometric coverage in the mid-infrared for 
known cool stars and brown dwarfs. Second, as already discussed, the
publication of the {\it BT-Settl} model atmospheres with revised solar
metallicities has provided opportunities to match
observational data to fundamental atmospheric parameters with
unprecedented accuracy ($\S$\ref{sec:tefflum}).
Despite these recent advances, it is still useful to compare our results with
earlier pioneering work in the field of effective temperature
determination for cool atmospheres. \citet{Golimowskietal2004a}
computed effective temperatures for 42 M, L, and T, dwarfs based on
observations in the {\it L$^{\prime}$} (3.4$-$4.1 $\micron$) and {\it M$^{\prime}$}
(4.6$-$4.8 $\micron$) bands. They first used photometry to calculate
bolometric fluxes based on observed spectra, and then used evolutionary
models \citep{Burrowsetal1997} to determine a range of effective
temperatures based on bolometric luminosities and radii with the
assumption of an age range of 0.1 to 10 Gyr as well as a unique value
for 3 Gyr.  \citet{Cushingetal2008} determined the effective
temperatures of nine L and T dwarfs by fitting observed flux-calibrated
spectra in the wavelength range 0.6$-$14.5 $\micron$ to their own
model atmospheres. Their technique, like ours, has the advantage of
relying solely on atmospheric models as opposed to the significantly
more uncertain evolutionary models, as discussed in detail in
$\S$\ref{subsec:models}.  Finally, \citet{Rajpurohitetal2013} have
recently compared optical spectra (0.52$-$1.0 $\micron$) for 152 M
dwarfs to the same {\it BT-Settl} models we use in this study.
Twenty-five of their M dwarfs have spectral types of M6V or later.

Table 6 compares our results to overlapping objects in these three
studies. While it is difficult to generalize from the small overlap
amongst the different samples, there is a tendency for our results to
be $\sim$100K cooler than the others. The cause of this discrepancy is
not clear. In the case of \citet{Golimowskietal2004a} the most likely
explanation is that their assumed mean age of 3 Gyr may not be
representative of our sample.  An age mismatch combined with the
significant uncertainty in the evolutionary models could easily
account for this temperature difference.  Out of the five objects in
common between this study and \citet{Rajpurohitetal2013}, the
effective temperature for one object agrees well while three objects
have mismatches of $\sim$100K, and another has a significantly larger
mismatch.  While we do not know what is causing the different values,
we note that the comparison of radii derived with our methodology with
empirically measured radii ($\S$\ref{sec:tefflum}, Figure 3) makes
systematic error in our measurements an unlikely explanation. A
temperature difference of $\sim$100K would produce a systematic radius
difference of 5\% to 10\% in the temperature range under
consideration, and yet our derived radii have a mean absolute residual
of only 3.4\% in a random scatter.  Because \citet{Rajpurohitetal2013}
base their calculations on optical spectra alone, we speculate that
the discrepancy may be due to the stronger effects of metallicity in
altering the optical colors of late M dwarfs; a small change in
metallicity can significantly change the slope of the blue end of the
{\it SED}.  Because our method uses twenty different colors composed
of optical, near infrared, and mid infrared bands, the selective
effect of metallicitry in optical colors is ameliorated in our
calculations.


\begin{deluxetable}{rllccccc}
\tablecolumns{8}
\tabletypesize{\scriptsize}
\tablecaption{Comparison of Effective Temperatures from Different Studies}
\setlength{\tabcolsep}{0.15in}
\tablehead{
           \colhead{ID}               &
           \colhead{Name}             &
           \colhead{Spectral}         &
           \colhead{This}             &
           \colhead{G2004\tablenotemark{a}}   &
           \colhead{G2004\tablenotemark{a}}               &
           \colhead{C2008\tablenotemark{a}}       &
           \colhead{R2013\tablenotemark{a}}    \\
           \colhead{ }                &
           \colhead{ }                &
           \colhead{Type }                &
           \colhead{Work}             &
           \colhead{Range}            &
           \colhead{3 Gyr}            &
           \colhead{ }                &
           \colhead{ }                }
\startdata
30   & LHS 292            &  M6.0V    & 2588$\pm$32     & 2475$-$2750 & 2725    &  \nodata & 2700        \\
32   & GJ 406             &  M6.0V    & 2700$\pm$56     & 2650$-$2900 & 2900    &  \nodata & \nodata     \\            
40   & LEHPM2-0174        &  M6.5V    & 2598$\pm$25     & \nodata     & \nodata &  \nodata & 2700        \\
47   & LHS 3003           &  M7.0V    & 2581$\pm$17     & 2350$-$2650 & 2600    &  \nodata & \nodata     \\
38   & LP 851-346         &  M7.5V    & 2595$\pm$28     & \nodata     & \nodata &  \nodata & 2600        \\
 7   & LHS 132            &  M8.0V    & 2513$\pm$29     & \nodata     & \nodata &  \nodata & 2600        \\
27   & LHS 2065           &  M9.0V    & 2324$\pm$27     & 2150$-$2425 & 2400    &  \nodata & \nodata     \\
58   & SIPS J2045-6332    &  M9.0V    & 2179$\pm$111    & \nodata     & \nodata &  \nodata & 2500        \\
 4   & BRI B0021-0214     &  M9.5V    & 2315$\pm$54     & 2150$-$2475 & 2425    &  \nodata & \nodata     \\
20   & 2MASS J0746+2000AB &  L0.0J    & 2310$\pm$51     & 1900$-$2225 & 2200    &  \nodata & \nodata     \\  
44   & 2MASS J1439+1929   &  L1.0     & 2186$\pm$100    & 1950$-$2275 & 2250    &  \nodata & \nodata     \\
41   & Kelu-1AB           &  L2.0J    & 2026$\pm$45     & 2100$-$2350 & 2300    &  \nodata & \nodata     \\
33   & DENIS J1058-1548   &  L3.0     & 1804$\pm$13     & 1600$-$1950 & 1900    &  \nodata & \nodata     \\
 5   & 2MASS J0036+1821   &  L3.5     & 1796$\pm$33     & 1650$-$1975 & 1900    &  1700    & \nodata     \\
 1   & GJ 1001 BC         &  L4.5     & 1725$\pm$21     & 1750$-$1975 & 1850    &  \nodata & \nodata     \\
60   & 2MASS J2224-0158   &  L4.5     & 1567$\pm$88     & 1475$-$1800 & 1750    &  1700    & \nodata     \\
\enddata
\tablenotetext{a}{G2004 \citet{Golimowskietal2004a}; C2008 \citet{Cushingetal2008}; R2013 \citet{Rajpurohitetal2013}}
\end{deluxetable}

\clearpage


In addition to comparisons to other studies with objects in common to
ours, we compare the general trends of our {\it HR} diagram (Figure 4)
with the values derived by \citet{Konopackyetal2010}. That study used
Keck AO resolved near infrared photometry of M and L binaries as well
as {\it HST} resolved optical photometry to derive effective
temperatures and luminosities. Twenty-two of their targets fall in the
temperature range of our study, but because theirs was a high
resolution AO study there are no targets in common. Figure 7 shows
their results over-plotted on our {\it HR} diagram. The large
uncertainties in \citet{Konopackyetal2010} make their data difficult
to interpret, and are probably a result of the lack of mid infrared
photometry in their methodology.  There is good agreement between
their results and ours at cooler temperatures, but the two trends
steadily diverge for temperatures above $\sim$2000K, with
\citet{Konopackyetal2010} predicting temperatures as much as 500K
cooler for a given luminosity. The discrepancy is probably a result of
atmospheric modeling. While the {\it BT-Settl} models used in our
study predict the rate of atmospheric dust formation and sedimentation
for a wide range of temperatures, the ``DUSTY'' models
\citep{Allardetal2001} used by \citet{Konopackyetal2010} assume the
extreme case where grains do not settle below the photosphere, thus
providing a strong source of opacity. The ``DUSTY'' models replicate
the conditions of L dwarf atmospheres well but gradually become
inadequate at hotter temperatures where grain formation is less
relevant \citep{Allardetal2013}. The additional source of opacity
then causes the M dwarfs to appear cooler and larger than they really
are.


\begin{figure}[!ht]
\begin{center}
      \includegraphics[scale=0.5, angle=0]{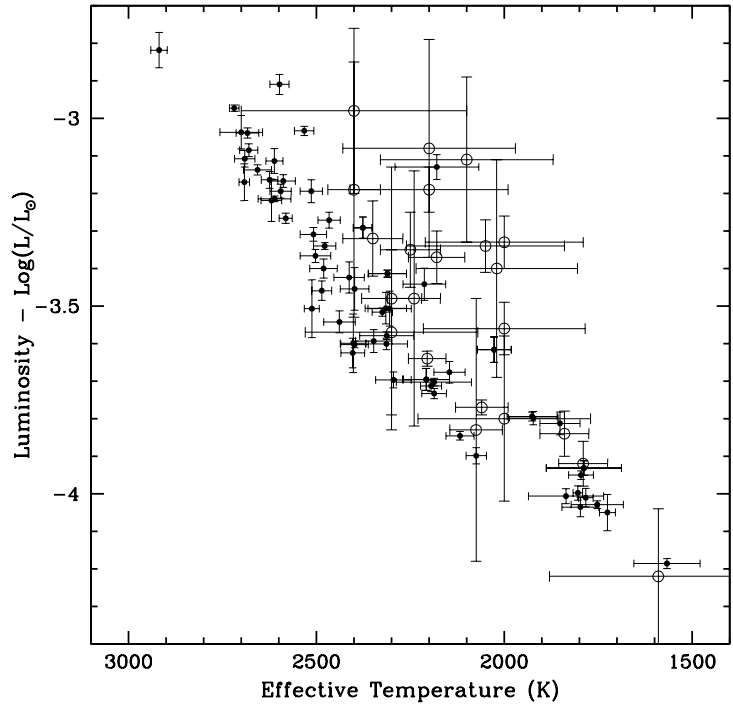}
      \figcaption[Figure 7]{\scriptsize {\it HR} diagram of Figure 5
with data from \citet{Konopackyetal2010} over-plotted with open circles. 
The data agree well at low temperatures, but steadily diverge at higher temperatures.
Both data sets have the minimum radius at $\sim$2075K.}
\end{center}
\end{figure}

\clearpage

\subsection{Color-Magnitude Relations}
\label{subsec:relations}
 
Color-absolute magnitude relations are often the first tools used in
estimating the distance to a star or brown dwarf. Determining useful
relations using only near infrared colors is challenging for
late M and L dwarfs due to the degenerate nature of the near infrared
color-magnitude sequence.  One possible solution is the use of spectral
type-magnitude relations \citep[e.g.,][]{Cruzetal2003}; however, such relations
require accurate knowledge of spectral types in a consistent system
and are subject to the uncertainties inherent to any discrete
classification system. Here we present new color-magnitude relations
based on the optical photometry, {\it 2MASS} photometry, and
trigonometric parallaxes reported in Table 1.  Table 7 presents third
order polynomial fits for all color-magnitude combinations of the
filters {\it VRIJHK$_s$} except for those with the color $R-I$, which
becomes degenerate (i.e. nearly vertical) for $R-I > 2.5$.
 As an example, the first line of Table 7 should be written
algebraically as
$$M_V = 0.21509(V-R)^3 -2.81698(V-R)^2 + 14.16273(V-R)
 -1.45226\; (\pm 0.53)\;\; ; \; 1.61\le(V-R)\le3.64$$
The relations are an extension of those
published in \citet{Henryetal2004} into the very red optical regime.
They are also complementary to the {\it izJ} relations of
\citet{Schmidtetal2010}.  Figure 8 shows the color-absolute magnitude diagrams
and polynomial fits for  $M_v\;vs.\;(V-K_s)$ and $M_{Ks}\;vs.\;(R-K_s)$. Known binaries as well as objects
that are otherwise elevated in the color-magnitude diagrams were
excluded when computing the polynomial fits. The $1\sigma$
uncertainties vary widely by color, and are as small as 0.24
magnitudes for colors that combine the {\it V} filter with the {\it
JHK$_s$} filters.

\clearpage
\begin{deluxetable}{llcccccc}
\tablecolumns{8}
\tabletypesize{\scriptsize}
\tablecaption{Coefficients for Color-Magnitude Polynomial Fits}
\setlength{\tabcolsep}{0.2in}
\tablehead{
           \colhead{abs. mag.}               &
	   \colhead{color}                   &
           \colhead{3$^{rd}$ order}          &
           \colhead{2$^{nd}$ order}          &
           \colhead{1$^{st}$ order}          &
           \colhead{constant}                &
           \colhead{range}                   &
           \colhead{$\sigma$}                }
\startdata
  $M_V$ & $V-R$ &     0.21509 &  $-$2.81698 &     14.16273 &   $-$1.45226 &  1.61$-$3.64  &  0.53 \\
  $M_V$ & $V-I$ &  $-$0.48431 &     7.02913 &  $-$30.24232 &     55.89960 &  3.44$-$6.24  &  0.64 \\
  $M_V$ & $V-J$ &  $-$0.05553 &     1.34699 &   $-$8.78095 &     32.19850 &  5.28$-$9.75  &  0.30 \\
  $M_V$ & $V-H$ &  $-$0.03062 &     0.79529 &   $-$5.04776 &     23.53283 &  5.91$-$11.00 &  0.24 \\
  $M_V$ & $V-K$ &  $-$0.02006 &     0.54283 &   $-$3.22970 &     19.05263 &  6.23$-$11.80 &  0.25 \\
  $M_V$ & $R-J$ &     0.38685 &  $-$4.78978 &     21.67522 &  $-$19.02625 &  3.67$-$6.19  &  0.39 \\
  $M_V$ & $R-H$ &  $-$0.03466 &     0.96656 &   $-$4.96743 &     21.51357 &  4.30$-$7.44  &  0.37 \\
  $M_V$ & $R-K$ &  $-$0.06296 &     1.36944 &   $-$7.28721 &     25.94882 &  5.30$-$8.24  &  0.40 \\
  $M_V$ & $I-J$ &     1.18205 &  $-$9.28970 &     27.59574 &  $-$11.71022 &  1.84$-$3.70  &  0.42 \\
  $M_V$ & $I-H$ &     0.24541 &  $-$2.81568 &     13.65365 &   $-$4.94381 &  2.47$-$4.95  &  0.48 \\
  $M_V$ & $I-K$ &     0.09183 &  $-$1.32390 &      8.75709 &   $-$0.69280 &  2.79$-$5.75  &  0.51 \\
  $M_V$ & $J-H$ &     5.05439 & $-$19.22739 &     30.20127 &      5.70728 &  0.51$-$1.25  &  1.11 \\
  $M_V$ & $J-K$ &     4.35996 & $-$22.11834 &     40.93688 &   $-$5.24138 &  0.80$-$2.05  &  0.92 \\
  $M_V$ & $H-K$ &    23.11303 & $-$54.44877 &     51.74136 &      4.69129 &  0.29$-$0.80  &  0.84 \\
  $M_R$ & $V-R$ &     0.21509 &  $-$2.81698 &     13.16273 &   $-$1.45226 &  1.61$-$3.64  &  0.53 \\
  $M_R$ & $V-I$ &  $-$0.39598 &     5.59585 &  $-$23.46994 &     44.27366 &  3.44$-$6.24  &  0.60 \\
  $M_R$ & $V-J$ &  $-$0.06508 &     1.48971 &   $-$9.70545 &     32.85954 &  5.28$-$9.75  &  0.33 \\
  $M_R$ & $V-H$ &  $-$0.03213 &     0.78557 &   $-$4.94969 &     21.92657 &  5.91$-$11.00 &  0.28 \\
  $M_R$ & $V-K$ &  $-$0.01882 &     0.47360 &   $-$2.65145 &     16.13934 &  6.23$-$11.80 &  0.27 \\
  $M_R$ & $R-J$ &     0.10246 &  $-$0.87144 &      3.43087 &      7.64284 &  3.67$-$6.19  &  0.31 \\
  $M_R$ & $R-H$ &  $-$0.09460 &     1.86774 &   $-$9.85431 &     28.99232 &  4.30$-$7.44  &  0.26 \\
  $M_R$ & $R-K$ &  $-$0.08589 &     1.71800 &   $-$9.39445 &     28.84437 &  5.30$-$8.24  &  0.28 \\
  $M_R$ & $I-J$ &     0.56097 &  $-$4.48907 &     14.76241 &   $-$2.05088 &  1.84$-$3.70  &  0.30 \\
  $M_R$ & $I-H$ &     0.16178 &  $-$1.99780 &     10.43314 &   $-$2.43140 &  2.47$-$4.95  &  0.32 \\
  $M_R$ & $I-K$ &     0.05698 &  $-$0.92853 &      6.77139 &      0.79636 &  2.79$-$5.75  &  0.35 \\
  $M_R$ & $J-H$ &     6.18765 & $-$22.77418 &     31.75342 &      3.61753 &  0.51$-$1.25  &  0.84 \\
  $M_R$ & $J-K$ &     3.81245 & $-$19.70377 &     36.29580 &   $-$4.63378 &  0.80$-$2.05  &  0.69 \\
  $M_R$ & $H-K$ &    26.23045 & $-$59.75888 &     51.53576 &      3.25785 &  0.29$-$0.80  &  0.65 \\
  $M_I$ & $V-R$ &  $-$0.30086 &     1.51360 &      1.22679 &      6.92302 &  1.61$-$3.64  &  0.55 \\
  $M_I$ & $V-I$ &  $-$0.48431 &     7.02913 &  $-$31.24230 &     55.89957 &  3.44$-$6.24  &  0.64 \\
  $M_I$ & $V-J$ &  $-$0.08775 &     2.06323 &  $-$14.53807 &     43.98302 &  5.28$-$9.75  &  0.37 \\
  $M_I$ & $V-H$ &  $-$0.05264 &     1.35632 &  $-$10.21586 &     35.60933 &  5.91$-$11.00 &  0.32 \\
  $M_I$ & $V-K$ &  $-$0.03540 &     0.96451 &   $-$7.46468 &     29.33155 &  6.23$-$11.80 &  0.30 \\
  $M_I$ & $R-J$ &     0.05949 &  $-$0.06127 &   $-$1.61308 &     15.59044 &  3.67$-$6.19  &  0.37 \\
  $M_I$ & $R-H$ &  $-$0.15245 &     3.02932 &  $-$17.57707 &     43.57051 &  4.30$-$7.44  &  0.29 \\
  $M_I$ & $R-K$ &  $-$0.13466 &     2.76845 &  $-$16.87739 &     44.05227 &  5.30$-$8.24  &  0.29 \\
  $M_I$ & $I-J$ &     0.34947 &  $-$2.28896 &      7.32700 &      3.70661 &  1.84$-$3.70  &  0.26 \\
  $M_I$ & $I-H$ &  $-$0.05625 &     0.65099 &   $-$0.12146 &      8.93958 &  2.47$-$4.95  &  0.27 \\
  $M_I$ & $I-K$ &  $-$0.07979 &     0.96350 &   $-$1.79721 &     11.08030 &  2.79$-$5.75  &  0.30 \\
  $M_I$ & $J-H$ &     1.00902 &  $-$9.06504 &     20.09129 &      4.41802 &  0.51$-$1.25  &  0.76 \\
  $M_I$ & $J-K$ &     2.30591 & $-$13.03453 &     26.79644 &   $-$2.67247 &  0.80$-$2.05  &  0.62 \\
  $M_I$ & $H-K$ &     6.36493 & $-$24.92761 &     31.89872 &      4.41508 &  0.29$-$0.80  &  0.60 \\
  $M_J$ & $V-R$ &  $-$0.33337 &     2.21935 &   $-$2.66206 &      9.68513 &  1.61$-$3.64  &  0.39 \\
  $M_J$ & $V-I$ &  $-$0.36037 &     5.32260 &  $-$24.29187 &     45.23880 &  3.44$-$6.24  &  0.45 \\
  $M_J$ & $V-J$ &  $-$0.05553 &     1.34699 &   $-$9.78094 &     32.19847 &  5.28$-$9.75  &  0.30 \\
  $M_J$ & $V-H$ &  $-$0.03301 &     0.88633 &   $-$6.95483 &     26.83274 &  5.91$-$11.00 &  0.26 \\
  $M_J$ & $V-K$ &  $-$0.02220 &     0.63644 &   $-$5.19321 &     22.87309 &  6.23$-$11.80 &  0.25 \\
  $M_J$ & $R-J$ &     0.10247 &  $-$0.87144 &      2.43089 &      7.64280 &  3.67$-$6.19  &  0.31 \\
  $M_J$ & $R-H$ &  $-$0.07510 &     1.64461 &  $-$10.09617 &     28.93402 &  4.30$-$7.44  &  0.25 \\
  $M_J$ & $R-K$ &  $-$0.07709 &     1.67484 &  $-$10.64410 &     31.00300 &  5.30$-$8.24  &  0.26 \\
  $M_J$ & $I-J$ &     0.34947 &  $-$2.28897 &      6.32701 &      3.70660 &  1.84$-$3.70  &  0.26 \\
  $M_J$ & $I-H$ &  $-$0.01826 &     0.36899 &   $-$0.39111 &      8.86504 &  2.47$-$4.95  &  0.25 \\
  $M_J$ & $I-K$ &  $-$0.05043 &     0.71480 &   $-$1.94540 &     10.88528 &  2.79$-$5.75  &  0.26 \\
  $M_J$ & $J-H$ &  $-$1.75450 &     1.46627 &      6.33858 &      6.88885 &  0.51$-$1.25  &  0.52 \\
  $M_J$ & $J-K$ &     0.89335 &  $-$5.42563 &     12.79557 &      2.55729 &  0.80$-$2.05  &  0.44 \\
  $M_J$ & $H-K$ &  $-$0.71691 &  $-$5.04246 &     13.53644 &      6.37982 &  0.29$-$0.80  &  0.44 \\
  $M_H$ & $V-R$ &  $-$0.14308 &     0.68377 &      1.11084 &      6.14662 &  1.61$-$3.64  &  0.33 \\
  $M_H$ & $V-I$ &  $-$0.27640 &     4.04822 &  $-$18.12088 &     34.99090 &  3.44$-$6.24  &  0.38 \\
  $M_H$ & $V-J$ &  $-$0.04698 &     1.11998 &   $-$7.94481 &     26.92333 &  5.28$-$9.75  &  0.25 \\
  $M_H$ & $V-H$ &  $-$0.03062 &     0.79529 &   $-$6.04775 &     23.53280 &  5.91$-$11.00 &  0.24 \\
  $M_H$ & $V-K$ &  $-$0.02177 &     0.59741 &   $-$4.70539 &     20.62932 &  6.23$-$11.80 &  0.24 \\
  $M_H$ & $R-J$ &     0.05414 &  $-$0.30135 &      0.23514 &      9.78688 &  3.67$-$6.19  &  0.27 \\
  $M_H$ & $R-H$ &  $-$0.09460 &     1.86773 &  $-$10.85430 &     28.99230 &  4.30$-$7.44  &  0.26 \\
  $M_H$ & $R-K$ &  $-$0.08348 &     1.70954 &  $-$10.45700 &     29.39801 &  5.30$-$8.24  &  0.27 \\
  $M_H$ & $I-J$ &     0.19024 &  $-$1.18253 &      3.72420 &      5.16641 &  1.84$-$3.70  &  0.24 \\
  $M_H$ & $I-H$ &  $-$0.05625 &     0.65099 &   $-$1.12146 &      8.93959 &  2.47$-$4.95  &  0.27 \\
  $M_H$ & $I-K$ &  $-$0.06209 &     0.76792 &   $-$1.96050 &     10.17762 &  2.79$-$5.75  &  0.28 \\
  $M_H$ & $J-H$ &  $-$1.75451 &     1.46627 &      5.33858 &      6.88885 &  0.51$-$1.25  &  0.52 \\
  $M_H$ & $J-K$ &     1.03555 &  $-$6.09029 &     13.19045 &      2.07593 &  0.80$-$2.05  &  0.45 \\
  $M_H$ & $H-K$ &     0.29861 &  $-$8.11097 &     14.59934 &      5.74926 &  0.29$-$0.80  &  0.43 \\
  $M_K$ & $V-R$ &  $-$0.00121 &  $-$0.37537 &      3.46600 &      4.16422 &  1.61$-$3.64  &  0.31 \\
  $M_K$ & $V-I$ &  $-$0.20466 &     3.00271 &  $-$13.26362 &     27.36341 &  3.44$-$6.24  &  0.34 \\
  $M_K$ & $V-J$ &  $-$0.03630 &     0.86706 &   $-$6.07434 &     22.17402 &  5.28$-$9.75  &  0.24 \\
  $M_K$ & $V-H$ &  $-$0.02655 &     0.68316 &   $-$5.12699 &     20.83821 &  5.91$-$11.00 &  0.24 \\
  $M_K$ & $V-K$ &  $-$0.02006 &     0.54283 &   $-$4.22970 &     19.05261 &  6.23$-$11.80 &  0.25 \\
  $M_K$ & $R-J$ &     0.04475 &  $-$0.23500 &      0.13773 &      9.38717 &  3.67$-$6.19  &  0.26 \\
  $M_K$ & $R-H$ &  $-$0.10097 &     1.92794 &  $-$11.03651 &     28.84667 &  4.30$-$7.44  &  0.27 \\
  $M_K$ & $R-K$ &  $-$0.08589 &     1.71800 &  $-$10.39445 &     28.84436 &  5.30$-$8.24  &  0.28 \\
  $M_K$ & $I-J$ &     0.14572 &  $-$0.88536 &      2.93380 &      5.59314 &  1.84$-$3.70  &  0.25 \\
  $M_K$ & $I-H$ &  $-$0.09030 &     0.98749 &   $-$2.35957 &     10.16948 &  2.47$-$4.95  &  0.28 \\
  $M_K$ & $I-K$ &  $-$0.07979 &     0.96350 &   $-$2.79722 &     11.08030 &  2.79$-$5.75  &  0.30 \\
  $M_K$ & $J-H$ &  $-$1.86948 &     1.75166 &      4.48358 &      6.92764 &  0.51$-$1.25  &  0.49 \\
  $M_K$ & $J-K$ &  $-$1.86948 &     1.75166 &      4.48358 &      6.92764 &  0.51$-$1.25  &  0.49 \\
  $M_K$ & $H-K$ &     0.29864 &  $-$8.11102 &     13.59936 &      5.74925 &  0.29$-$0.80  &  0.43 \\
\enddata
\end{deluxetable}

\clearpage

\addtolength{\voffset}{-0.5in}
\begin{landscape}
\begin{figure}[ht]
       \subfigure[]
      {\includegraphics[scale=0.55]{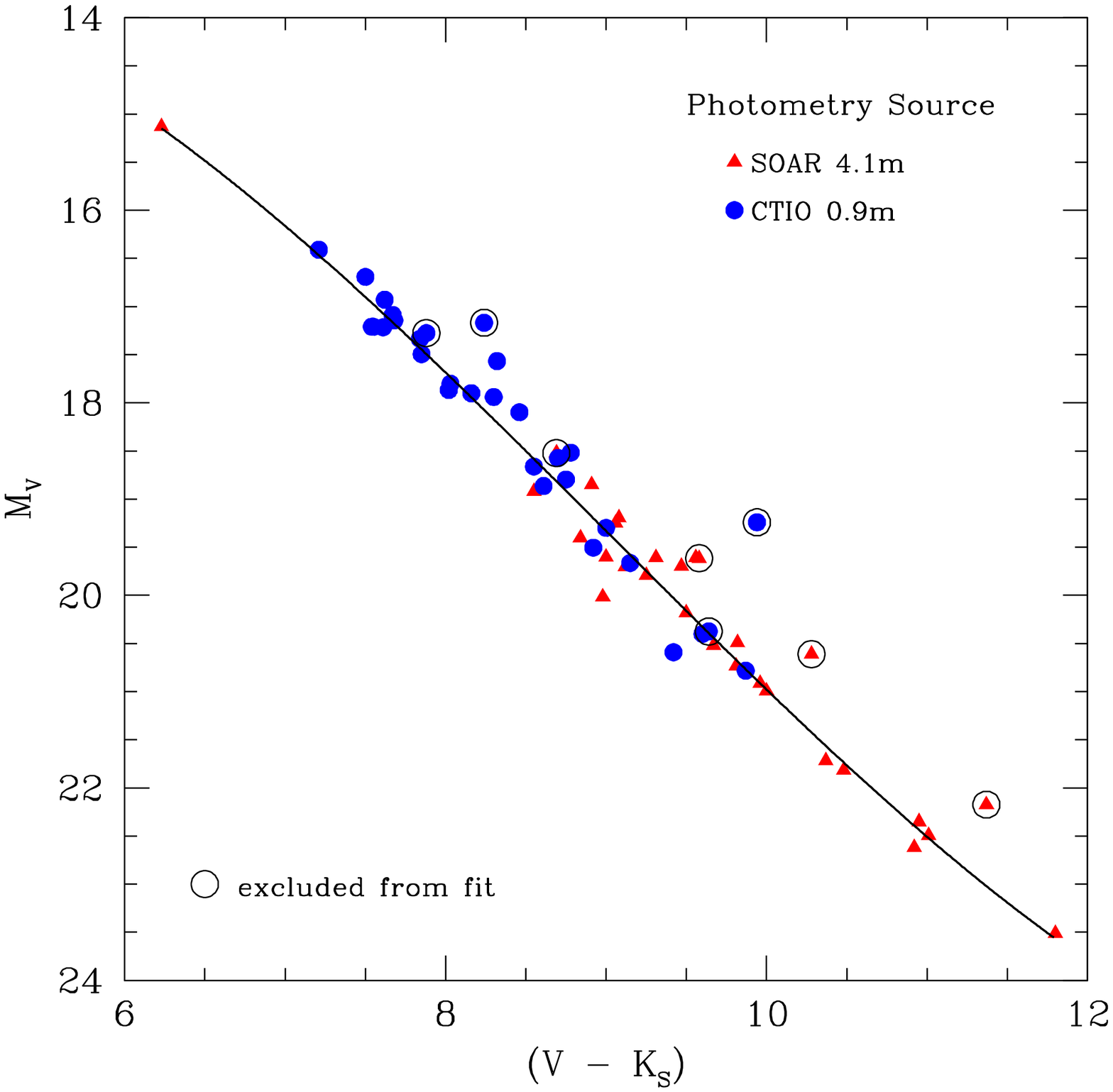}}
\subfigure[]
      {\includegraphics[scale=0.55]{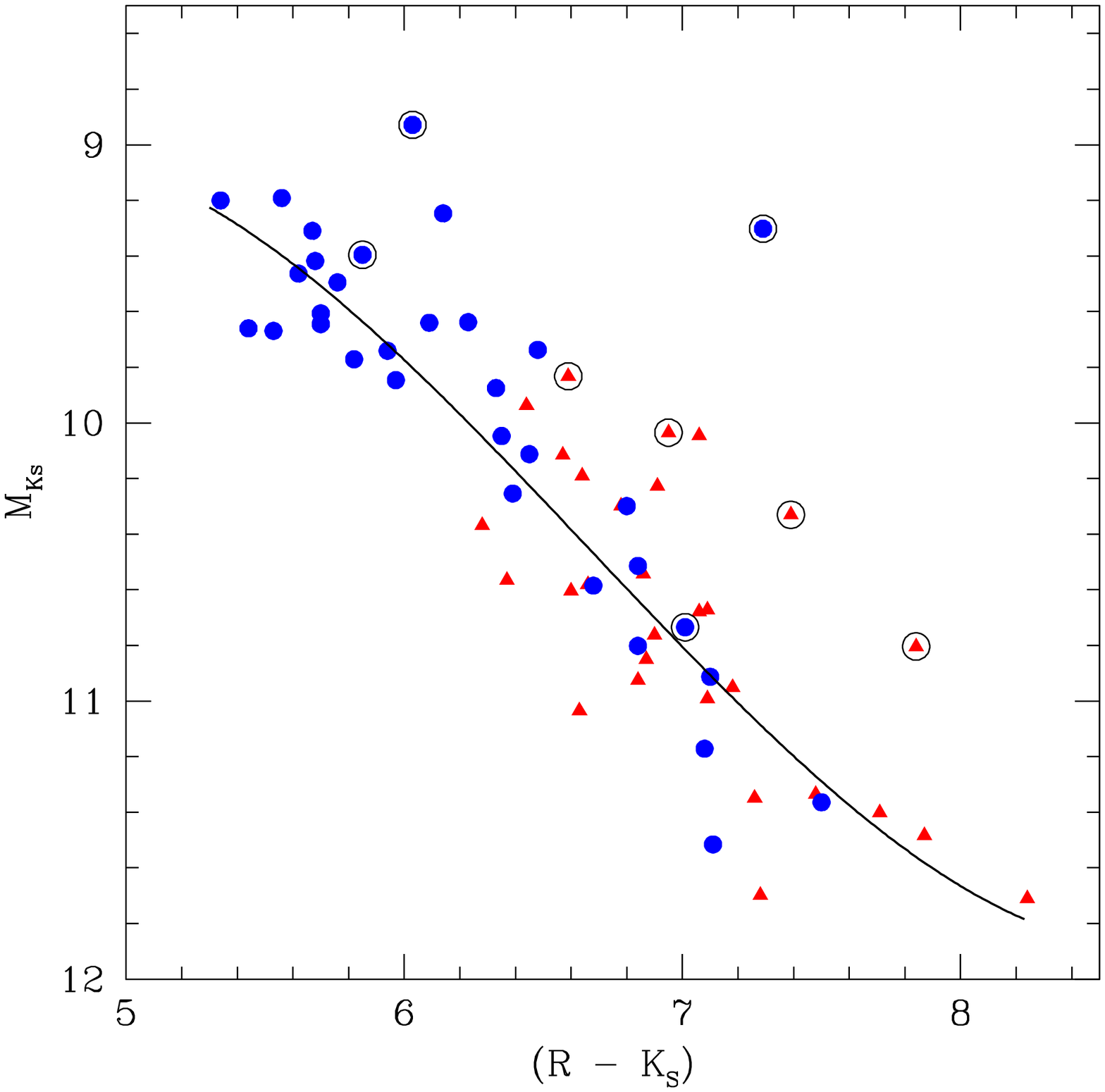}}
       \figcaption[Figure 6]{\scriptsize Example color-absolute magnitude diagrams
     over-plotted with third order polynomial fits. The M$_v$ $\times$ (V$-$K) 
     relation shown in (a) has a particularly low uncertainty ($\sigma = 0.25$ mag)
     due to the steep decrease in $V$ band flux in the late M and L
     dwarf sequence. The M$_k$ $\times$ (R$-$K) relation shown in (b)
     has a slightly higher uncertainty ($\sigma = 0.28$ mag) but is more
     practical from an observational point of view due to the difficulty
     in obtaining $V$ band photometry for L dwarfs.  Binary or otherwise 
     elevated objects were excluded from the polynomial fits and are
     shown enclosed with open circles. Figure b uses the same labeling scheme
     as figure a.}

\end{figure}
\end{landscape}
\clearpage

\addtolength{\voffset}{0.5in}


\subsection{Optical Variability}
\label{subsec:variability}

Photometric variability in very low mass stars and brown dwarfs has
lately become an active area of research because variability can serve
as a probe of many aspects of an object's atmosphere \citep[e.g,][]{Heinzeetal2013}.
 The leading candidate mechanisms thought to cause
photometric variability are non-uniform cloud coverage in L and early T dwarfs
\citep[e.g.,][]{Radiganetal2012, Apaietal2013},
optical emission due to magnetic activity, and the existence of cooler
star spots due to localized magnetic activity. The period
of variability is often thought to correspond to the object's period of
rotation. \citet{Hardingetal2011} have suggested that the link between
optical variability and radio variability in two L dwarfs is
indicative of auroral emission analogous to that seen in
Jupiter. \citet{Khandrikaetal2013} report an overall variability
fraction of 36$^{+7}_{-6}$\% for objects with spectral types ranging
from L0 to L5 based on their own observations as well as 
six previous studies \citep{BailerJonesandMundt2001,Gelinoetal2002,
Koen2003,Enochetal2003,Koenetal2004,Koen2005}. The threshold for variability of
these studies ranged from 10 to 36 milli-magnitudes and were conducted using various photometric bands.

We have measured {\it I} band photometric variability as part of
our parallax observations. Differential photometry of the parallax
target is measured with respect to the astrometric reference
stars. Any reference star found to be variable to more than 50 milli-magnitudes is
discarded and the remaining stars are used to determine the baseline
variability for the field. Details of the procedure are discussed in
\citet{Jaoetal2011}. Figure 9 shows the $1\sigma$ variability for 36
parallax targets\footnote{GJ 1001BC is photometrically contaminated by
the much brighter A component, and was therefore excluded from the
variability study.}. Because the parallax targets were mostly fainter
than the reference stars, photometric signal-to-noise of the target objects is the limiting factor
for sensitivity to variability. This limit becomes more pronounced
for cooler and fainter stars, thus creating the upward linear trend
for the least variable objects in Figure 9. Because of this trend, we have 
conservatively set the threshold for deeming a target variable at 15 
milli-magnitudes, as indicated by the dashed line in Figure 9. 
We detect 13 variable objects out of 36, corresponding to an overall
variability of 36$^{+9}_{-7}$\% where the uncertainties are calculated using
binomial statistics\footnote{A review of binomial statistics as applied to stellar
populations can be found in the appendix of \citet{Burgasseretal2003}}.
 While this result is in excellent agreement to that of
\citet{Khandrikaetal2013} (36$^{+7}_{-6}$\%), we note that our sample
includes spectral types M6V to L4, while theirs ranges from L0 to L5.
Targets found to be significantly variable are labeled in Figure 9 with their
ID numbers. The objects 2MASS J0451-3402 (L0.5, ID 15), 2MASS J1705-0516AB
(L0.5 joint type, ID 56), and SIPS J2045-6332 (M9.0V, ID 58) stand out
as being much more variable than the rest of the sample. We defer discussion
of these objects until $\S$\ref{sec:individual}.

\begin{figure}[ht!]
\begin{center}
      \includegraphics[scale=0.5, angle=0]{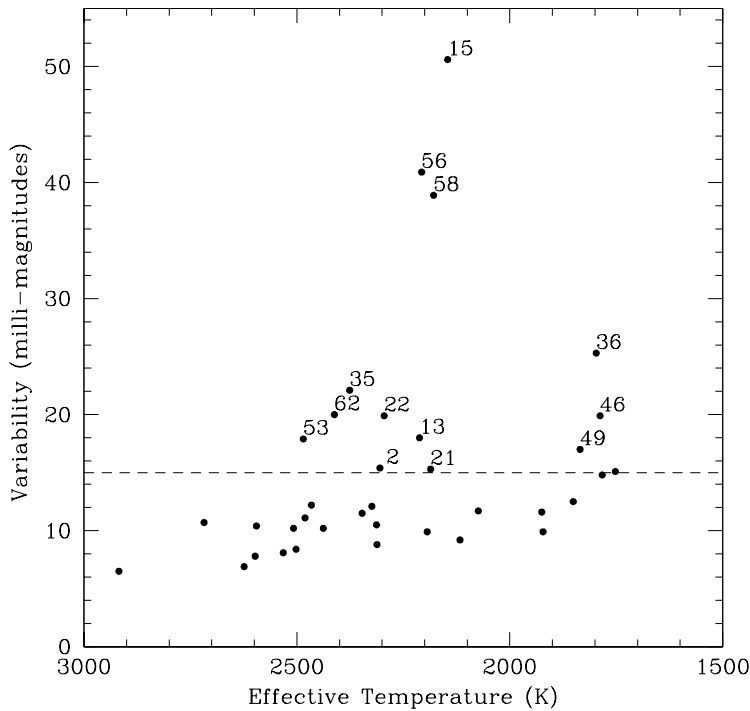}
      \figcaption[Figure 7]{\scriptsize {\it I} band photometric variability
derived from trigonometric parallax observations. The linear increase
in minimum variability with decreasing temperature is most likely 
not real and caused by lower signal-to-noise for fainter targets.
To account for this trend we established the threshold for deeming a
target variable at 15 milli-magnitudes, indicated by the dashed line.
Thirteen out of a sample of
36 targets are above the threshold and are labeled wit the ID number used
in Tables 1 and 4. See $\S$\ref{sec:individual} for a discussion of the three most variable
targets.}

\end{center}
\end{figure}
\clearpage


\subsection{ DENIS J1454-6604AB $-$ A New Astrometric Binary}
\label{subsec:den1454}
DENIS J1454-6604AB is an L3.5 dwarf first identified by
\citet{Phan-Baoetal2008}.  We report a trigonometric parallax of
84.88$\pm$1.71 mas, corresponding to a distance of
11.78$^{+0.24}_{-0.23}$ pc. Figure 10 shows the residuals to the
trigonometric parallax solution, denoting the motion of the object's
photocenter once the parallax reflex motion and the proper motion have been
subtracted. The sinusoidal trend in the RA axis is strong indication
of an unseen companion that is causing the system's photocenter to
move with respect to the reference stars. The absence of a
discernible trend in the declination axis indicates that the system
must be nearly edge-on and its orbit  has an orientation that is
predominantly East-West. At this stage it is not possible to determine
the system's period or component masses. While it may appear in Figure 10 that the
system has completed nearly half an orbital cycle in the $\sim$4 years
that we have been monitoring it, unconstrained eccentricity means that
the system may take any amount of time to complete the remainder of its orbit.

Once the full orbit of a photocentric astrometric system is mapped,
determining the mass and luminosity ratio of the system is a
degenerate problem. The same perturbation can be produced by either a
system where the companion has a much lower mass and luminosity than
the primary or by a system where the components have {\it nearly} the
same mass and luminosity.
We note that a system where the two components are
{\it exactly} equal would be symmetric about the barycenter and would
therefore produce no perturbation at all. The fact that the system
appears elevated in the {\it HR} diagram (Figure 4) is an indication that the
secondary component is contributing considerable light and that
therefore the nearly equal mass scenario is more likely. As described
in \citet{Dieterichetal2011}, a single high resolution image where
both components are resolved is enough to determine the flux ratio of
the components and therefore determine individual dynamical masses once the full
photocentric orbit has been mapped.

We will closely monitor this system with the goal of reporting the component masses
once orbital mapping is complete.


\begin{figure}[ht]
\begin{center}
       \includegraphics[scale=0.4, angle=90]{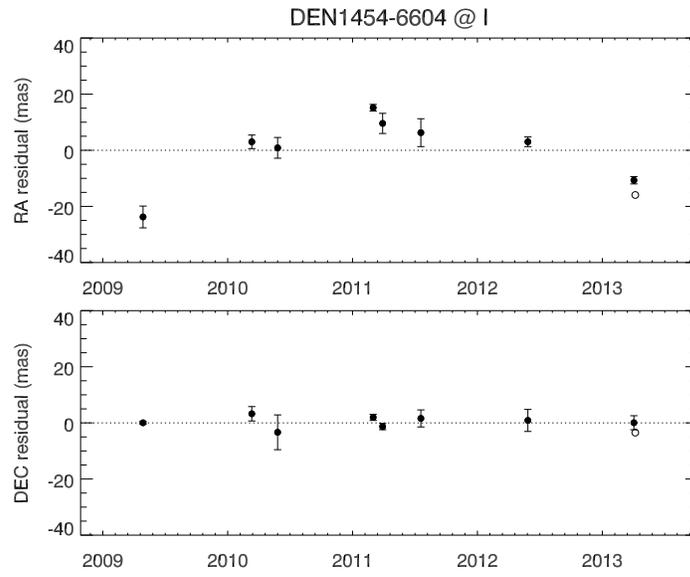}
       \figcaption[Figure10]{\scriptsize Astrometric residuals indicating  a perturbation on the photocenter position
for DENIS J1454-6604 with data taken in the $I$ band. The dots correspond to the
positions of the system's photocenter once proper motion and parallax reflex motion are
removed. Each solid dot is the nightly average of typically three consecutive observations.
The open dot represents a night with a single observation. The lack of a discernible
perturbation in the declination axis indicates that the system is viewed nearly edge-on
and that its orbital orientation is primarily East$-$West.}
\end{center}
\end{figure}
\clearpage

\begin{figure}[ht]
\begin{center}
   \subfigure[]
      {\includegraphics[scale=0.4, angle=-90]{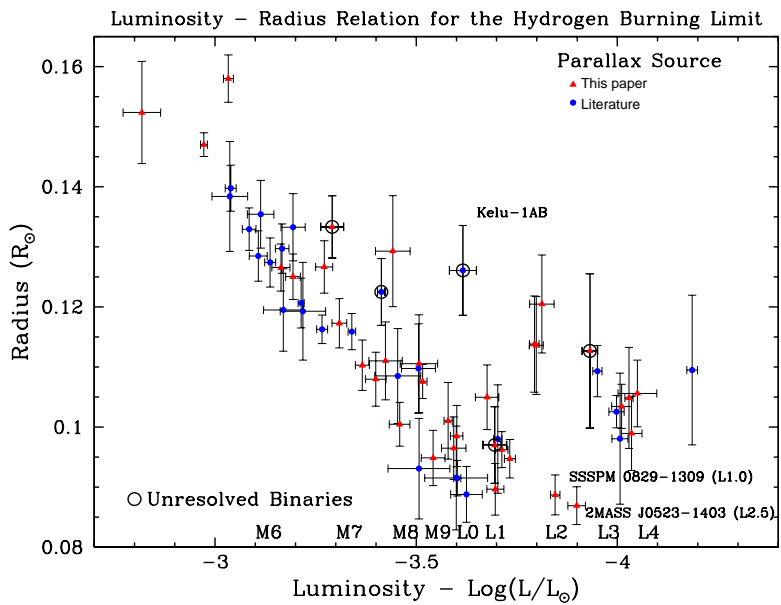}}
\subfigure[]
      {\includegraphics[scale=0.4, angle=-90]{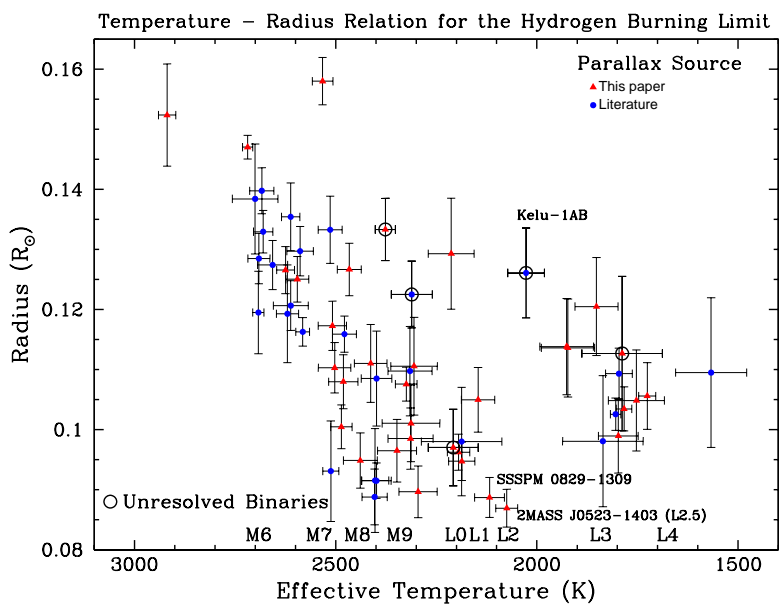}}
 \figcaption[figure11]{\scriptsize (a) Luminosity$-$Radius and (b)
Temperature$-$Radius diagrams for the observed sample. The targets are
the same as in Figure 5 except for LEHPM2-0174 and SIPS J2045-6332,
which were excluded for scaling purposes due to their large radii and are discussed
in $\S$\ref{sec:individual}. These diagrams provide the same
fundamental information as an {\it HR} diagram, but make radius easier
to visualize. Once known and suspected binaries are excluded, the
radius trends have a minimum about 2MASS J0523-1403 (L2.5),
indicating the onset of core electron degeneracy for cooler
objects.  The location and relevance
of Kelu-1AB is discussed in $\S$\ref{sec:individual}.
Versions of these diagrams that
use the ID labels in Table 1 and spectral types for plotting symbols are available as
supplemental online material.}
\end{center}
\end{figure}

\clearpage

\section{Discussion $-$ The End of the Stellar Main Sequence}
\label{sec:discussion}
One of the most remarkable facts about {\it VLM} stars is the fact
that a small change in mass or metallicity can bring about profound
changes to the basic physics of the object's interior, if the change in
mass or metallicity places the object in the realm of the brown
dwarfs, on the other side of the hydrogen burning minimum mass limit.
If the object is unable to reach thermodynamic equilibrium through
sustained nuclear fusion, the object's collapse will be halted by
non-thermal electron degeneracy pressure. The macroscopic properties
of ({\it sub})-stellar matter are then ruled by different physics and
obey a different equation of state
\citep[e.g.,][]{SaumonChabrierandvanHorn1995}.  Once electron
degeneracy sets in at the core, the greater gravitational force of a
more massive object will cause a larger fraction of the brown dwarf to
become degenerate, causing it to have a smaller radius.  The
mass-radius relation therefore has a pronounced local minimum near the
critical mass attained by the most massive brown dwarfs
\citep{Chabrieretal2009,Burrowsetal2011}.  Identifying
the stellar/substellar boundary by locating the minimum radius in the
stellar/substellar sequence has an advantage over other search
methods (e.g., a dynamical mass search): while the {\it values}
associated with the locus of minimum radius depend on the
unconstrained details of evolutionary models
($\S$\ref{subsec:models}), its {\it existence} is a matter of basic
physics and is therefore largely model independent.

In Figures 11 (a) and (b) we re-arrange the {\it HR} diagram of Figure
5 to make radius an explicit function of luminosity (a), and
effective temperature (b). We do not plot the data for LEHPM2-0174 and
SIPS J2045-6332, which have abnormally elevated radii and would have made
the figures difficult to read. These two objects are discussed individually
in $\S$\ref{sec:individual}.
After excluding the objects marked as known
binaries and a few other elevated objects that we suspect are binary
or young objects, 
both diagrams show the inversion of the radius trend near the location of
the L2.5 dwarf 2MASS J0523-1403. 
Figures 11 (a) and (b) can be
compared to Figures 3, 4, and 5 of
\citet{Burrowsetal2011} and Figures 1 and 2 of
\citet{Chabrieretal2009} for insight into how our data fit the
predicted local minimum in the radius trends. While these works
examine radius as a theoretical function of mass at given isochrones, there is a remarkable
similarity between the overall shape of the theoretical mass-radius
trend and the luminosity-radius and temperature-radius trends we
detect empirically. The real data are likely best represented by a combination
of isochrones that is biased towards the ages at which high mass substellar objects
shine as early L dwarfs (See $\S$\ref{subsec:discontinuity} and Figures 12$-$15). 
While Figure 1 of \citet{Chabrieretal2009} indicates radii as small as $\sim0.075R_{\sun}$
for the 10 Gyr isochrone, we should not expect to see radii this small in this study
because substellar objects with that age are no longer in the luminosity range we
observed (M6V to L4, $\S$\ref{sec:sample}). The same argument is valid for 
the figures in \citet{Burrowsetal2011}.
Indeed, because luminosity and temperature are
primarily functions of mass for stars and primarily functions of mass
and age for brown dwarfs, our plots in Figure 11 essentially replicate
the morphology of the mass-radius relation with added dispersion
caused by the observed sample's finite ranges of metallicity and
age.

2MASS J0523-1403 has $T_{eff}=2074\pm27K$,
$log(L/L_{\odot})=-3.898\pm0.021$,
$(R/R_{\sun})=0.086\pm0.0031$, and $V - K = 9.42$. While we cannot exclude the possibility
of finding a stellar object with smaller radius, it is unlikely that such an
object would be far from the immediate vicinity of 2MASS J0523-1403 in
these diagrams.  If cooler and smaller radii stars exist,
they should be more abundant than the brown dwarfs forming the
upward radius trend at temperatures cooler than 1900 K in Figure 11b
because such stars would spend the vast majority of their lives on the
main sequence, where their positions in the diagrams would be almost
constant, whereas brown dwarfs would constantly cool and fade, thus moving to the right in the plots.
We detect no such objects. We note that while the brown dwarfs cooler than 1900
K in Figure 11b are brighter than any putative lower radius star of
the same temperature, the difference would amount to only $\sim$0.3
magnitudes, which is not enough to generate a selection effect in
our sample definition.

\subsection{A Discontinuity at the End of the Main Sequence}
\label{subsec:discontinuity}

Figure 11 shows a relative paucity of objects forming a gap at temperatures
(luminosities) immediately cooler (fainter) than 2MASS J0523-1403.
This gap is then followed by a densely populated region where radius
has an increasing trend in both diagrams. Although we cannot at
this point exclude the hypothesis that this gap is due to statistics
of small numbers, we note that the existence of such a gap is
consistent with the onset of the brown dwarf cooling curve. The
stellar sequence to the left-hand-side of 2MASS J0523-1403 is well
populated because {\it VLM} stars have extremely long main
sequence lives, therefore holding their positions in the diagrams. The
space immediately to the right-hand-side of 2MASS J0523-1403 is
sparsely populated because objects in that region must be in a very
narrow mass and age range. They must be very high mass brown dwarfs
that  stay in that high luminosity (temperature)
region for a relatively brief period of time before they fade and cool.
 The population density increases again to the right of
this gap because that region can be occupied by brown dwarfs with
several combinations of age and mass.

The space density as functions of luminosity and effective temperature
inferred from Figure 11 can be compared to theoretical mass and
luminosity functions, while keeping in mind the important caveat that
our observed sample is not volume complete ($\S$\ref{sec:sample}). The
mass functions of \citet{Burgasser2004} and \citet{Allenetal2005} both
predict a shallow local minimum in the space distribution of dwarfs at
temperatures $\sim$2000 K. In particular, Figure 6 of
\citet{Burgasser2004} predicts a relatively sharp drop in space
density at 2000 K, in a manner similar to our results.  However, the
subsequent increase in space density at cooler temperatures is
predicted to be gradual in both \citet{Burgasser2004} and
\citet{Allenetal2005} (Figure 2).  Neither mass function predicts the
sudden increase in space density we see at $\sim$1800 K in Figure
11b. This last point is particularly noteworthy because our sample
selection criteria ($\S$\ref{sec:sample}) aims to evenly sample the
spectral type sequence.  Our selection effect works {\it against} the
detection of any variation in space density as a function of mass and
luminosity, and yet we still detect a sharper gap between $\sim$2000 K
and $\sim$1800 K than what is expected from the mass functions.
\citet{Burgasser2004} also predicts a population with significant stellar content
down to temperatures of $\sim$1900 K, whereas the
temperature-radius and luminosity-radius trends indicate that the
coolest stellar object in our sample is 2MASS J0523-1403, with
$T_{eff}=2074\pm27K$.  In summary, one may say that the current mass
functions are useful in replicating the overall morphology of our
observed distribution, but do not fully explain the detailed structure
we notice at the end of the stellar main sequence. As we discuss in 
$\S$\ref{sec:conclusion}, only observing a truly volume-complete sample
will yield definite answers about population properties such as the mass
function.
 
The discontinuity is even more pronounced in terms of radius: whereas
radius decreases steadily with decreasing temperature until the
sequence reaches 2MASS J0523$-$1403 ($R=0.086R_{\sun}$), it then not
only starts increasing, but jumps abruptly to a group of objects with $R\sim0.1R_{\sun}$.
The discontinuity in
radius is also visible as an offset in the {\it HR} diagram
(Figure 4). This discontinuity is further evidence of the end of the
stellar main sequence and has a simple explanation: whereas stars
achieve their minimum radius at the zero age main sequence, brown
dwarfs continue to contract slightly as they cool
\citep{Burrowsetal1997,Baraffeetal1998,Chabrieretal2000,Baraffeetal2003}.
Substellar objects with radii falling in the discontinuity should
therefore be high mass late L, T, or Y dwarfs and fall outside the
luminosity range of our sample (M6V to L4, $\S$\ref{sec:sample}). We note that this sudden
increase in radius is a different effect than the previously mentioned sudden
decreases in luminosity and temperature, and indeed counteracts the
decrease in luminosity. The fact that these discontinuities occur at
the same location and can be explained by consequences of the
stellar/substellar boundary provides strong evidence that we have
indeed detected the boundary.

The above argument for the causes of the discontinuity also lend
credence to the idea that 2MASS J0523-1403 is a star despite the fact
that it has the smallest radius in the sample. We note that 2MASS
J0523-1403 and the L1.0 dwarf SSSPM J0829-1309 located immediately to
its left fit nicely within the linear stellar sequences in Figures 5 and 11.
As already discussed, we would also expect stars to be
more prevalent around the locus of minimum radius due to the limited
amount of time in which a massive brown dwarf would occupy that
parameter space.  Most importantly, there is a difference between the
{\it local} minimum in the radius trends and the {\it absolute}
minimum. While theory predicts that the object with the smallest
radius should be the most massive {\it brown dwarf}
\citep{Burrowsetal2011}, such an object would not attain
its minimum radius until it cools down and enters the T and Y dwarf
regime, and therefore drifts beyond the luminosity range of this study.

\subsection{Comparison of the {\it HR} Diagram to Evolutionary Models}
\label{subsec:models}
We now compare our results to the predictions of the four most
prevalent evolutionary models encompassing the stellar/substellar
boundary
\citep{Burrowsetal1993,Burrowsetal1997,Baraffeetal1998,Chabrieretal2000,
Baraffeetal2003}\footnote{We note that while \citet{Burrowsetal1997}
is well known for presenting a unified theory of brown dwarf and giant
planet evolution, data in that paper concerning the hydrogen burning
limit are from \citet{Burrowsetal1993}.}. All of these models are the
combination of an interior structure model and an atmospheric model
used as a boundary condition. As previously discussed, atmospheric
models have become highly sophisticated and achieved a great degree of
success over the last several years. On the other hand, the
evolutionary models we discuss here are at least a decade old, and
none of them currently incorporates the state-of-the-art in
atmospheric models. The discrepancy is due in part to the lack of
observational constraints for evolutionary models. While an
atmospheric model may be fully tested against an observed spectrum,
testing an evolutionary model requires accurate knowledge of age and
mass. The available evolutionary models are also hindered by the fact that
none of them incorporate the latest revised solar abundances that
are used to translate observed metallicity diagnostic features into
the number densities for different species used by the models. The
current accepted values for solar abundances \citep{Caffauetal2011}
constitutes a reduction of 22\% when compared to the original
values used by the evolutionary models we discuss here
\citep[e.g.][]{Grevesseetal1993}.  We therefore cannot expect any of the
models we consider here to be strictly correct, but comparing their
predictions to our results is nevertheless a useful endeavor.

Figures 12 through 15 show several evolutionary tracks from these models
over-plotted on our luminosity-radius and temperature-radius diagrams.
Table 8 lists the properties predicted for the hydrogen burning minimum mass (HBMM)
tracks for the four models. We also include the zero metallicity
model of \citet{Burrowsetal1993}, which is listed to illustrate the effect of a
reduction in metallicity. All models except for the unrealistic
zero metallicity model predict the hydrogen burning limit at
significantly cooler temperatures and lower luminosities than our values.
The evolutionary tracks of \citet{Chabrieretal2000} and
\citet{Baraffeetal2003} have reasonable agreement with the
observations for $log(L/L_{\sun})\gtrsim-3.5$, where objects are
solidly in the stellar domain.  \citet{Chabrieretal2000} has also
achieved some success in reproducing the radii of brown dwarfs with
$log(L/L_{\sun})\sim-4.0$, but cannot account for the small radius
of 2MASS J0523-1403 and several other stellar objects.  And while
\citet{Burrowsetal1993,Burrowsetal1997} seems to accurately predict
the radius of the smallest stars, the model radii are too small everywhere
else in the sequence.  In sum, we see that at the level of accuracy
needed to predict the entirety of our observations these models are
for the most part mutually exclusive.

While the differences between our results and model predictions (Table
7) may at first seem large, they must be examined in the context of
the recently revised solar abundances that are 22\% lower than the
metallicities used to compute the models
\citep{Caffauetal2011}. Lowering the metal content of a (sub)stellar
object has the effect of decreasing opacities both in the atmosphere
and in the interior. The net effect is a facilitation of radiative
transfer from the object's core to space and thus a decrease in
the temperature gradient between the core and the atmosphere. Because
in the low metallicity scenario energy escapes the stellar core more
easily, maintaining the minimal core temperature necessary for
sustained hydrogen burning requires a higher rate of energy
generation.  As shown by the $Z/Z_{\sun}=0$ model of Table 8, the
minimum stellar mass, minimum effective temperature, and minimum
luminosity all increase as a result of a decrease in metallicity.
 The effect of metallicity on the minimum luminosity is
particularly strong. When compared to the Burrows $Z/Z_{\sun}=1.28$
model, the $Z/Z_{\sun}=0.00$ model produces a minimum luminosity that
is greater by a factor of 20.4. Our results suggest a minimum
luminosity that is greater than that predicted by the
$Z/Z_{\sun}=1.28$ models by a factor ranging from $\sim$2.0 to $\sim$3.2,
depending upon the model chosen.  From Figure 4 of
\citet{Burrowsetal2011}, a lower metallicity would also cause a more
pronounced local minimum in the radius trends we detect in our Figure 11.

It is interesting to note that if we accept the masses of the several
evolutionary tracks shown in Figures 12 through 15, then three out of
the four models
\citep{Burrowsetal1997,Baraffeetal1998,Baraffeetal2003} show a jump
from stellar masses at $log(L/L_{\sun})\sim3.9$ ($T_{eff}\sim2075\;K$) to
masses $\lesssim 0.050\; M_{\sun}$ for cooler objects. The
\citet{Chabrieretal2000} models show a slightly smaller jump to masses
$\lesssim 0.060 M_{\sun}$.  This interpretation is difficult to
reconcile with the idea of a continuous mass function for substellar
objects. Because more massive objects cool more slowly, we would
expect to see more brown dwarfs in the mass range of
$\sim$0.070$-$0.050 $M_{\sun}$ than less massive objects occupying the
same temperature range.  As an example, the mass function of
\citet{Allenetal2005} predicts the mean mass of spectral type L5 to be
0.067 $M_{\sun}$, and yet comparing our results to evolutionary models
shows masses $\lesssim 0.050 M_{\sun}$ in the L3 temperature range. A
discontinuous mass function that produces objects of stellar mass and
then jumps to such low masses without producing the intermediate mass
objects is not likely. Observations and theory could be reconciled by
either increasing the masses associated with the evolutionary tracks
or decreasing the radii predicted by our {\it SED} fitting
technique. We note however that a systemic over-prediction of radius
values by our fitting technique would likely manifest itself in a
manner independent of spectral type, and would therefore also be
noticeable in the stellar part of Figures 12 through 15 and in our
comparison to interferometric radii (Figure 3).

Finally, we note that while our observations do not address the
minimum mass for hydrogen burning, higher values for mass should also
be expected as a result of the downward revision in solar
abundances. Independent confirmation of this effect through a
dynamical mass study would further enhance the body of evidence we
have presented for the end of the stellar main sequence at values close
to those of 2MASS J0523-1402 (L2.5): $T_{eff}\sim2075K$, $log(L/L_{\sun})\sim-3.9$,
$(R/R_{\sun})\sim0.086$, and $V - K = 9.42$.

\clearpage
 
 \begin{deluxetable}{lcccccl}
 \tablecolumns{7}
 \tabletypesize{\scriptsize}
 \tablecaption{Properties of Evolutionary Models}
 \setlength{\tabcolsep}{0.03in}
 \tablehead{
            \colhead{Model}               &
 	   \colhead{H. Burning}          &
            \colhead{H. Burning}          &
            \colhead{H. Burning}          &
            \colhead{Metallicity\tablenotemark{a}}         &
            \colhead{min. Stellar }       &
 	   \colhead{Atmospheric}         \\
            \colhead{ }                   &
            \colhead{min. Mass ($M_{\sun}$)} &
 	   \colhead{min. $T_{eff}$ (K) }    &
            \colhead{min. $Log(L/L_{\odot})$} &
            \colhead{$Z/Z_{\sun}$}           &
            \colhead{Radius ($R/R_{\sun}$) }   &            
            \colhead{Properties}          }
 \startdata
 \citet{Burrowsetal1993,Burrowsetal1997}  &  0.0767       & 1747        & $-$4.21       & 1.28    & 0.085          & gray with grains \\
 \citet{Burrowsetal1993}                  &  0.094        & 3630        & $-$2.90       & 0.00    & 0.090          & metal free  \\
 \citet{Baraffeetal1998}                  &  $\sim$0.072  & 1700        & $-$4.26       & 1.28    & 0.085          & non-gray without grains \\
 \citet{Chabrieretal2000}                 &  $\sim$0.070  & 1550        & $-$4.42       & 1.28    & 0.086          & ``DUSTY'' grains do not settle \\
 \citet{Baraffeetal2003}                  &  $\sim$0.072  & 1560        & $-$4.47       & 1.28    & 0.081          & ``COND'' clear \& metal depleted \\
  Our Results                             &  \nodata      & $\sim$2075  & $\sim$ $-$3.9 & \nodata & $\sim$0.086    & \nodata        \\
 \enddata
 
 \tablenotetext{a}{Models with $Z/Z_{\sun}=1.28$ were originally meant as solar metallicity models. The new value takes into account
 the revised solar metallicities of \citet{Caffauetal2011}.}
 \end{deluxetable}
\clearpage
 

\begin{figure}[ht]
\begin{center}
   \subfigure[]
      {\includegraphics[scale=0.4, angle=-90]{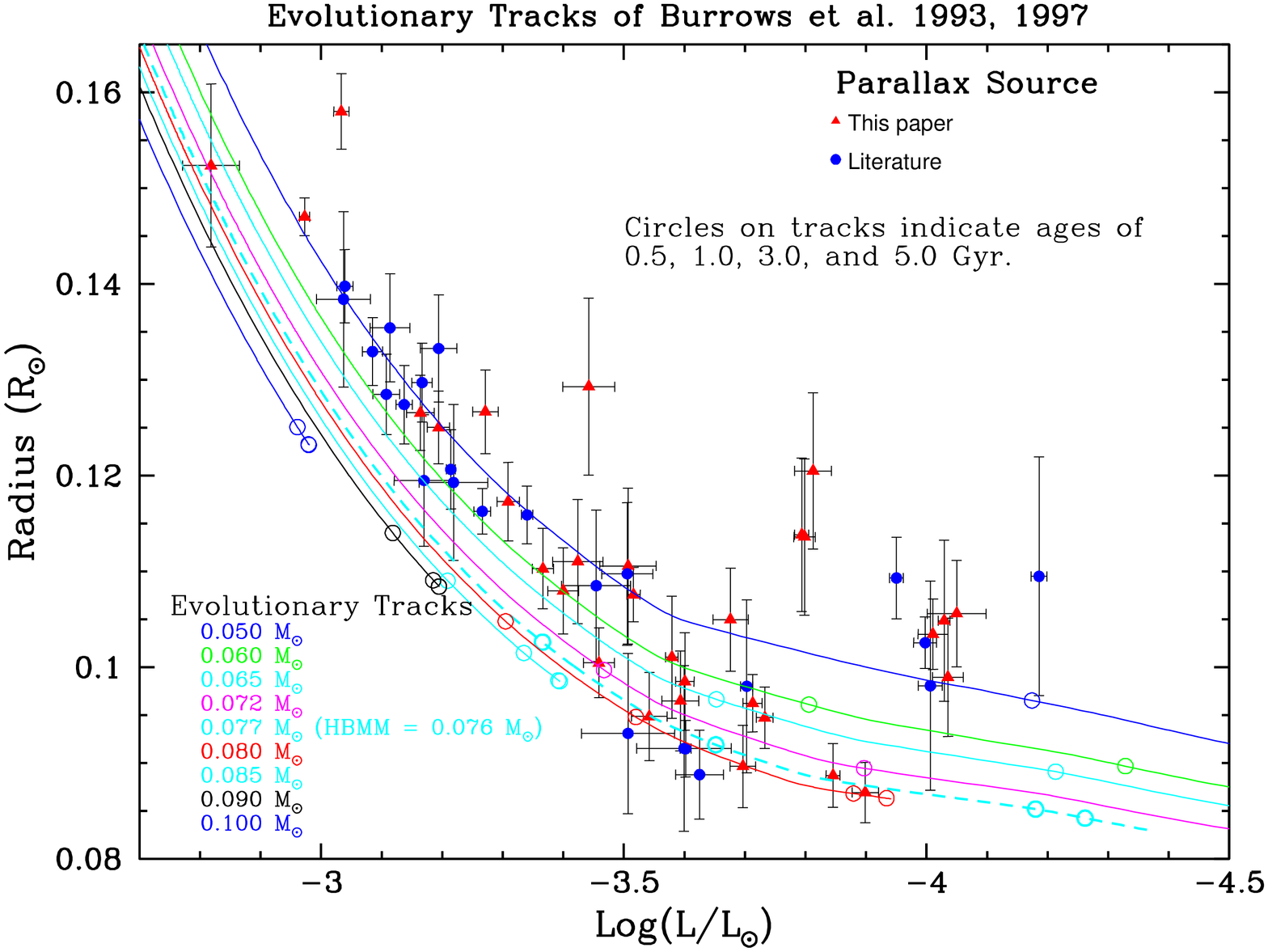}}
\subfigure[]
      {\includegraphics[scale=0.4, angle=-90]{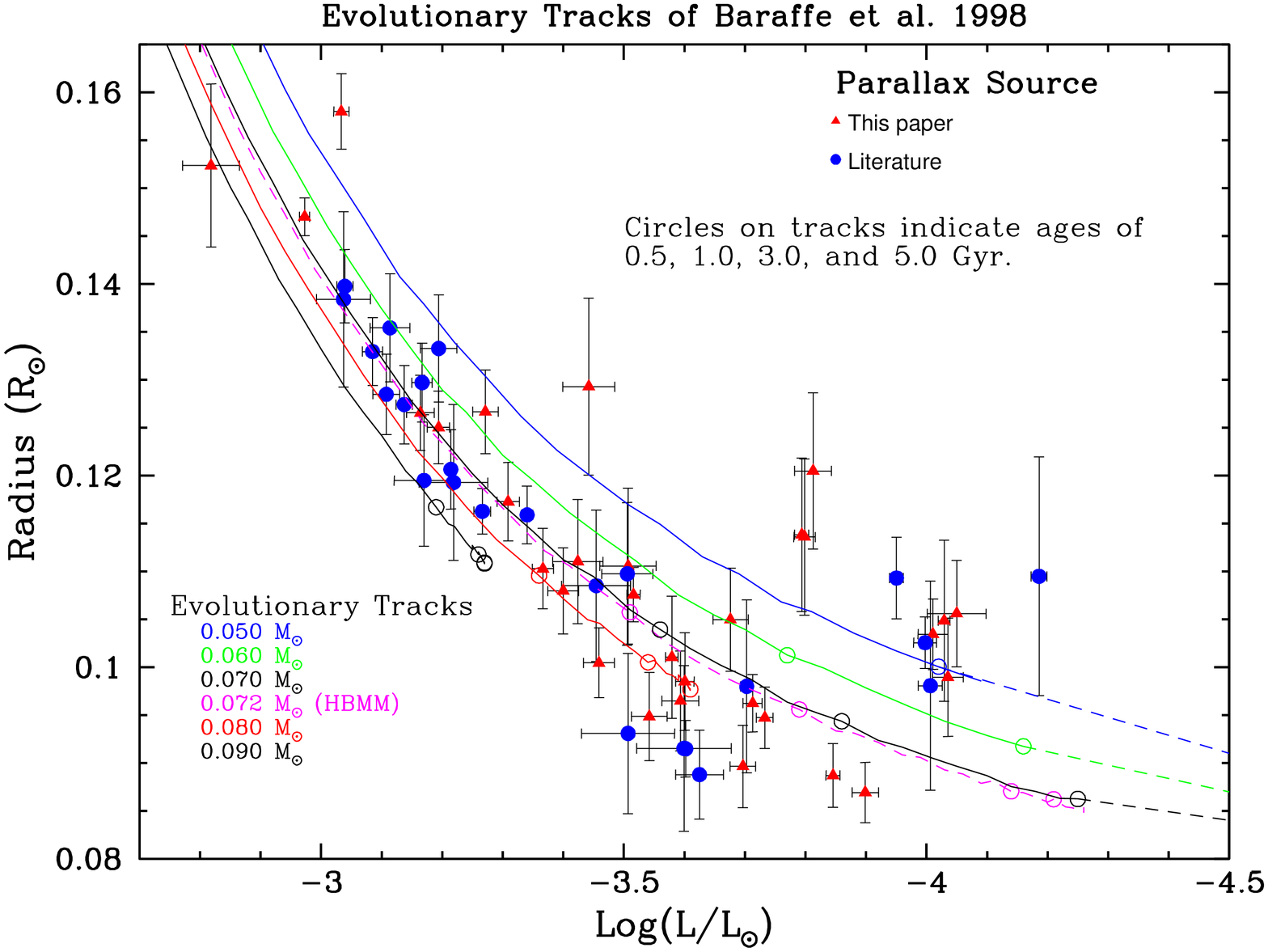}}
 \figcaption[figure10]{\scriptsize  Evolutionary tracks for the models of
(a) \citet{Burrowsetal1997,Burrowsetal1993}  and  (b) \citet{Baraffeetal1998}
over-plotted on the luminosity-radius diagram.  Dashed lines indicate the
continuation of substellar evolutionary tracks where no data are available.
The open circles on the evolutionary tracks represent ages of 0.5, 1.0, 3.0,
and 5.0 Gyr from left to right, with the circles for older ages not in plotting range in some of the
substellar tracks. The circles for older ages overlap each other in the stellar
tracks because there is little evolution at those ages. 
The track corresponding to the hydrogen burning minimum mass is plotted
with a dashed line and has its properties summarized in Table 8.
These diagrams are best seen in color in the online version of the journal.}
\end{center}
\end{figure}
\clearpage

\begin{figure}[ht]
\begin{center}
   \subfigure[]
      {\includegraphics[scale=0.4, angle=-90]{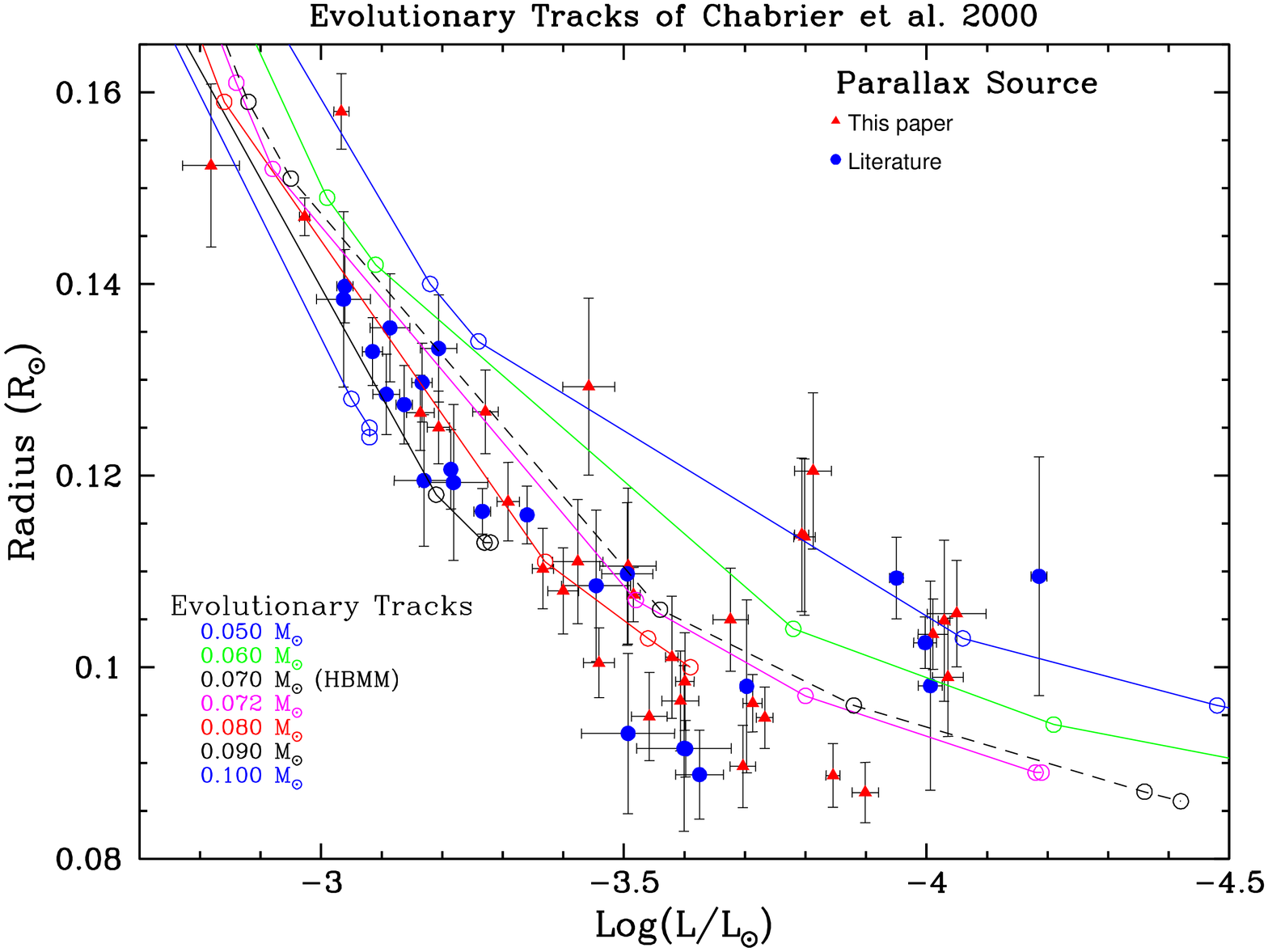}}
\subfigure[]
      {\includegraphics[scale=0.4, angle=-90]{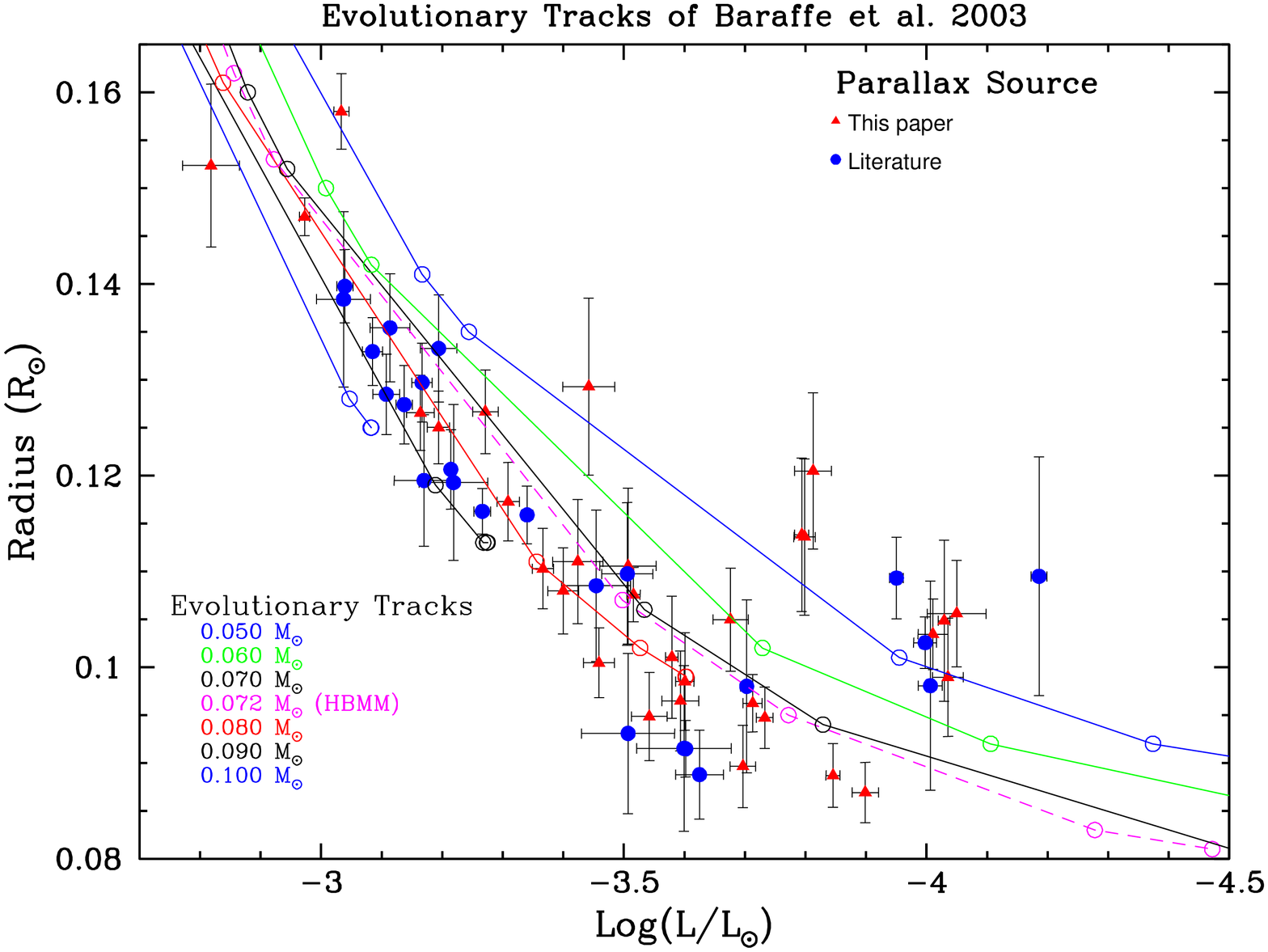}}
 \figcaption[figure10]{\scriptsize  Evolutionary tracks for the models of (a)
\citet{Chabrieretal2000} and (b) \citet{Baraffeetal2003}
over-plotted on the luminosity-radius diagram. Open dots represent ages of 0.05, 0.10, 0.12,
0.50, 1.00, and 10.0 Gyr, except for the 0.10 $M_{\sun}$ track, which starts at 0.10 Gyr.
The circles for older ages are not in the plotting range in some of the
substellar tracks. The circles for older ages overlap each other in the stellar
tracks because there is little evolution at those ages. 
The track corresponding to the hydrogen burning minimum mass is plotted
with a dashed line and has its properties summarized in Table 8.
The models were computed only at the values where open dot are plotted,
with lines connecting the open dots for visualization purposes only.
This diagram is best visualized in color in the online version of the journal.
}
\end{center}
\end{figure}

\clearpage


\begin{figure}[ht]
\begin{center}
   \subfigure[]
      {\includegraphics[scale=0.4, angle=-90]{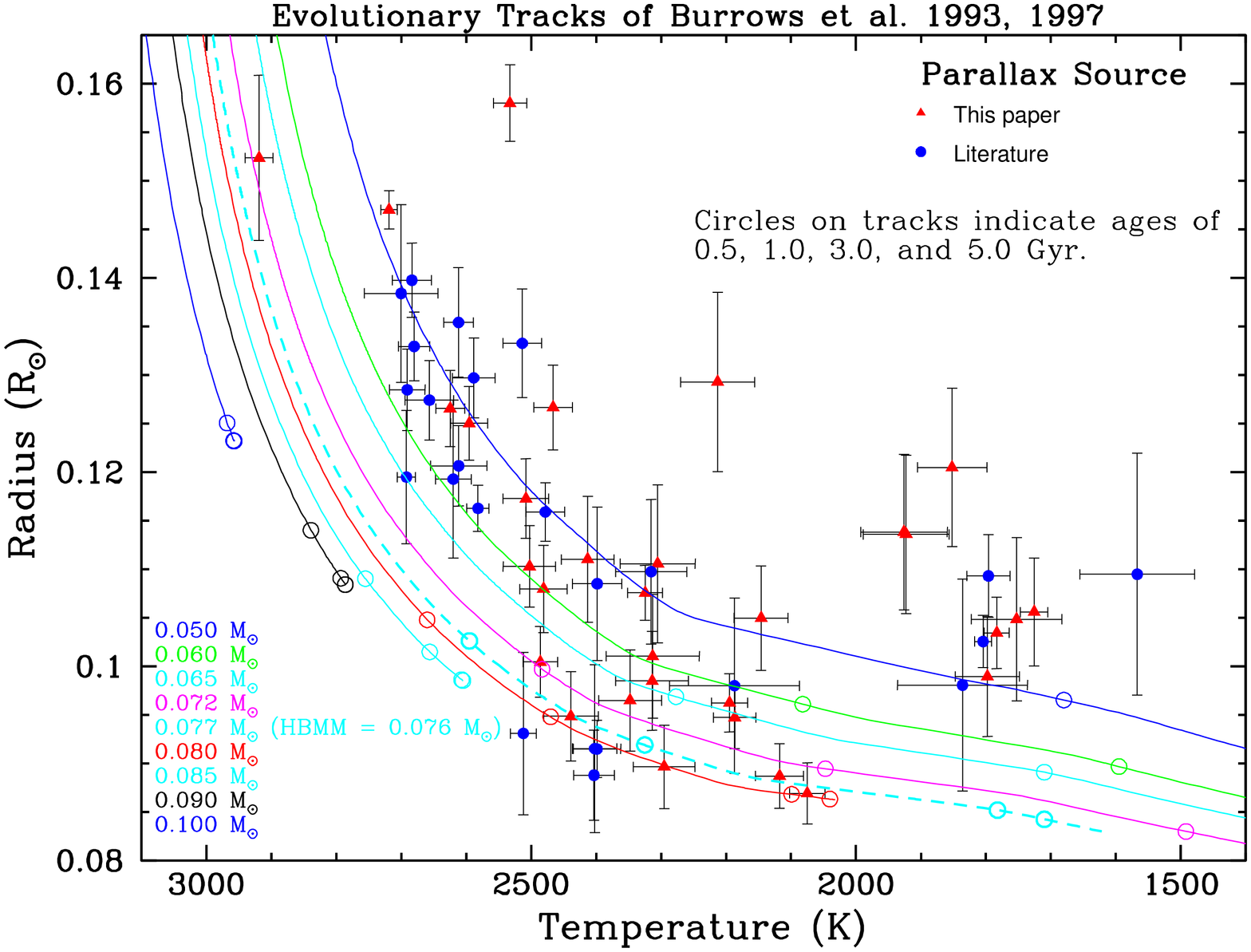}}
\subfigure[]
      {\includegraphics[scale=0.4, angle=-90]{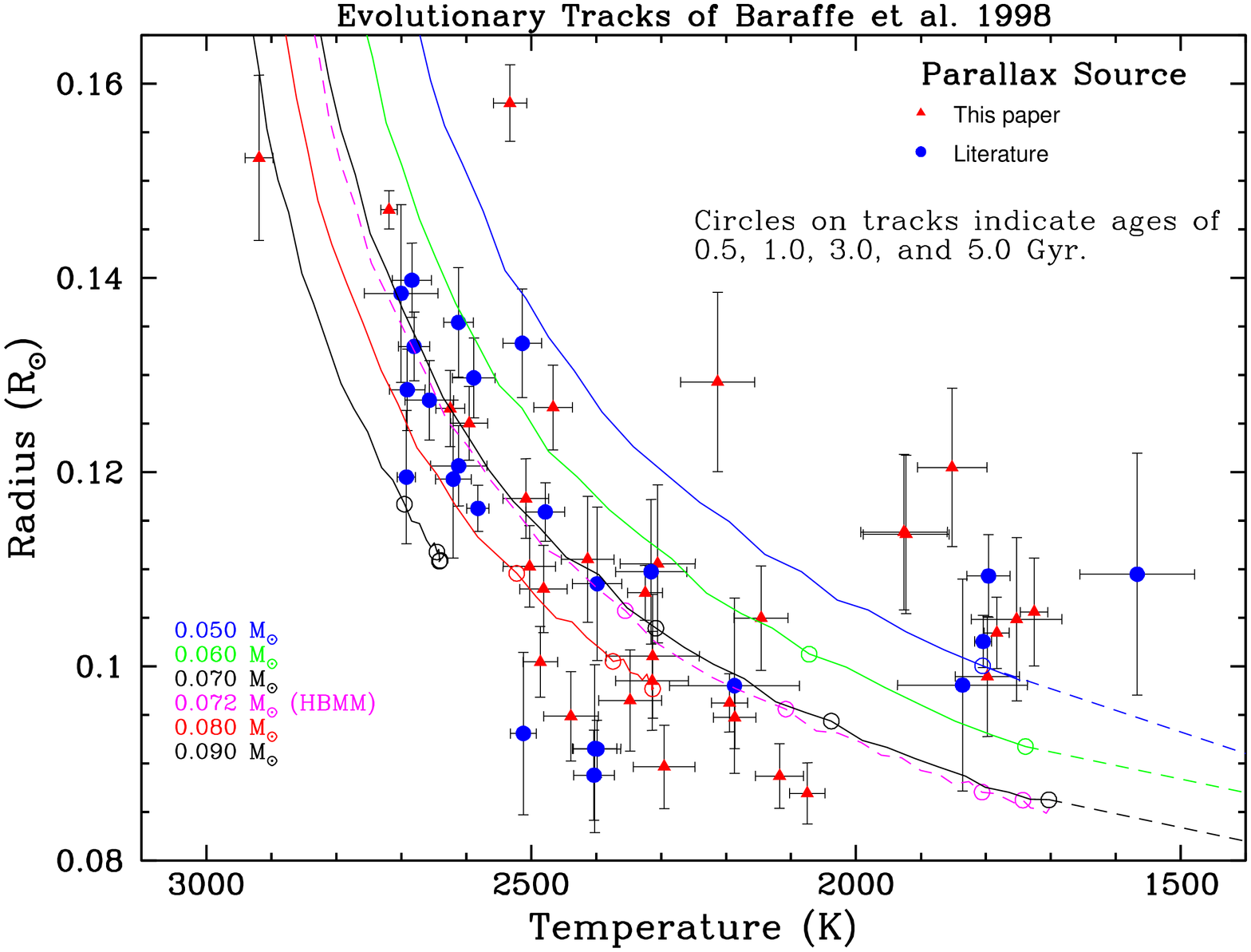}}
 \figcaption[figure10]{\scriptsize  Evolutionary tracks for the models of
(a) \citet{Burrowsetal1997,Burrowsetal1993}  and (b) \citet{Baraffeetal1998}
over-plotted on the temperature-radius diagram.  Dashed lines indicate the
continuation of substellar evolutionary tracks where no data are available.
The open circles on the evolutionary tracks represent ages of 0.5, 1.0, 3.0,
and 5.0 Gyr from left to right, with the circles for older ages not in the plotting range in some of the
substellar tracks. The circles for older ages overlap each other in the stellar
tracks because there is little evolution at those ages.
The track corresponding to the hydrogen burning minimum mass is plotted
with a dashed line and has its properties summarized in Table 8. 
These diagrams are best seen in color in the online version of the journal.}
\end{center}
\end{figure}
\clearpage


\begin{figure}[ht]
\begin{center}
   \subfigure[]
      {\includegraphics[scale=0.4, angle=-90]{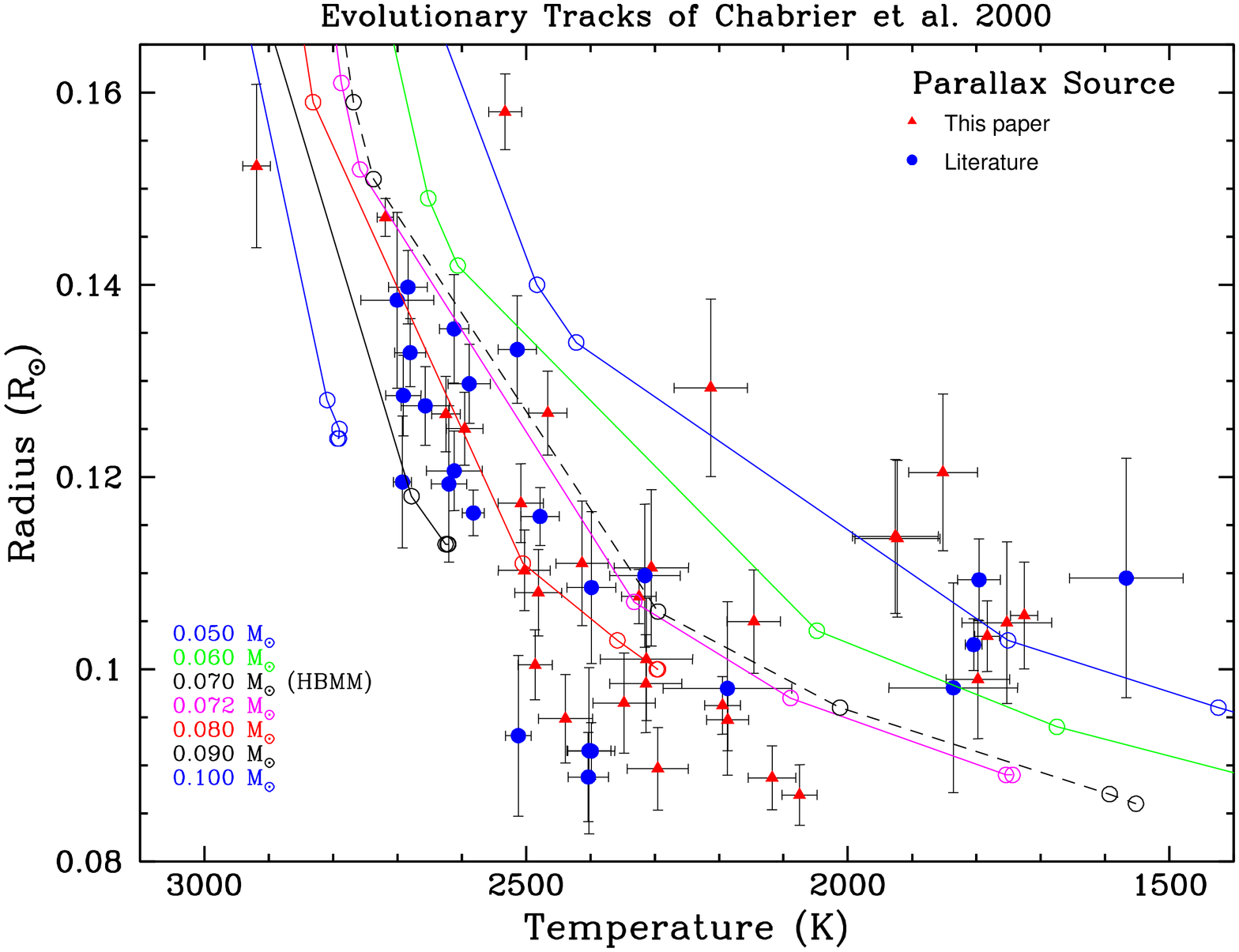}}
\subfigure[]
      {\includegraphics[scale=0.4, angle=-90]{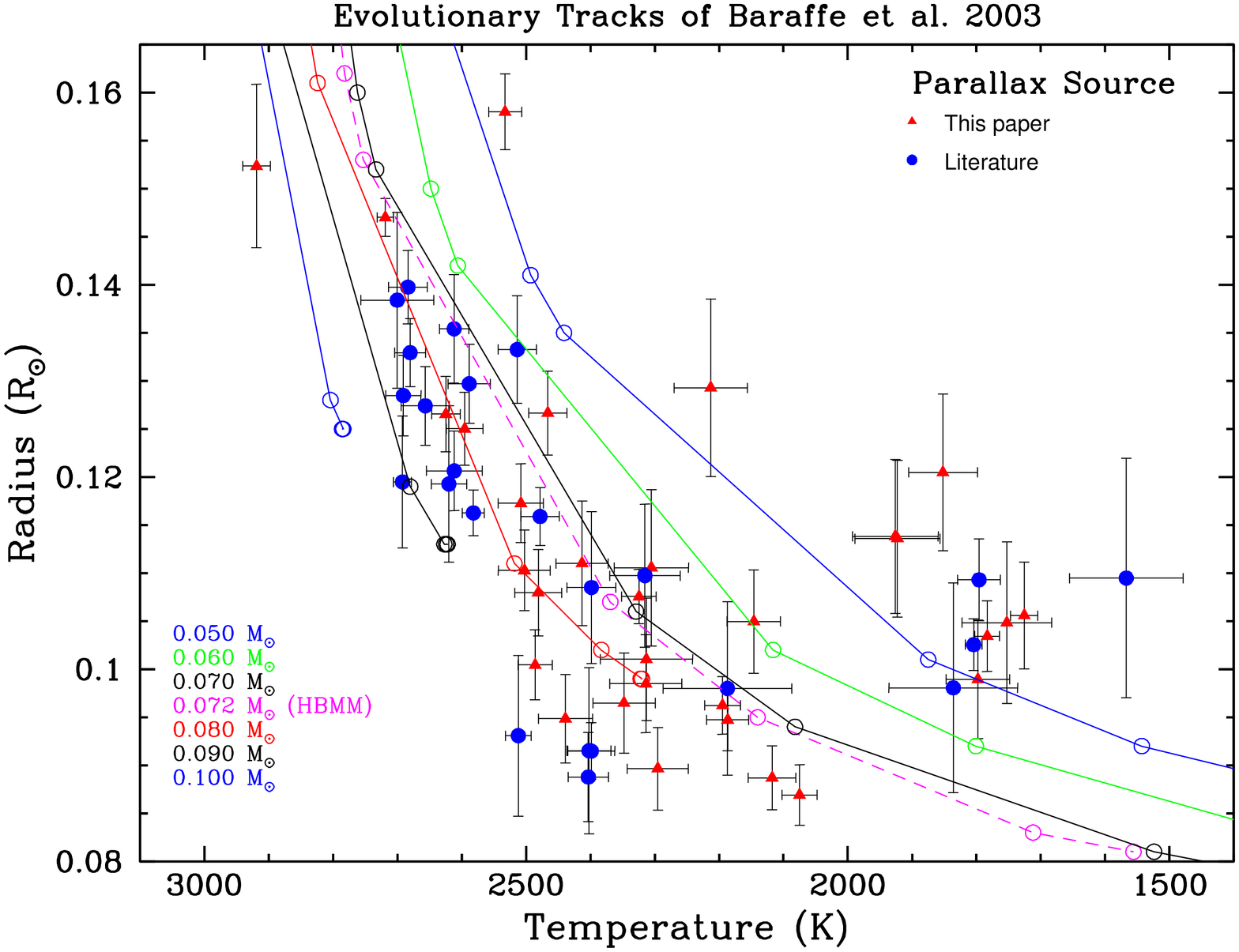}}
 \figcaption[figure10]{\scriptsize  Evolutionary tracks for the models of
(a) \citet{Chabrieretal2000} and (b) \citet{Baraffeetal2003}
over-plotted on the temperature-radius diagram. Open dots represent ages of 0.05, 0.10, 0.12,
0.50, 1.00, and 10.0 Gyr, except for the 0.10 $M_{\sun}$ track, which starts at 0.10 Gyr.
The circles for older ages are not in the plotting range in some of the
substellar tracks. The circles for older ages overlap each other in the stellar
tracks because there is little evolution at those ages. 
The track corresponding to the hydrogen burning minimum mass is plotted
with a dashed line and has its properties summarized on Table 8.
The models were computed only at the values where open dot are plotted,
with lines connecting the open dots for visualization purposes only.
This diagram is best visualized in color in the online version of the journal.
}
\end{center}
\end{figure}

\clearpage

\subsection{Comparison of Radii With Other Studies}
\label{subsec:radiuscomp}

Unfortunately, there are only a few other observational studies that
directly measure or calculate radii for objects in the temperature
range considered here.  These objects are too faint to be observed by
the {\it Kepler} mission except as companions to more massive stars.
Their faintness also means that they are likely to remain outside the
domain of long baseline optical interferometry for the foreseeable
future.  There are nevertheless several examples of {\it VLM}
eclipsing binary companions where the primary star in the system is an early M
dwarf or a solar analogue \citep[e.g.,][and references
therein]{Burrowsetal2011}. Such systems are valuable for comparisons
regarding mass and radius, but lack the photometric coverage needed to
calibrate the {\it SED} and derive the luminosity in a manner
analogous to this work. We note that the only known eclipsing system
where both members are brown dwarfs \citep{Stassunetal2006} is a member of
the Orion star forming region, and is therefore only a few million years old.
\citet{Stassunetal2006} measure radii of 0.669$\pm$0.034$R_{\sun}$ and 
0.511$\pm$0.026$R_{\sun}$ for the two components. At such a young age and such
large radii, this system is a valuable probe of early substellar evolution,
but should not be compared to the much older objects we discuss in this
study.

There have been two recent studies that derive the stellar parameters
needed for placing objects in the {\it HR} diagram. As already
mentioned, \citet{Konopackyetal2010} derived effective temperatures
that agree with our values for early L dwarfs but steadily diverge as
the temperature increases ($\S$\ref{subsec:Teff}, Figure 7), and their errors are $\sim$200K.  Although
their data are limited at temperatures cooler than $\sim$2000K for
the determination of a robust radius trend, they also have a local
minimum in radius at $T_{eff} = 2075K$, for 2MASS J2140+16B, consistent with
our results.  More recently, \citet{Sorahanaetal2013}
derived radii for several L and T dwarfs based on {\it AKARI} near
infrared spectra. They report a sharp radius minimum of 0.064
R$_{\sun}$ at 1800K. Figure 16 shows their results over-plotted in our
temperature-radius plot.  There are several reasons why the results
of \citet{Sorahanaetal2013} deserve further scrutiny. First, it is
difficult to imagine the cause of such a sudden drop in radius at L
dwarf temperatures, and there are questions as to whether or not such
high densities can be accommodated by any reasonable equation of state
\citep[e.g.,][]{SaumonChabrierandvanHorn1995}. Also, they attempt to
derive the {\it SED} based on near infrared spectra alone, covering
wavelengths from 1$\micron$ to 5$\micron$ only. Finally, we note that
their effective temperatures agree to other studies for most objects
but are higher by as much as a few hundred K when compared to
\citet{Golimowskietal2004a} and \citet{Cushingetal2008} for objects
corresponding to the sharp drop in radius. In $\S$\ref{subsec:Teff} we
discussed the importance of optical and mid-infrared data when
deriving effective temperatures.  Because \citet{Sorahanaetal2013}
do not use mid-infrared or optical data, their results should be
approached with caution.

 
 \begin{figure}[ht]
 \begin{center}
        \includegraphics[scale=0.5, angle=-90]{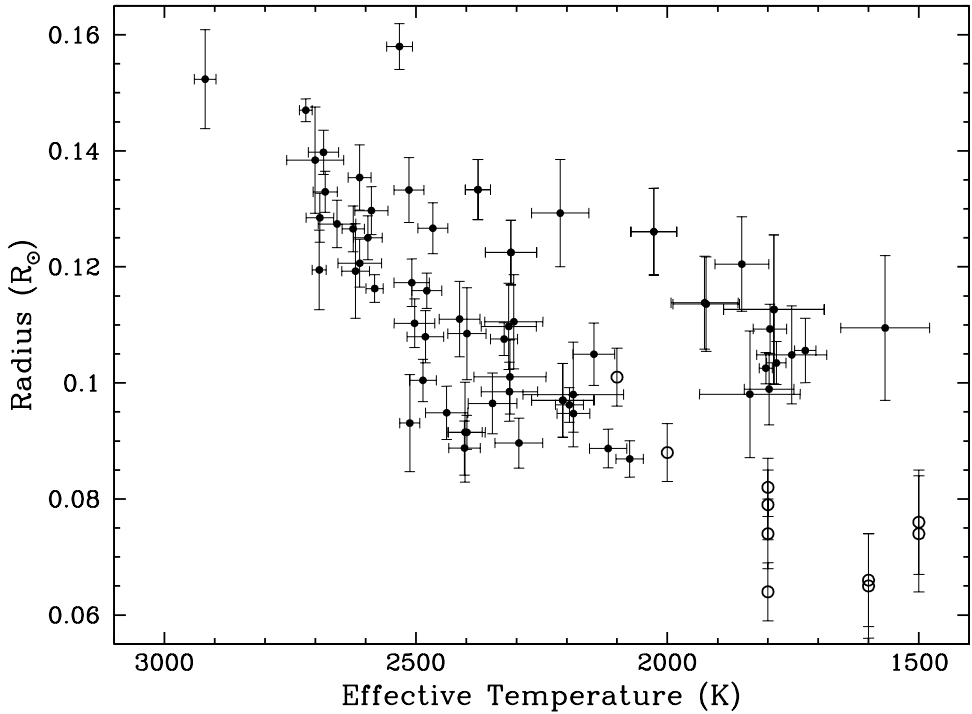}
        \figcaption[Figure16]{\scriptsize Data from \citet{Sorahanaetal2013}
 (open circles) over-plotted on our temperature-radius diagram. Their radius
 minimum at 1800K is probably a result of their unrealistically high temperatures
 for these objects.}
 \end{center}
 \end{figure}
 
\section{Notes on Individual Objects}
\label{sec:individual}

{\bf GJ 1001BC (L4.5 ID\# 1)} is a binary L dwarf with nearly equal
luminosity components
\citep{Golimowskietal2004b}. \citet{Golimowskietal2006,Golimowskietal2007}
report a preliminary total system dynamical mass of 0.10$M_{\sun}$
based on orbital mapping using {\it HST} and {\it VLT}.  The conservative assumption
of a mass ratio $\geq$3:2 based on nearly equal luminosity would make
individual masses range between 0.04$M_{\sun}$ and 0.06$M_{\sun}$, thus
placing both objects in the brown dwarf regime. We derive
$T_{eff}=1725\pm21$ and $log(L/L_{\sun})=-4.049\pm0.48$ for each
component, assuming the two objects are identical. These numbers are generally above the
 hydrogen burning limit
numbers predicted by models but below our numbers (Table 7).  This
inconsistency is further evidence that the hydrogen burning limit must
happen at higher luminosities and temperatures than what is predicted
by the currently accepted models.

{\bf LEHPM1-0494 A (M6.0V ID\# 3)} and {\bf B (M9.5V ID\# 2)} are
reported by \citet{Caballero2007} to be a wide common
proper motion binary with separation of 78$\arcsec$.  We report trigonometric parallaxes for both
components based on individual reductions of the same field of view,
and derive distances of 26.88$^{+1.51}_{-1.36}$ pc for the A component
and 25.14$^{+1.40}_{-1.26}$ pc for the B component for a projected separation
of $\sim$2100 AU. These
trigonometric distances are in good agreement with Caballero's
distance estimate of 23$\pm$2 pc and support his claim of a physical
association between these two objects.

{\bf LHS 1604 (M7.5V ID\# 12)} was first reported by
\citet{Cruzetal2007} as being over-luminous by $\sim$0.6 magnitudes in
$J$. They suggested that the near-infrared photometry is consistent with
an unresolved M7.5V/M9.0V binary. LHS 1604 is the only star in our
sample for which we were not able to calculate $T_{eff}$ or perform an
{\it SED} fit using the procedures outlined in $\S$\ref{sec:tefflum} $-$ the fits diverged due to 
a large infrared excess.
We observed LHS 1604 using high resolution laser guide star adaptive
optics on Gemini North and preliminary results do not show a resolved
companion. We defer a thorough analysis of this target to a future
publication where we discuss our high resolution observations and use
them to place limits on the properties of the putative companion
(Dieterich et al. in preparation).  We are also monitoring LHS 1604 for
astrometric perturbations but it is too early to notice any trends.

{\bf 2MASS J0451-3402 (L0.5 ID\# 15)} has the highest photometric
variability in our sample.  It was first noted as a photometrically
variable target by \citet{Koen2004}, who reported a sinusoidal trend with
a period of 3.454 days and mean amplitude of $\sim$1\% (10 milli-magnitudes), though varying
to as high as 4\% (40 milli-magnitudes).  While our observations do not have the cadance
necessary to obtain phase information, the variability of 51
milli-magnitudes in the {\it I} band we detect is in agreement, if not
somewhat higher, to that of \citet{Koen2004}. It is interesting to
note the spike in variability around $T_{eff}\sim2100\;K$ in Figure
9. further investigation is needed to determine whether this trend has
a physical cause associated with that temperature range or whether
this is a coincidence.

{\bf 2MASS J0523-1403 ( L2.5 ID\# 17)} is discussed throughout this
paper as the object closest to the local minimum in the
luminosity-radius and temperature-radius trends (Figure 11). As we
discussed in $\S$\ref{sec:discussion}, there is strong evidence
indicating that the end of the stellar main sequence must lie in its
proximity in parameter space. The target has been described as having
variable radio and H$\alpha$ emission
\citep{Berger2002,Antonovaetal2007, Bergeretal2010}. Despite the
common association between H$\alpha$ emission and youth, we note that
it is difficult to conceive of a target with such a small radius
($R/R_{\sun} = 0.086\pm.0031$) as being young. As discussed in
$\S$\ref{subsec:variability}, radio emission is often used as a probe
of magnetic fields, and may be accompanied by optical variability if
they result in auroral phenomena. We detect no significant $I$ band
variability for 2MASS J0523-1403 (upper limit $\sim$11.7 milli-magnitudes ), 
meaning that either the star was in
a mostly quiescent state during the $\sim$3 years for which we
monitored the target (2010.98$-$2013.12) or that the link between
radio emission and $I$ band variability is not universal.

{\bf SSSPM J0829-1309 ( L1.0 ID\# 23)} is an object very similar to
2MASS J0523-1403 but slightly more luminous. The two objects have
1$\sigma$ uncertainties that overlap in radius and $T_{eff}$,
but not luminosity.  As shown in Figure 11, the location of SSSPM
J0829-1309 is crucial for establishing 2MASS J0523-1403 as being close
to the minimum of the radius trends. Taken together, 2MASS J0523-1403
and SSSPM J0829-1309 show that the radius trends in Figure 11 are real
and therefore the conclusions we draw in this paper are not the result
of one isolated odd object (i.e. 2MASS J0523-1403).

{\bf LHS 2397aAB (M8.5V (joint) ID\# 35)} is an
M8.0V/L7.5\footnote{infrared spectral type for secondary} binary
\citep{FreedCloseandSiegler2003}. \citet{DupuyLiuandIreland2009}
report a total system dynamical mass of 0.146$^{+0.015}_{-0.013}$
$M_{\sun}$.  \citet{Konopackyetal2010} derive individual dynamical
masses of 0.09$\pm$0.06 $M_{\sun}$ for the primary and 0.06$\pm$0.05
$M_{\sun}$ for the secondary.  The system is therefore an important
probe of the hydrogen burning mass limit because two coeval components
presumably with the same metallicity lie on opposite sides of the
stellar/substellar boundary. We are mapping the astrometric orbit for
this system in a manner similar to that discussed in
$\S$\ref{subsec:den1454} for DENIS J1454-6604AB and will publish
refined individual dynamical masses as soon as orbital mapping is
complete.

{\bf LEHPM2-0174 (M6.5V ID\# 40)} appears over-luminous in Figure 4.
It is most likely an unresolved multiple, a young object, or both. We note
that we could not determine a reliable source for the spectral type
of this object, thus leaving open the possibility that it has been miss-characterized
as an M6.5V. LEHPM2-0174 is excluded from Figure 11 because scaling the figure
to fit its radius (0.173$R_{\sun}$) would make the figure difficult to read.

{\bf Kelu-1AB (L2.0 (joint) ID\# 41)} is a well known L2/L4 binary
\citep{LiuandLegget2005}. That study notes that the presence of
Li$\lambda$6708 makes both components substellar with masses
$\lesssim$ 0.06 $M_{\sun}$ according to the lithium test of
\citet{Reboloetal1992}, although they note that the Li$\lambda$6708
detection is tenuous. Deconvolution of this system would
provide important information about the hydrogen burning limit due to
its location in the temperature-radius trend (Figure 11b). If we assume
that the system is an equal luminosity binary, then the deconvolved
radii of the components are $\sim$0.089 $M_{\sun}$. That number would
further constrain the position of 2MASS J0523-1309 as being in the
minimum of the radius trend.  However, because the components of
Kelu1-AB do {\it not} have equal luminosities, we can expect the A
component to be a more massive brown dwarf or a stellar component with
mass just above the hydrogen burning limit.  In either case, the A
component would have a smaller radius than the B component.
Determining the precise radius, $T_{eff}$ and luminosity of the A
component is crucial for determining the exact location of the point
of minimal radius in Figure 11. 

{\bf 2MASS J1705-0516AB (L0.5 (joint) ID\# 56)} was first reported as
an M9V/L3 binary by \citet{Reidetal2006}. The system's position in
the midst of the main sequence in the {\it HR} diagram (Figure 4) shows that
the system is dominated by the A component in luminosity.
Our parallax observations detect a clear astrometric
perturbation. We are working on mapping the system's orbit and will
soon be able to publish dynamical masses for the individual
components.  Like LHS 2397aAB, this system will serve as a crucial
benchmark system with components likely residing on either side of
the stellar/substellar boundary. As indicated in Figure 9, this target has one
of the largest optical variabilities in the sample, at 41 milli-magnitudes in $I$.
We defer a more thorough discussion of 2MASS J1705-0516AB to a future paper (Dieterich et
al. in preparation).

{\bf SIPS J2045-6332 (M9.0V ID\# 58)} is an extremely over-luminous
object (Figure 4).  We note that unresolved equal luminosity duplicity
alone cannot explain the over-luminosity.  The object is also highly
variable at 39 milli-magnitudes in {\it I}, as shown in Figure 9.
 The variability suggests that youth may
play a role in explaining the over-luminosity of SIPS J2045-6332.

{\bf LHS 4039C (M9.0V ID\# 62)} is a member of a triple system with an
M4V$-$M9V binary with separation 6\arcsec\~ and a DA white dwarf 103\arcsec\~ away
\citep{Scholzetal2004,Subasavageetal2009}. \citet{Subasavageetal2009}
reported the trigonometric parallax for the white dwarf component as
42.82$\pm$2.40 mas. In this paper we have reduced the same data using
LHS 4039C as the science target and measure a parallax of
44.38$\pm$2.09, thus supporting the physical association of the
system. The intriguing combination of a white dwarf and a {\it VLM}
star in the same system allows us to constrain the properties of LHS
4039C based on the better understood models of white dwarf
evolution. Based on the white dwarf cooling time of 0.81$\pm$0.05 Gyr
\citep{Subasavageetal2009} and the progenitor age of 4.4$\pm$3.7 Gyr \citep{IbenandLaughlin1989}
assuming a progenitor mass of 1.17$\pm$0.26$M_{\sun}$ \citep{Williamsetal2009},
we infer a total system age of 5.2$\pm$3.7 Gyr. 
Assuming the system to be coeval, LHS4039C is then a main
sequence star with no remaining traces of youth.  Its locus on the {\it HR}
diagram is therefore an
indication of  where the {\it VLM} stellar main sequence lies\footnote{
It was not possible to label this object in Figure 4 due to crowding of the diagram.
The reader is referred to the online supplements where diagrams are plotted using ID numbers.}.

\section{Conclusions and Future Work}
\label{sec:conclusion}
We have determined fundamental properties (effective
temperatures, luminosities, and radii) based on a photometric and
astrometric survey of 63 targets with spectral types ranging from M6V
to L4, and used the data to construct an {\it HR} diagram of the
stellar/substellar boundary. We find strong evidence for the local
minimum in radius signaling the stellar/substellar boundary close to
the locus of 2MASS J0523-1403 at $T_{eff}\sim2075K$,
$(R/R_{\sun})\sim0.086$, and $log(L/L_{\sun})\sim-3.9$. The two panels of Figure 11
present the evidence for the local minimum in the radius trends as functions of luminosity and temperature.
While our sample is not volume complete, it covers the photometric
color range from M6V to L4 in a continuous manner.  As we discussed in
$\S$\ref{sec:discussion}, our interpretation of the radius trends leaves
little chance for the discovery of stellar objects at temperatures
cooler than $\sim$2,000K.

Our plans for the future include making the sample volume-complete so
that population properties such as the mass and luminosity functions
can be studied with more rigor. We have already started a volume-complete
astrometric search of all southern systems with primaries having  spectral
types ranging from M3V to L5 within 15 pc and would like to extend the
search to 20 pc and to L7. One of the fundamental questions that this larger
volume-complete search will answer is whether or not the gap we see in
Figure 11 after 2MASS J0523-1403 is real or whether it is an effect of
statistics of small numbers.  As we discussed in
$\S$\ref{sec:discussion}, the existence of a gap immediately after the
onset of the brown dwarf cooling curve is a natural consequence of the
fact that only very massive brown dwarfs can occupy that parameter
space and do so for a small fraction of their lifetimes. As discussed in $\S$\ref{sec:sample},
19 targets are still undergoing parallax observations and will be ready
for placement in the {\it HR} diagram during the next few years. These targets
are mostly L dwarfs. These additional targets constitute a powerful test
of the ideas we discussed in this work $-$ if they follow the same trends,
they will provide independent confirmation of our conclusions.
We also plan to perform higher cadence variability studies on targets
with effective temperatures $\gtrsim$2100K to investigate the spike in variability we notice
for targets just above the hydrogen burning limit ($\S\S$\ref{subsec:variability}, 
\ref{sec:individual}, Figure 9). 

In this paper we have addressed the question: {\it ``What do objects at the stellar/substellar
boundary look like to an observer?''} We next plan to 
populate the {\it HR} diagram with dynamical masses for systems such
as GJ 1001BC, LHS 2397aAB, 2MASS J1705-0516AB ($\S$\ref{sec:individual})
as well as the newly discovered binary DENIS J1454-6604AB ($\S$\ref{subsec:den1454}).
Only then we will be able to answer the question:
{\it ``What are the masses of objects at the stellar/substellar boundary?''}
Along with the answer to the first question, we hope that this work will
bring us closer to a complete and thorough understanding of (sub)stellar
structure and evolution at the stellar/substellar boundary.

This research was supported by NSF grants AST-0507711, AST-09-08402, and AST-11-09445.
{\it CTIO} 0.9m observations were made possible through the {\it SMARTS} Consortium.
S. B. D. acknowledges travel support for {\it SOAR} observing runs from 
NOAO's graduate student dissertation support program.
We thank Michael Bessell, Adam Burrows, Adam Burgasser, Douglas Gies, and Russel White
for useful discussions. We also thank the anonymous referee for suggestions
that increased the quality of the paper.
The authors thank the staff of Cerro Tololo Inter-American Observatory
for welcoming us into their country and for their continuous help and support.
We are especially grateful to Sean Points, {\it SOAR/SOI} Instrument Scientist,
for his help and to Steven Heathcote, {\it SOAR} Director, for allowing us
to stand by on engineering nights and use any remaining time after engineering
activities were concluded.
This publication makes use of data products from the Two Micron All
Sky Survey, which is a joint project of the University of
Massachusetts and the Infrared Processing and Analysis
Center/California Institute of Technology, funded by the National
Aeronautics and Space Administration and the National Science
Foundation. This publication makes use of data products from the
Wide-field Infrared Survey Explorer, which is a joint project of the
University of California, Los Angeles, and the Jet Propulsion
Laboratory/California Institute of Technology, funded by the National
Aeronautics and Space Administration.  This research has made use of
the SIMBAD database, operated at CDS, Strasbourg, France. This
research has made use of NASA's Astrophysics Data System.
This research has benefited from the M, L, T, and Y dwarf 
compendium housed at DwarfArchives.org.

\bibliographystyle{apj}
\bibliography{HRreferences}

\end{document}